\documentclass[epj,nopacs]{svjour}
\usepackage{amsmath,amsbsy}
\usepackage{graphicx}
\usepackage{pennames}
\usepackage{fancyhdr}
\usepackage{longtable}
\begin{document}
%
\newcommand{\mv}{\,\text{mV}}
\newcommand{\V}{\,\text{V}}
\newcommand{\mG}{\,\text{mG}}
\newcommand{\nm}{\,\text{nm}}
\newcommand{\mm}{\,\text{mm}}
\newcommand{\cm}{\,\text{cm}}
\newcommand{\m}{\,\text{m}}
\newcommand{\km}{\,\text{km}}
\newcommand{\mpc}{\,\text{Mpc}}
\newcommand{\kel}{\,^\circ K}
\newcommand{\sr}{\,\text{sr}}
\newcommand{\pmt}{\it PMT}
\newcommand{\phe}{\it pe}
\newcommand{\y}{\,\text{y}}
\newcommand{\days}{\,\text{d}}
\newcommand{\hou}{\,\text{h}}
\newcommand{\minu}{\,\text{m}}
\newcommand{\s}{\,\text{s}}
\newcommand{\usec}{\,\mu{\text s}}
\newcommand{\msec}{\,\text{ms}}
\newcommand{\nsec}{\,\text{ns}}
\newcommand{\psec}{\,\text{ps}}
\newcommand{\hz}{\,\text{Hz}}
\newcommand{\khz}{\,\text{kHz}}
\newcommand{\Mhz}{\,\text{MHz}}
\newcommand{\mg}{\,\text{mg}}
\newcommand{\lit}{\,\text{l}}
\newcommand{\gra}{\,\text{g}}
\newcommand{\kg}{\,\text{kg}}
\newcommand{\ev}{\,\text{eV}}
\newcommand{\kev}{\,\text{keV}}
\newcommand{\mev}{\,\text{MeV}}
\newcommand{\gev}{\,\text{GeV}}
\newcommand{\mw}{\,\text{MW}}
\newcommand{\gw}{\,\text{GW}}
\newcommand{\celsi}{\,^\circ\text{C}}
\newcommand{\qq}{\text q}
\newcommand{\qu}{\text u}
\newcommand{\qd}{\text d}
\newcommand{\nux}{\nu_x}
\newcommand{\anux}{\bar{\nu}_x}
\def\isotope#1{\mbox{${}^{#1}$}}                  
\def\ra{\rightarrow}
\def\lra{\leftrightarrow}
\def\units#1{\hbox{$\,{\rm #1}$}}                
\def\dmsq{\delta m^{2}}
\def\sinsq{{\rm sin}^2 2\theta}
\def\Reines{$\overline{\nu}_e\,+\,p\,\rightarrow\,e^+\,+\,n$}
\newcommand{\vect}[1]{\overrightarrow{\sf #1}}
\newcommand{\system}[1]{\left\{\matrix{#1}\right.}
\newcommand{\displayfrac}[2]{\frac{\displaystyle #1}{\displaystyle #2}}
\newcommand{\nucl}[2]{{}^{#1}\mbox{#2}}
\newcommand{\diff}{{\rm\,d}}
\newcommand{\lsim}{\,\lower .5ex\hbox{$\buildrel < \over {\sim}$}\,}
\newcommand{\gsim}{\,\lower .5ex\hbox{$\buildrel > \over {\sim}$}\,}
\newcommand{\apm}[2]{{}^{#1}_{#2}}
\newcommand{\fact}[1]{#1{\mbox !}}
\newcommand{\varr}[2]{\lb\begin{array}{c} #1 \\ #2 \end{array}\rb}
\newcommand{\bd}[1]{\mbox{\mathbf{${#1}$}}}
\newcommand{\alf}[1]{\mbox{#1}}
\newcommand{\rb}{\right)}
\newcommand{\lb}{\left(}
\newcommand{\be}{\begin{equation}}
\newcommand{\ee}{\end{equation}}
\newcommand{\bn}{\bd{\nabla}}
\newcommand{\ii}{\mbox{i}}
\newcommand{\betm}{\beta^-}
\newcommand{\betp}{\beta^+}
\newcommand{\ie}{{\em i.e.} }
\newcommand{\ea}{{\em et al.}}
%
\renewcommand{\sectionmark}[1]{\markright{\thesection.\ #1}}
\lhead[\rmfamily \thepage]{\fancyplain{}{\bfseries \rightmark}}
\rhead[\fancyplain{}{\bfseries \leftmark}]{\rmfamily \thepage}
\cfoot{}
%
\renewcommand{\textfraction}{0.05}
\renewcommand{\floatpagefraction}{0.35}
\renewcommand{\topfraction}{0.95}
\renewcommand{\bottomfraction}{0.95}
\setcounter{totalnumber}{5}
\title{Search for neutrino oscillations on a long base-line at the CHOOZ 
nuclear power station}
\author{
M.~Apollonio\inst{3},
A.~Baldini\inst{2},
C.~Bemporad\inst{2},
E.~Caffau\inst{3},
F.~Cei\inst{2},
Y.~D\'eclais\inst{5}%
\thanks{\emph{Present address:} IPNL-IN2P3-CNRS Lyon}%
H.~de~Kerret\inst{6},
B.~Dieterle\inst{8},
A.~Etenko\inst{4},
L.~Foresti\inst{3},
J.~George\inst{8},
G.~Giannini\inst{3},
M.~Grassi\inst{2},
Y.~Kozlov\inst{4},
W.~Kropp\inst{7},
D.~Kryn\inst{6},
M.~Laiman\inst{5},
C.~E.~Lane\inst{1},
B.~Lefi\`evre\inst{6},
I.~Machulin\inst{4},
A.~Martemyanov\inst{4},
V.~Martemyanov\inst{4},
L.~Mikaelyan\inst{4},
D.~Nicol\`o\inst{2},
M.~Obolensky\inst{6},
R.~Pazzi\inst{2},
G.~Pieri\inst{2},
L.~Price\inst{7},
S.~Riley\inst{7},
R.~Reeder\inst{8},
A.~Sabelnikov\inst{4},
G.~Santin\inst{3},
M.~Skorokhvatov\inst{4},
H.~Sobel\inst{7},
J.~Steele\inst{1},
R.~Steinberg\inst{1},
S.~Sukhotin\inst{4},
S.~Tomshaw\inst{1},
D.~Veron\inst{7},
V.~Vyrodov\inst{7}}
\institute{
Drexel University
\and
INFN and University of Pisa
\and
INFN and University of Trieste
\and
Kurchatov Institute
\and
LAPP-IN2P3-CNRS Annecy
\and
PCC-IN2P3-CNRS Coll\`ege de France
\and
University of California, Irvine
\and
University of New Mexico, Albuquerque}
\date{Received: 18 November 2002 / Revised version: }
%
\abstract{This final article about the CHOOZ experiment 
presents a complete description of the $\Pagne$ source and detector,
the calibration methods and stability checks, the 
event reconstruction procedures and the Monte Carlo simulation. The data 
analysis, systematic effects and the methods used to reach our conclusions are 
fully discussed. 
Some new remarks are presented on the deduction of the confidence limits and on
the correct treatment of systematic errors.}
%
%
\maketitle
\section{Introduction}
Neutrino oscillation experiments are sensitive probes of the possible existence
of a finite neutrino mass and provide a way to study physics
beyond the Standard Model of electroweak interactions~\cite{Sbilenky}. In fact,
lepton flavour
violation and the existence of nonzero neutrino masses can give rise to 
neutrino oscillations, as first pointed out by Pontecorvo~\cite{Ponte1,Ponte2} 
and Maki \ea~\cite{Maki}.
Several experiments, studying solar~\cite{Kirsten2,sksolar,sno1,sno2} or 
atmospheric neutrinos~\cite{kamio1,kamio2,skatm1,skatm2,skatm3,skatmrev}, have 
measured fluxes consistently lower than expectations.
This can be interpreted as due to various forms of neutrino 
oscillations. In particular the so-called ``atmospheric neutrino anomaly'' is 
the observation of a $\Pgngm /\Pgne$ ratio which is roughly one half of what 
expected and its possible explanation might be due to
either oscillation of $\Pgngm \lra \Pgngt$ or to $\Pgngm \lra \Pgne$.
In a model with two neutrino eigenstates of mass $m_1$ and $m_2$ which mix to 
form two flavour states, a pure beam of electron--flavoured neutrinos has a
survival probability which oscillates due to the $m_1-m_2$ mass
difference. For a single neutrino energy $E_\nu(\rm{MeV})$ and a
distance from the source $L$ (meters), the survival probability can be
written in terms of the mixing parameter $\sinsq$ and the difference of
the squared masses $\dmsq$ = $\left| {m_2^2-m_1^2}\right| $ as follows:
\begin{equation}
   P(\Pagne \ra \Pagne) =
   1 - {\rm  sin}^2 2\theta\ {\rm sin}^2
   \left(
   \frac{1.27\,\dmsq ({\rm eV}^2)\,L({\rm m})} {E_\nu({\rm MeV})}
   \right).
\label{prob:trans}
\end{equation}
Atmospheric neutrino results give a $\dmsq$ from $10^{-2}$ to
$10^{-3}\units{eV^2}$. 
Long base-line (L-B) reactor neutrino 
experiments~\cite{Carlo} have been one of the most powerful ways to investigate
$\Pagne \ra \Pagngm$ neutrino oscillations (or, more generally, 
$\Pagne \ra \overline{\nu}_{\rm x}$ oscillations). The 
CHOOZ~\cite{Chooz98,Chooz99} and PALO VERDE\cite{Palo3} experiments utilized 
the high intensity and purity of the reactor core flux to achieve high 
sensitivity.

The CHOOZ experiment had an average value of $L/E\sim300$ 
($L \sim1\units{km}$, $E\sim3 \units{MeV}$), 
an intense and nearly pure neutrino flavour
composition ($\sim 100\%\, \Pagne$) and an intensity known to better than
$2\%$. It could therefore make a definitive contribution to solving the 
problem of the atmospheric neutrino anomaly. 
CHOOZ removed the possibility of explaining the atmospheric neutrino anomaly
by $\Pgne \leftrightarrow \Pgngm$ oscillations and Super--Kamiokande showed 
that $\Pgngm \leftrightarrow \Pgngt$ caused the effect~\cite{Fogli}.

The experiment was designed to detect reactor $\Pagne$'s via the inverse
$\beta$-decay reaction 
\begin{equation}
\Pagne + \Pp \ra \Pep + \Pn
\label{invbet}
\end{equation}
The signature is a delayed coincidence between the prompt $\Pep$ signal 
(boosted by the two $511-\kev$ annihilation $\gamma$ rays) and the signal from
the neutron capture. The target material is a Hydrogen-rich (free protons) 
paraffin--based liquid scintillator loaded with Gadolinium, which was chosen 
due to its large neutron capture cross section and to the high $\gamma$-ray 
energy released after n-capture ($\sim 8\mev$, well above the natural 
radioactivity).

In this final paper we present a complete description of the 
experiment, calibration methods and stability checks, event 
reconstruction procedures and the Monte Carlo simulation. 

Three different analyses, which have already been 
published\cite{Chooz99}, are more 
extensively discussed in this paper. 
The main one is based on all the available 
information: the measured number of positron events as a function of 
energy, separately obtained from each reactor. 
It uses the two spectral shapes, as well as the absolute 
normalizations. The second result is based only on the comparison of the 
positron spectra from the two, different-distance  reactors.
This analysis is largely unaffected by the absolute value of the $\Pagne$ 
flux, the cross section, the number of target protons and the detector 
efficiencies, and is therefore dominated by statistical errors. 
The sensitivity in this case is limited to 
$\dmsq \gsim 2 \cdot 10^{-3} \units{eV^2}$ 
due to the small distance,  $\Delta L = 116.7$ m, between the reactors.
The explored $(\dmsq,\sin^2 2 \theta)$ parameter space still matches 
well the region of the atmospheric neutrino anomaly.
The third analysis is similar to the first, but does not include the knowledge 
of the absolute normalizations.

Finally, some new remarks are presented, concerning the deduction of the 
confidence limits and the correct treatment of the systematic errors.
\section{The $\Pagne$ source and the expected interaction rate}
Nuclear reactors were the first sources used to search for neutrino
oscillations and, in general, for systematic studies of neutrino properties. 
In past experiments (\ie up to the 1980s), the knowledge of the reactor 
neutrino flux and spectrum  ( $\approx 10\%$ accurate at that time) limited
the sensitivity to oscillations. Oscillation tests were performed by 
comparing the neutrino event rate at different distances
(up to $\sim 100\m$) from the reactor. This approach eliminated the systematic 
uncertainty associated with the absolute neutrino flux, often much greater than
the experiment statistical error. Since then, our knowledge base of
fission reactors, in particular of pressurized water reactors (PWR), has much
improved, thanks to the previous reactor experiments and to direct studies
of fission reactions on several elements. Present uncertainty on the 
neutrino flux and spectrum induces systematic errors on the event rate which 
are lower than $3\%$. It is therefore possible to perform experiments, at one 
fixed distance from a reactor only, relying on the adequate knowledge of the 
neutrino source. The reactor $\Pagne$ flux is perfectly isotropic and 
essentially with no contamination from other neutrino types, since the $\Pgne$ 
emission rate is lower than the one of $\Pagne$ by a factor $> 10^{5}$ and may 
therefore be discarded~\cite{Schreck0}.
\subsection{Description}
The CHOOZ nuclear power plant is located in the village of the 
same name, standing on the banks of the River Meuse in the north of France 
close to the border with Belgium.

The power plant 
consists of two twin pressurized-water (PWR) reactors belonging to a 
recently developed generation (N4) in France.
The main innovation consists in an improved
power yield (4.25 GWth, 1.45 GWe at full operating conditions), larger than 
any other PWR reactor. 
The first reactor reached full
power in May $1997$, the second in August $1997$.
Tab.~\ref{tab:dataacq} summarizes data acquisition (from April 7, 1997 to July 
20, 1998).
\begin{table}[htbp]
\caption{\small Summary of the CHOOZ data acquisition cycle
         between April $1997$ and July $1998$.}
\label{tab:dataacq}
\begin{center}
\begin{tabular}{lcc}
\hline
                    & Time (h)  & $\int W \diff t$ (\gw h)\\
\hline
Run                 & 8761.7    & \\
Live time           & 8209.3    & \\
Dead time           & 552.4     & \\
Reactor 1 only ON   & 2058.0    & 8295 \\
Reactor 2 only ON   & 1187.8    & 4136 \\
Reactors 1 \& 2 ON  & 1543.1    & 8841 \\
Reactors 1 \& 2 OFF & 3420.4    & \\
\hline
\end{tabular}
\end{center}
\end{table}
The schedule was quite suitable for 
separating the individual reactor contributions and for determining the 
reactor-OFF background.

The core of both reactors consists of an assembly of 205 fuel elements 
bound to the socket plate of the reactor vessel. The vessel is filled 
with pressurized water ($ p = 155$ bars) at a temperature ranging from $280 
\celsi $ at the entrance to about $320 \celsi$ at the exit. The water, which 
acts as a neutron moderator and cooling element, circulates through four 
independent systems.
Each of these involves a primary 
pump and a steam generator comprising 5416 tubes immersed in the water of a 
secondary loop at a much lower pressure ($56$ bars) than
primary loop pressure. As soon as the primary water passes through these 
tubes, the secondary water is vaporized and the steam produced travels to the
turbine-alternator unit connected to the electric system~\cite{dossier}. 
%
%

The water in the primary loop is slightly doped by Boron, a strong
absorber of thermal neutrons. Boron activity yields 
information on the trend of the chain reaction in the 
core. 
\subsection{Reactor power monitor}
Two methods were developed by the E.D.F. technical staff to monitor the
total thermal power produced by each reactor. The first one is based on the
heat balance of the steam generators. Measurements are made of parameters 
(water flow rate, vapour pressure and temperature in the secondary loop) needed
to determine the 
amount of heat supplied to the steam generators. The resulting values are
periodically available on a special computer network which 
records the data on an Excel file.
The overall precision on the calculated thermal power is claimed to be $0.6\%$.

\begin{figure}[htb]
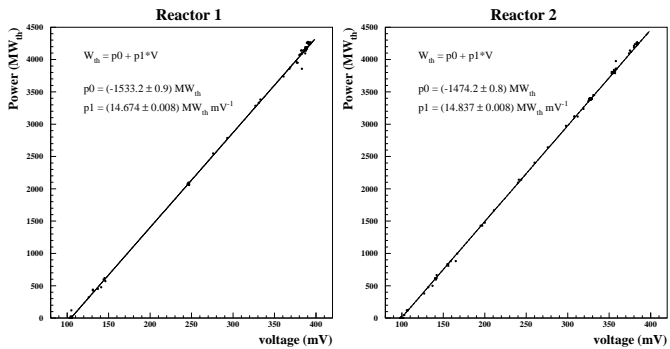

  \begin{center}
    \mbox{\hspace{-0.1cm}\includegraphics[width=0.55\linewidth]{fig.1a}
          \hspace{-0.5cm}\includegraphics[width=0.55\linewidth]{fig.1b}}
    \caption{\small Voltmeter calibration for reactor 1 (left) and 2 (right).}
    \label{fig:calivol}
  \end{center}
\end{figure}

The second set of thermal power data is provided by the external neutron flux 
measurement. This flux is expected to be directly proportional to the fission
rate inside the core and, as a consequence, to the released thermal power.
For each reactor, four different neutron detectors (one 
proportional counters plus three ionization chambers) are located at the 
opposite sides of the reactor vessel. 
This method is less precise than the other one
because of the spread in the energy release per fission 
of the different fissile isotopes. Accuracy is estimated to be about
$1.5\%$. However, this method has the advantage of operating continuously and 
is used simultaneously as a power monitor and sensor for the reactor safety 
system. Neutron detector outputs are 
fed to the computer network and written twice a
day (or more frequently if the power ramps up or down) to another Excel 
file. The computer is also programmed to drive a direct current generator whose
amplitude is proportional to the neutron 
detection rate.
Such a signal was measured in the CHOOZ experiment
by a CAMAC  voltmeter across a $5\Omega$ resistor 
(connected in series to each loop).
This provided 
a direct, instantaneous information about the 
thermal power for both reactors.  
The voltmeter calibration is
shown in Fig.~\ref{fig:calivol}, where 
thermal power values, provided by E.D.F.,
are plotted versus the corresponding voltmeter data.
The calibration parameters
obtained by a linear fit are also given.
\subsection{Map of the reactor core}
\begin{figure}[htb]
  \begin{center}
    \mbox{\includegraphics[width=0.9\linewidth]{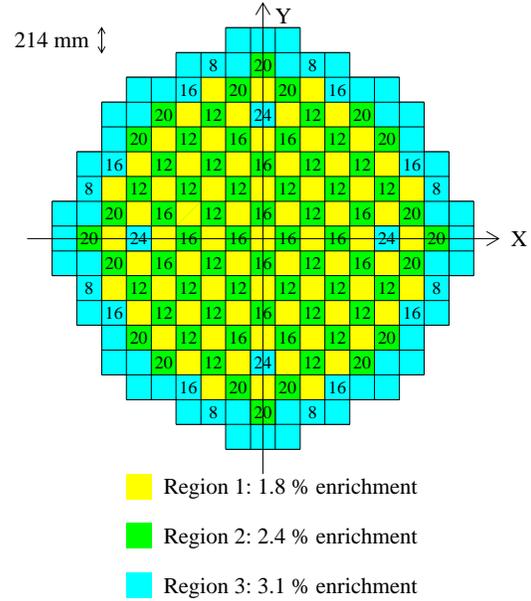}}
    \caption{\small Schematic view of the fuel rods in the core for the first 
      cycle of the CHOOZ reactors. The number of Boron poison rods assembled 
      with each fuel element is also indicated.}
    \label{fig:mapfuel}
  \end{center}
\end{figure}

The nuclear fuel consists of $110$ T of uranium oxide tablets (diameter
$ = 8.2\mm$) enriched with $\nucl{235}{U}$ and stacked in $4\m$ long, $1\cm$ 
wide assemblies. The standard enrichment for this type of reactor is 
$3.1\%$. Each fuel element contains 264 assemblies.
About 1/3 of the 205 fuel elements are changed at the end of each cycle. The 
remainder is
displaced towards the centre and new fuel elements are arranged in the outer 
part of the core, so as to get the fuel burning as uniformly as possible. For 
the start-up of the CHOOZ reactors, brand-new fuel rods were used. In
order to reproduce the behaviour of reactors at equilibrium, where partially 
burned-up fuel is used, 137 less enriched elements are located at the centre 
(68 at $2.4\%$ and 69 at $1.8\%$, while the standard enrichment is $3.1\%$). A 
schematic map of the reactor core is drawn in 
Fig.~\ref{fig:mapfuel}.

As shown in Fig.~\ref{fig:mapfuel}, the fuel assemblies of the most enriched 
elements are surrounded by a number of $12.7\%$ Boron-doped steel rods, termed 
``poison'' rods. The rods absorb the thermal neutron 
excess, thus accumulating $\nucl{11}{B}$, which has a negligible neutron 
absorption cross section, inside the core. 
Therefore the neutron absorption power of the poison rods diminishes with time
partly compensating the fuel burn-up.
The number of rods per fuel element varies according to the fuel loading as 
well as to the element position in the core. This was taken into account 
when computing fuel evolution.
\subsection{Fuel evolution}
The unit used to describe the aging of the fuel at nuclear reactors is the 
$\mw \days$/T, which measures the amount of energy per ton extracted from the
nuclear fuel after its introduction into the reactor core. This quantity is 
called ``burn-up'' and is closely related to the fissile isotope composition of
the fuel. 

\begin{figure}[htb]
  \begin{center}
    \mbox{\includegraphics[bb=0 50 567 567,width=0.9\linewidth]{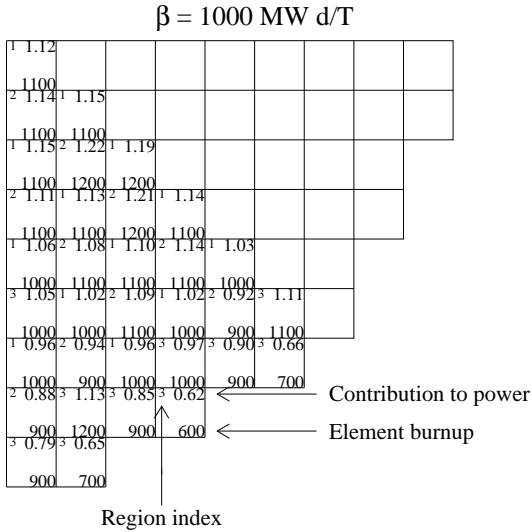}}
    \caption{\small Power distribution and burn-up values for the fuel elements
             in an octant of the CHOOZ reactor core at a certain step 
             ($\beta =1000$ of the first cycle). The contribution to the power 
             of each element is normalized to have a mean value equal to one.}
    \label{fig:mapburn}
  \end{center}
\end{figure}

For any PWR reactor, the procedure to compute the evolution of the fuel in the 
core needs daily information provided by the reactor technical staff. This 
includes:
\begin{itemize}
\item 
the daily cumulative burn-up, given as
\begin{equation}
  \beta(t) \equiv \frac{1}{M_U} \int_0^{t} W_{th}(t') \diff t',
  \label{burnday}
\end{equation}
$W$ being the thermal power, $t$ the time since the start of reactor operation 
and $M_U = 110.26$~T the total amount of uranium
in the core:
\item
the burn-up $\beta_i$ and the relative contribution $\alpha_i$ to the power 
from the i-th fuel element, at several 
stages of the reactor cycle (an example is shown in Fig.~\ref{fig:mapburn}).
\end{itemize}
These two inputs determine the daily burn-up of each fuel element.
An example of the tables provided by E.D.F. for these quantities is given
in Fig.~\ref{fig:mapburn}.
E.D.F. also provides another set of tables (at several burn-up
stages) in which the relative power contribution  $f_k^i$ from the k-th
fissile isotope for the i-th fuel element is given.

The number $n_k^i$ of fissions per second for the i-th element for each 
isotope $k$ can then be computed with the equation
\begin{equation}
  n_k^i(\beta)=\displayfrac{\alpha_i(\beta) f_k^i(\beta) W(t)}{\sum_k 
               f_k^i(\beta)E_k},
  \label{ratefiss}
\end{equation}
where $E_k$ is the energy release per fission for the $k$-th isotope, whose 
values are listed in Tab.~\ref{tab:enefis}.
\begin{table}[htb]
  \begin{center}
    \caption{\small Energy release per fission of the main fissile isotopes 
      (from ref.~\cite{enefis}).}
    \begin{tabular}{||l|c||}
      \hline
      \hline
      isotope & energy (MeV)\\
      \hline
      $\nucl{235}{U}$ & $201.7\pm 0.6$\\
      $\nucl{238}{U}$ & $205.0\pm 0.9$\\
      $\nucl{239}{Pu}$ & $210.0\pm 0.9$\\
      $\nucl{241}{Pu}$ & $212.4\pm 1.0$\\
      \hline
      \hline
    \end{tabular}
    \label{tab:enefis}
  \end{center}
\end{table}
Adding the contribution of all the fuel elements yields 
the average number $N_k$ of
fissions per second for the $k$-th isotope. 
Contributions from other fissioning isotopes, such as 
$\nucl{236}{U}$, $\nucl{240}{Pu}$, $\nucl{242}{Pu}$, etc. amount to less than 
$0.1\%$ and are therefore neglected.

To obtain the source spectrum, in addition to the average
fission rate $N_k$ of each of the four isotopes,
we need the corresponding differential
neutrino yield per fission $S_k(E_\nu)$. 
The next section explains how these spectra are evaluated.
\subsection{The expected neutrino spectra}
We used the so-called ``conversion'' approach which is 
the most reliable and recent method to determine 
the $\Pagne$ spectrum at reactors. 
This method utilizes measurements of the electron 
spectrum emitted by a layer of fissile material activated by thermal neutrons. 
The experimental electron spectrum is then converted into the $\Pagne$ 
one~\cite{Carter,Davis1,Borovoi}. The most recent and precise
measurements were reported by Schreckenbach 
\ea~\cite{Schreck1,Schreck2,Schreck3}. In the latter case thin foils enriched 
with the main fissile nuclei (about $1$~mg) were exposed to an intense 
thermal neutron flux ($\approx 3\cdot 10^{14}\s^{-1}$). A high-resolution
($\delta p/p = 3.5  \cdot 10^{-4}$) $\beta$-spectrometer was used to measure
the momentum of the emerging electrons. 

The $\betm$ spectrum for each fissile isotope is approximated by the 
superposition of a set of hypothetical allowed branches with amplitude $a_i$ 
and end-point $E_0^i$:
\begin{equation}
S_\beta(E_\beta) = \sum_i a_i S_\beta^i [E_\beta,E_0^i,\overline{Z}(E_0^i)],
\label{bmspe}
\end{equation}
where $S_\beta^i$ is the spectrum shape of the 
i-th branch and the summation is over the branches with end-point larger than 
$E_\beta$. The spectrum weakly depends also on the charge $\overline{Z}$
(average charge of the $\beta$-decaying nuclei with end-point $E_0^i$) because 
of the Coulomb interaction in the final state.
The measured electron spectrum is then used to determine the set of values
$\{a_i,E_0^i\}$ by means of a fit procedure. Eq.~(\ref{bmspe}), with the 
introduction of the best-fit parameters, reproduces the measured spectrum 
to better than $1\%$.

The $\betm$ spectrum for each individual hypothetical branch is
then converted into the correlated $\Pagne$ spectrum under 
the assumption that both the electron and the antineutrino share the total 
available energy $E_0^i$. Thus, for each branch with end-point 
$E_0^i$, the probability of emitting an electron with energy $E_\beta$ is equal
to the probability of having a $\Pagne$ of energy $E_0^i-E_\beta$. Inserting 
the fit parameters into (\ref{bmspe}), one obtains
\begin{equation}
S_\nu(E_\nu) = \sum_i a_i S_\beta^i [(E_0^i-E_\nu),E_0^i,\overline{Z}(E_0^i)]
\label{nuspe}
\end{equation}
These yields contain a normalization error of $1.9\%$ stemming from the 
error on the neutron flux and from the absolute calibration uncertainty of the 
spectrometer. The conversion procedure also introduces a global 
shape uncertainty 
into the neutrino spectrum, beyond the inherent experimental errors. The main 
sources of this additional uncertainty, ranging from $1.34\%$ at $3\mev$ to 
$9.2\%$ at $8\mev$, are the scattering in the nuclear charge distribution and 
the higher-order corrections (higher Coulomb terms and weak magnetism, for 
which an uncertainty of the order of the correction term itself was assumed).

\begin{figure}[htb]
  \begin{center}
    \mbox{\includegraphics[width=0.9\linewidth]{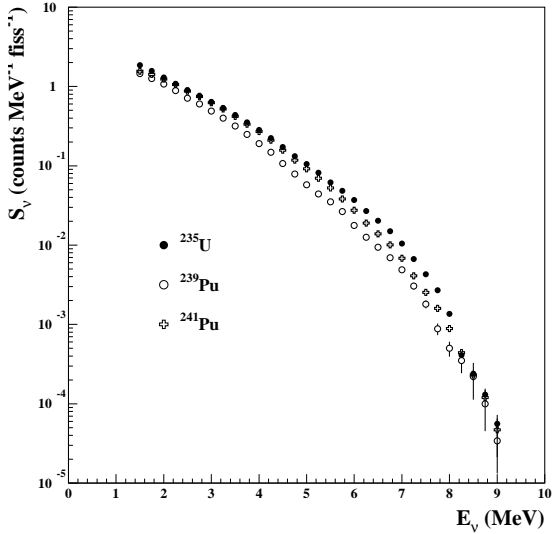}}
    \caption{\small Neutrino yield per fission of the listed isotopes, as 
      determined
     by converting the measured $\beta$  spectra~\cite{Schreck2,Schreck3}.}
    \label{fig:nuspe}
  \end{center}
\end{figure}

This method was applied to obtain the neutrino yields of 
the $\nucl{235}{U}$, $\nucl{239}{Pu}$ and $\nucl{241}{Pu}$ fissions. The 
resulting spectra are presented in Fig.~\ref{fig:nuspe}.
Unfortunately, no experimental data is available for $\nucl{238}{U}$
which cannot be fissioned by thermal neutrons. 
We must therefore rely on
the theoretical predictions~\cite{VSMS} to estimate
the contribution to the $\Pagne$ spectrum by the $\nucl{238}{U}$ fission 
products. Although these predictions are less 
reliable than direct measurements,
it should be noted that the contribution to the number of fissions, due to 
this isotope, is quite stable and never higher than $8\%$. Thus any possible 
discrepancy between the predicted and the real spectrum should not lead to 
significant errors.
\subsection{Systematic uncertainties of the neutrino spectrum}
\label{sec:uncert}
In past experiments at reactors, e.g. at G\"osgen~\cite{Gosgen} and 
Bugey~\cite{Bugey}, more than $3\times 
10^4$ neutrino events were recorded at several reactor--detector distances $R$.
Since no evidence 
for  oscillations was observed and rates were found to 
be consistent with a $1/R^{2}$ 
law, these experiments can be interpreted as a check that a
reactor is a neutrino source of characteristics known at a few percent level. 
The statistical accuracy obtained in these experiments 
makes it possible to use these results
to discriminate between the existing models of reactor neutrino
spectra.

\begin{figure}[htb]
  \begin{center}
    \mbox{\includegraphics[width=0.8\linewidth]{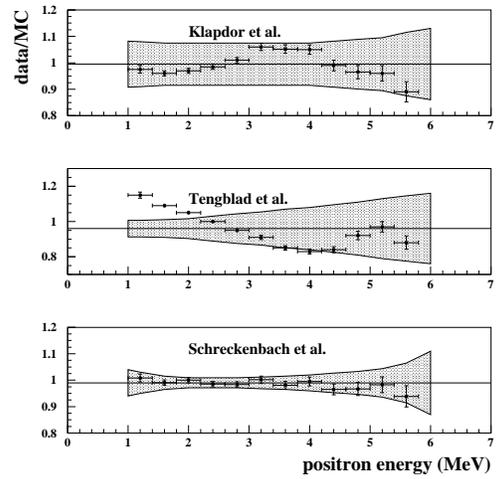}}
    \caption{\small Comparison of Bugey 3 data with three different reactor 
           spectrum 
        models. The error bars include only statistical uncertainties.
        The dashed lines are the quadratic sum of the quoted error of the 
        models and the error due to the energy calibration.}
    \label{fig:Bugeycomp}
  \end{center}
\end{figure}

The Bugey 3 collaboration~\cite{Achkar2} measured the positron energy spectrum
at $15$ and $40\m$ from the reactor core and compared its data with the
results of a Monte Carlo simulation using the neutrino spectrum models proposed
by~\cite{Schreck2,KM,Tengblad}. As can be seen in Fig.~\ref{fig:Bugeycomp},
the data perfectly fit with the measurements made at Institute Laue Langevin
at Grenoble (ILL), whereas there is a lower compatibility with other models.

%

An improved measurement of the integral neutrino flux was performed in 1993.
The measurement used an integral type detector previously employed at the Rovno
nuclear power plant~\cite{Kuvshinnikov} and subsequently moved to 
Bugey~\cite{Declais}. In that apparatus only neutrons were detected in 
$\nucl{3}{He}$ counters, while positrons were not. The apparatus was installed
at $15\m$ from the reactor core.
About $3\times 10^5$ neutrino events were collected so that
the reaction rate was determined with $0.67\%$ statistical accuracy. The 
neutrino event rate, $n_\nu$, corresponds to a certain average fuel composition
and is related to the cross section per fission $\sigma_f$ and the number of 
target protons $N_p$ by
\begin{equation}
n_\nu = \frac{1}{4 \pi R^2} \frac{W_{th}}{\langle E_f \rangle} N_p \varepsilon 
\sigma_f,
\label{rate:nu}
\end{equation}
The fuel composition, the thermal power $W$, the average energy per fission
$\langle E_f \rangle$ absorbed in the reactor core and the distance
$R$ were provided by the E.D.F.--Bugey technical
staff. The efficiency of the neutron detection $\varepsilon$ was carefully 
measured; the overall accuracy was estimated to be $1.4\%$. The experimental
result was then compared to the expected neutrino flux, which can be inferred 
by introducing into (\ref{rate:nu}) the neutrino spectra $S_k$ obtained at 
ILL\cite{Schreck1,Schreck2,Schreck3}, the cross section for the 
reaction (\ref{invbet}) and the reactor parameters in (\ref{rate:nu}). The 
expected cross section per fission for reaction (\ref{invbet}) is given by
\begin{equation}
\begin{split}
\sigma_f^{exp} 
& = \int_{0}^{\infty} \sigma(E_\nu) S(E_\nu) \diff E_\nu \\
& = \sum_k f_k \int_{0}^{\infty} \sigma(E_\nu) S_k(E_\nu) \diff E_\nu =
\sum_k f_k \sigma_k
\end{split}
\label{fiss:cross}
\end{equation}
where $f_k$ refers to the contribution of the main fissile nuclei to the total
number of fission, $S_k$ to their corresponding $\Pagne$ spectrum and 
$\sigma_k$ to their cross section per fission. The result of the measurement 
was $\sigma_f^{meas} = 5.752 \times 10^{-19}\mbox{barns/fission} \pm 1.4\%$, 
which perfectly agrees with the expected value 
($\sigma_f^{exp} = 5.824 \times 10^{-19}\text{barns/fission} \pm 2.7\%$), and 
is twice as accurate as the predictions based on the knowledge of the neutrino
spectrum.

We could therefore adopt the conversion procedure for the
shape of the neutrino spectra but normalize the total cross section
per fission to the Bugey measurement, \ie, after
taking all the different reactors conditions into account.

\begin{figure}[htb]
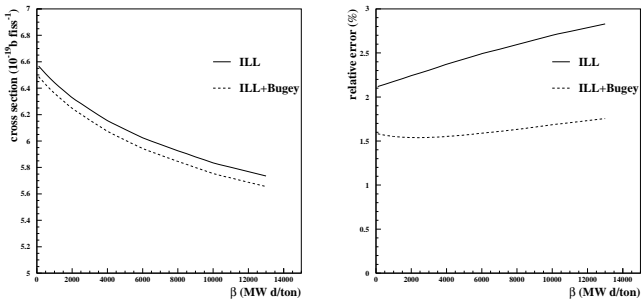

 \begin{center}
    \mbox{\includegraphics[width=0.5\linewidth]{fig.6a}
          \includegraphics[width=0.5\linewidth]{fig.6b}}
    \caption{\small Comparison of the combined (ILL+Bugey) reaction cross 
             section 
             with the ILL cross section (left) and their relative error (right)
             as a function of the first CHOOZ reactor cycle burn-up.}
    \label{fig:ILLBugey}
  \end{center}
\end{figure}

In figure \ref{fig:ILLBugey}(left) the ILL cross section 
(\ref{fiss:cross}) and the combined (ILL+Bugey) cross section  
obtained for the CHOOZ reactors are plotted vs. the reactor burn-up. The 
average ratio of the two curves amounts to $0.987$. By combining the 
uncertainty
on the neutrino spectra, on the cross section for the reaction (\ref{invbet}) 
and on the fission contributions $f_k$ (which are of the order of $5\%$), we
obtained the relative error on the neutrino detection rate, as a function of
the fuel burn-up. As shown in Fig. \ref{fig:ILLBugey}(right), the average error
decreases from $2.4\%$ (ILL data alone) to $1.6\%$
(ILL+Bugey). Other minor sources of errors come from the residual neutrino
emission from long-lived fission fragments (dealt with in the next
Section). However the reactor source can be considered to be known at 
a $2\%$ level. 
\subsection{Neutrino spectrum time relaxation and residual neutrino emission}
Another possible source of uncertainty of the neutrino flux is related to the 
residual emission due to the $\betm$ decay of long-lived fission fragments.  
Taking this further contribution into account, the linear relation
(\ref{rate:nu}) between the prompt $\Pagne$ interaction rate and the current
thermal power $W$ no longer holds. 
Nevertheless, the $\Pagne$ spectra determined at ILL were derived after about
$1.5\days$ exposure time, so that neutrinos from fission fragment decays of 
longer life
are not included. The expected neutrino rate based on this model may thus
be underestimated with respect to the experimental data.
Fortunately the maximum neutrino energy is above the
reaction (\ref{invbet}) threshold only in a few cases 
since, as can be expected, the longer the
lifetime, the lower the $Q$-value of the decay\footnote{
This effect might be much more relevant in the case of experiments looking for 
neutrino elastic scattering interactions (\ie, measurement of the neutrino 
magnetic moment) and needs a more careful treatment.}. 
This effect has been evaluated by using the cumulative yields of the known 
long-lived fission fragments~\cite{kopeikin}; the results for $\nucl{235}{U}$ 
and $\nucl{239}{Pu}$ are summarized in Tab.~\ref{tab:fisslong}.
\begin{table}[htb]
  \begin{center}
  \caption{\small Time evolution of neutrino spectra relative to infinite 
    irradiation time (from~\cite{kopeikin}).}
  \label{tab:fisslong}
\newcommand{\Rule}{\rule[-.7ex]{0ex}{2.9ex}}
    \leavevmode
    \begin{tabular}{|c|c|c|c|c|c|c|}
      \hline
      \hline
      $E_\nu$ (MeV) 
      & \multicolumn{3}{c}{$\nucl{235}{U}$} \vline
      & \multicolumn{3}{c}{$\nucl{239}{Pu}$} \vline \\
      \hline
      & $10^4$ s & $1.5$ d & $10^7$ & $10^4$ s & $1.5$ d & $10^7$ \\
      \cline{2-7}
      $1.5$ & $0.837$ & $0.946$ & $0.988$ & $0.861$ & $0.949$ & $0.990$ \\
      \hline
      $2$   & $0.897$ & $0.976$ & $0.992$ & $0.904$ & $0.968$ & $0.986$ \\
      \hline
      $2.5$ & $0.925$ & $0.981$ & $0.990$ & $0.939$ & $0.975$ & $0.986$ \\
      \hline 
      $3$   & $0.963$ & $0.997$ & $1.000$ & $0.967$ & $0.989$ & $0.993$ \\
      \hline
      $3.5$ & $0.967$ & $1.000$ & $1.000$ & $0.979$ & $0.997$ & $1.000$ \\
      \hline
      \hline
    \end{tabular}
  \end{center}
\end{table}
In particular, the reaction cross section $\sigma_f$ computed by using 
(\ref{fiss:cross}) is probably lower than the effective one by $\approx 0.3\%$.
This systematic shift affects the accuracy on the reaction cross section. We
will then assume an overall and conservative $1.9\%$ uncertainty on the 
integral neutrino rate.
\subsection{The inverse beta-decay reaction}
The detection of reactor antineutrinos is usually based on the
reaction (\ref{invbet}). This is the most suitable process since:
\begin{itemize}
\item[-] it has the highest cross section in the energy range of reactor 
antineutrinos:
\item[-] provides a convenient time correlated pair of positron and neutron 
signals, which allows us to reject most of the background.
\end{itemize}
The antineutrino and the positron energy are related by
\begin{equation}
E_{\Pagne} = E_\Pep + (M_\Pn - M_\Pp) + {\cal O}(E_{\Pagne}/M_\Pn),
\label{pos:energy}
\end{equation}
where the infinitesimal term corresponds to the neutron recoil. Thus a 
measurement of the 
positron energy allows an accurate determination of the energy of the incoming
antineutrino. The threshold for the reaction (\ref{invbet}) is
$1.804\mev$, equal to the nucleon mass difference plus the positron mass. In 
the low energy limit, the cross section for the reaction (\ref{invbet})
may be written as a function of outgoing positron energy as follows:
\begin{equation}
\sigma(E_\Pep) = \frac{2 \pi^{2} \hbar^{3}}{m_\Pe^{5} f \tau_\Pn} 
p_\Pep E_\Pep (1 + \delta_{rad} + \delta_{WM}) 
\label{inv:cross}
\end{equation}
The transition matrix element has been expressed in terms of the free neutron
decay phase-space factor $f = 1.71465(15)$ \cite{Wilkinson} and lifetime 
$\tau_n = (886.7 \pm 1.9)\s$~\cite{Schreck4}. Two higher-order correction terms
(both of $1\%$ order of magnitude) are also included:
\begin{description}
\item[(i)] a radiative correction of the order of $\alpha$, including an 
internal bremsstrahlung contribution, which can be approximated by
\begin{equation}
  \delta_{rad}(E_{e^+}) = 11.7 \times 10^{-3} (E_{e^+} - m_e)^{-0.3}
  \label{radcorr}
\end{equation}
with the positron energy expressed in $\mev$~\cite{Vogel} and
\item[(ii)] a correction for weak magnetism, arising from the difference 
$\mu = \mu_n-\mu_p = -4.705890(2) \mu_N$ between the anomalous magnetic moment 
of the neutron and the proton
\begin{equation}
  \delta_{WM}(E_{e^+}) = -2 \frac{\mu \lambda}{1 + 3\lambda^2} 
   (E_{e^+} + \Delta \beta p_{e^+})/M_p,
  \label{WMcorr}
\end{equation}
where $\lambda = g_A/g_V = 1.2601 \pm 0.0025$ is the ratio of axial-vector and 
vector coupling constants and $\Delta$ is the nucleon mass 
difference~\cite{Fayans}.
\end{description}
The knowledge of the cross section is therefore 
much more accurate than the one of the $\Pagne$
spectrum, the major limitation being related to the uncertainty on $\tau_n$ 
(whose relative error is at most $0.3\%$). 
\subsection{Simulation of the $\Pagne$ spectrum}
\label{sec:reasim}
The expected neutrino spectrum was obtained by means of a Monte Carlo
simulation of the reactor core in which we folded all the ingredients
described in the previous sections, that is:
\begin{itemize}
\item the daily variation of the flux due to the fissile elements
burn-up and to the live time of the apparatus;
\item the contribution of the individual fuel elements according to their
position inside the reactor core;
\item the individual contributions of the 
different fissile elements in each fuel
element to the neutrino flux.
\end{itemize}

\begin{figure}[htb]
  \begin{center}
   \mbox{\includegraphics[width=0.7\linewidth]{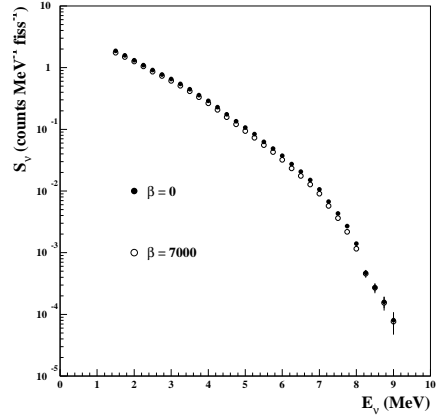}}
    \caption{\small Comparison between the neutrino spectrum at the beginning 
             and during the first cycle of the CHOOZ reactors.}
    \label{fig:nuyie}
  \end{center}
\end{figure}

\begin{figure}[htb]
  \begin{center}
    \mbox{\includegraphics[width=0.8\linewidth]{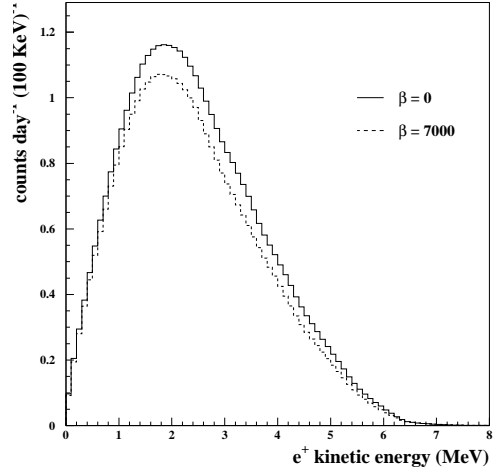}}
    \caption{\small Positron spectra at the startup and during the first cycle 
             of the CHOOZ reactors at maximum daily neutrino luminosity.}
    \label{fig:posyie}
  \end{center}
\end{figure}

\begin{figure}[htb]
  \begin{center}
    \mbox{\includegraphics[width=0.8\linewidth]{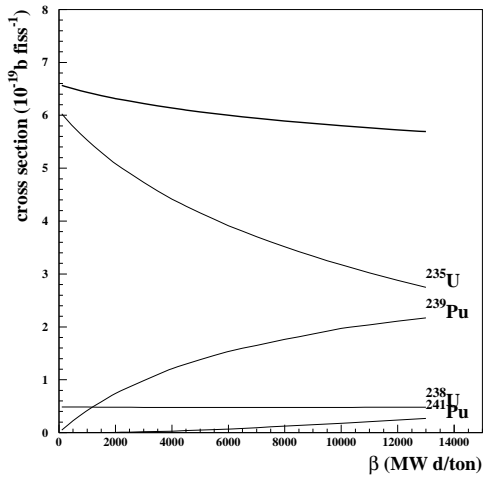}}
    \caption{\small Cross section per fission as a function of the reactor 
             burn-up. The contribution of each fissile isotope is also shown.}
    \label{fig:crofis}
  \end{center}
\end{figure}

Fig.~\ref{fig:nuyie} shows the neutrino spectrum
obtained at CHOOZ on the start-up day ($\beta=0$) and at an intermediate step
($\beta= 7000$) of the first reactor cycle. Due to the decrease of the 
$\nucl{235}{U}$ concentration, a reduction of the neutrino interaction rate
is observed and a softening of the neutrino spectrum is expected.
%
The energy spectrum of the positrons coming from (\ref{invbet}) is essentially 
the antineutrino spectrum shifted in energy and weighted by the cross section
(\ref{inv:cross}).
So, following (\ref{rate:nu}), each positron is assigned a weight given by
\begin{equation}
S_{e^+}(T_{e^+}) = \frac{N_p}{4\pi d^2} \sigma(E_\nu) S_\nu(E_\nu),
\label{posyie}
\end{equation}
where $d$ is the distance from the neutrino production point to the detector and
the positron kinetic energy $T_{e^+}$ is given by (\ref{pos:energy}). 
Fig.~\ref{fig:posyie} shows the positron yield obtained by generating the 
neutrino spectra drawn in Fig.~\ref{fig:nuyie} in one day of data taking with 
both reactors at full power. Although the generated neutrino luminosity is the
same, the decrease of the positron yield with the reactor operating time is
evident. 
The evolution of the positron spectrum must be followed very accurately in order
to reproduce the hardware threshold effects on the positron detection.
As an immediate consequence, also the integral neutrino interaction rate is 
expected to vary significantly during the reactor fuel cycle. A decrease of
about $10\%$ has been forecast 
for the cross section per fission (which is linear with
the interaction rate, according to (\ref{rate:nu})) during the first
cycle of the CHOOZ reactors, as shown in Fig.~\ref{fig:crofis}.
The measured neutrino rate as a function of the burn-up will be 
shown and compared to the expected behaviour, under the no-oscillation 
hypothesis.
%
\section{The experiment}
\subsection{The site}
\begin{figure}[htb]
  \begin{center}
    \mbox{\includegraphics[bb=0 100 600 700,width=\linewidth]{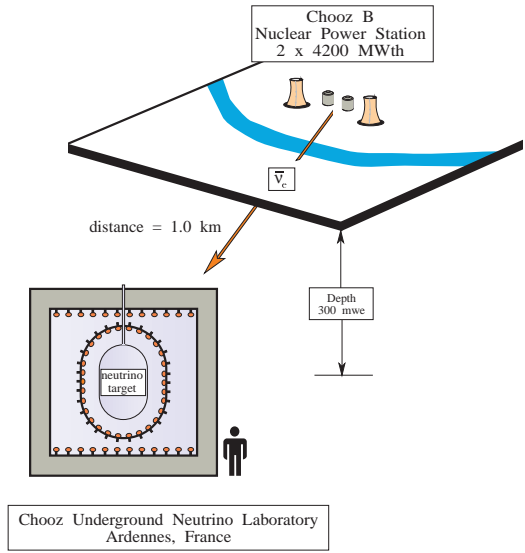}}
    \caption{\small Overview of the experiment site with indication of the 
	     source-detector distance and rock overburden.}
    \label{fig:view}
  \end{center}
\end{figure}

\begin{figure}[htb]
  \begin{center}
    \mbox{\includegraphics[width=0.7\linewidth]{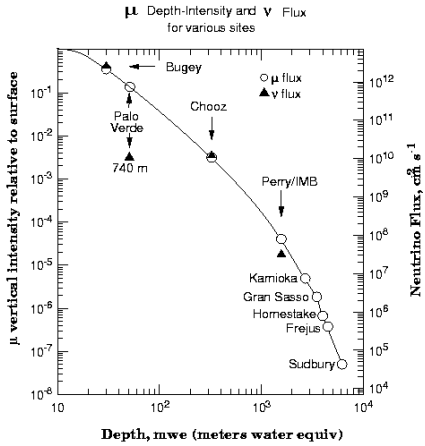}}
    \caption{\small Cosmic muon flux compared to the neutrino flux at the 
      different underground experimental sites. In the CHOOZ case the lower 
      neutrino flux is compensated by the reduction of the muon flux.}
    \label{fig:muflux}
  \end{center}
\end{figure}

The detector was located in an underground laboratory
about 1\units{km} from the neutrino source (see Fig.~\ref{fig:view}).
The 300\units{MWE} rock
overburden reduced the external cosmic ray muon flux by a factor of
$\sim300$ to a value of $0.4\units{m^{-2}s^{-1}}$, significantly
decreasing the most dangerous background, which is caused by fast
neutrons produced by muon--induced nuclear spallations in the materials
surrounding the detector. This cosmic ray shielding was an important
feature of the CHOOZ site. As shown in Fig.~\ref{fig:muflux}, 
the rock shielding
preserved the signal to noise ratio of previous reactor 
experiments, in spite of
a reduction by a factor $\sim 100$ of the neutrino flux due to the larger 
distance from the reactors.
The characteristics of the site were thoroughly assessed before and after the 
set--up of the experiment. We measured the cosmic ray flux and angular 
distribution and the results were compared with predictions based on the rock 
structure above and around the site. The natural 
radioactivity background, the radon level in the tunnel and the intensity and 
orientation of the local terrestrial magnetic field
were also measured. A geodesic survey was
performed to determine the exact location of the 
underground experiment with respect to the two 
reactors and its orientation. All 
the information obtained in the preliminary 
studies were used to guide 
improvements in the detector and in the detector simulation.

\begin{figure}[htb]
  \begin{center}
    \mbox{\includegraphics[width=0.7\linewidth]{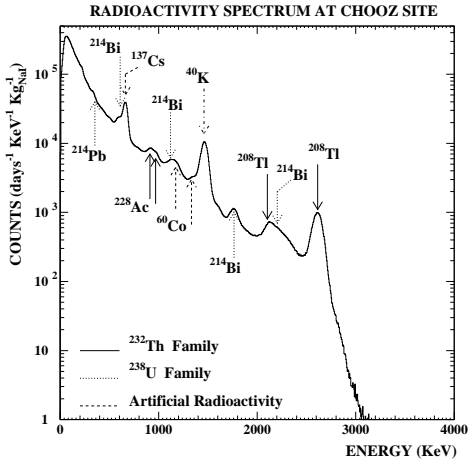}}
    \caption{\small Natural radioactivity spectrum 
             recorded by a NaI crystal in 
             the well hosting the detector.}
    \label{fig:radio}
  \end{center}
\end{figure}

The cosmic ray measurements were made for several days with a system of six 
Resistive Plate Chambers (RPC), each of area $1 \times 1 \units{m^{2}}$.
The comparison of the experimental and expected angular distributions 
was fairly good, but some discrepancies persisted. A further 
geological study  revealed the existence of several very high density rock 
layers ($3.1 \units{g/ cm^{3}}$ the normal density being  
$2.8 \units{g/ cm^{3}}$), whose positions and orientations fully explained the 
observed effects.
The natural radioactivity spectrum was measured, at the position where the 
experiment had to be installed, by a $3 \times 3 \units{inches^{2}}$ NaI 
crystal. Natural radioactivity was 
rather high and the spectrum shows 
the normal natural radioactivity lines but also some artificial radioactivity 
contributions (see Fig.~\ref{fig:radio}).

The magnetic field was measured by a rotating coil, pressurized-air device, and 
was found to be $B_\parallel=0.178 \pm 0.007 \units{G}$ and   
$B_\perp=0.388 \pm 0.007 \units{G}$. 
\subsection{The detector}
\begin{figure}[htb]
  \begin{center}
    \mbox{\includegraphics*[width=0.9\linewidth]{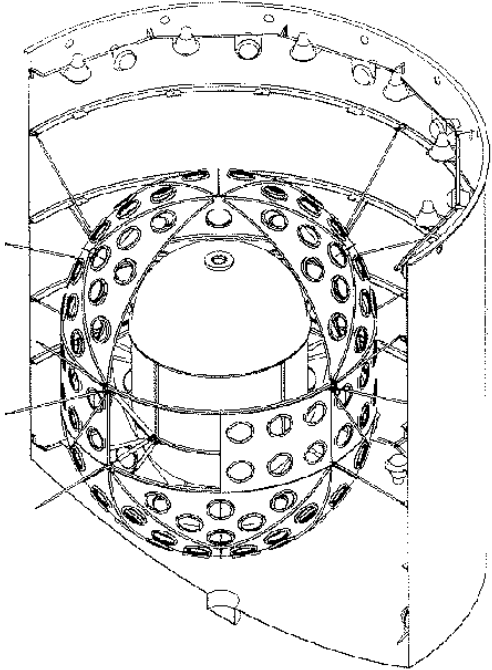}}
    \caption{\small Mechanical drawing of the detector; the visible holes on 
	the geode are for the PMT housing (from \cite{duverney}).}
    \label{fig:geode}
  \end{center}
\end{figure}

The detector (Fig.~\ref{fig:geode}) was installed 
in a welded cylindrical steel 
vessel 5.5\units{m} in diameter and 5.5\units{m} deep. 
The internal walls of the
vessel were painted with high--reflectivity white paint. The vessel was
placed in a pit 7\units{m} in diameter and 7\units{m} deep. To protect
the detector from the natural radioactivity of the rock, the steel
vessel was surrounded by 75\units{cm} of low radioactivity sand
(Comblanchien from Burgundy in France) and covered by 14\units{cm} of
cast iron.
The detector comprised three concentric regions:
\begin{itemize}
\item a central 5--ton target in a transparent Plexiglas container
(total mass = 117 kg) filled with a 0.09\% Gd--loaded scintillator 
(``Region I'');
\item an intermediate 17--ton region (70\units{cm} thick) equipped with 192
eight--inch PMT's (15\% surface coverage, $\sim 130$ photoelectrons/MeV), used 
to protect the target from PMT radioactivity and 
to contain the gamma rays from neutron capture (``Region II'');
\item an outer 90--ton optically separated active cosmic--ray muon veto
shield (80\units{cm} thick) equipped with two rings of 24 eight--inch PMT's
 (``Region III'').
\end{itemize}
%
%
The apparatus was conceived as a liquid scintillator 
low energy, high-resolution
calorimeter. The detector geometry (a central volume of scintillator surrounded
by photomultipliers) is common to the Borexino, LSND and SNO detectors.
The detector was simple and easily calibrated, while its behaviour could be
well checked.
Six laser flashers were installed in the three regions
together with calibration pipes to allow the introduction of radioactive
sources. The detector could be reliably simulated by the Monte Carlo
method.

The target region contained a Gd-loaded liquid scintillator. The neutrino 
detection was based on the
delayed coincidence between the prompt positron signal generated by reaction 
(\ref{invbet}), boosted by the annihilation $\gamma$-rays, and the signal 
associated with the $\gamma$-ray emission 
following the neutron capture reaction
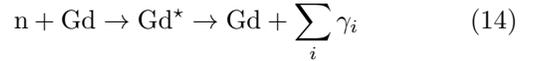
\begin{equation}
  \Pn + {\rm Gd}\rightarrow {\rm Gd^{\star}}\rightarrow {\rm Gd}+\sum_i\gamma_i
  \label{cattura}
\end{equation}
The choice of a Gd-doping was to maximize the neutron capture efficiency; 
Gadolinium has the highest thermal neutron 
cross section. Moreover, the large total $\gamma$-ray energy ($\approx 8\mev$, 
as shown in Tab.~\ref{tab:Gdiso}) easily discriminates 
the neutron capture from 
the natural radioactivity, whose energy does not exceed $3.5\mev$. 
\begin{table}[htb]
  \caption{\small Abundances and thermal neutron capture cross 
sections for the Gd isotopes.}
  \label{tab:Gdiso}
  \begin{center}
    \begin{tabular}{|c|c|c|c|c|}
      \hline
         Gd   & $\sum_i E_{\gamma_i}$ & Abundance & Cross section & Relative \\ 
      isotope &  (KeV) & (\%) & (barns) & intensity \\ 
      \hline
      \hline
      152     & 6247 & 0.20 & 735 & $3\cdot 10^{-5}$ \\
      \hline
      154     & 6438 & 2.18 & 85 & $3.8\cdot 10^{-5}$ \\
      \hline
      155     & 8536 & 14.80 & 60900 & $0.1848$ \\
      \hline
      156     & 6360 & 20.47 & 1.50 & $6\cdot 10^{-6}$ \\
      \hline
      157     & 7937 & 15.65 & 254000 & $0.8151$ \\
      \hline
      158     & 5942 & 24.84 & 2.20 & $1.1\cdot 10^{-5}$ \\
      \hline
      160     & 5635 & 21.86 & 0.77 & $3\cdot 10^{-6}$ \\
      \hline
    \end{tabular}
  \end{center}
\end{table}

Region II was filled with an undoped high-flash point liquid scintillator. It
provided a high-efficiency containment of the e.m. energy deposit; this 
was higher than $99 \%$ for positrons from $\Pagne$-interactions in Region I. 
The containment of the $\gamma$-rays due to the neutron capture on Gd was 
(on average) slightly lower than $95\%$ for an energy deposit
$E > 6\mev$. The intermediate volume was bounded by the ``geode'', an opaque
plastic structure serving as a support for the 192 inward-looking 
photomultiplier tubes (PMT from now on).

The outer volume, also filled with the undoped scintillator of Region II, was 
the  ``Veto'' region (Region III). 
An additional 48 PMT's, arranged in two circular 
rings located at the top and the bottom of the main tank, detected the 
scintillation light associated with through-going cosmic muons. The Veto
signal was used to tag and reject this major background source. The outer
scintillator layer was also thick enough to shield 
the neutrino target against the natural radioactivity 
from the surrounding materials.

The inner detector volume was separated from Region II by a transparent 
$8\mm$-thick vessel, a vertical cylindrical surface 
closed by two hemispherical end-caps. The outer radius of the cylinder and of
the end-caps was $90\cm$, the height of the cylinder was $100\cm$; 
the inner volume was $5.555\m^3$, while the mass was $150\kg$ (empty). 
The vessel was made of an acrylic polymer (Altuglass), chosen for its excellent
optical and mechanical properties and for its chemical resistance to 
aromatic compounds in the scintillator. The upper part of the vessel 
was fitted with a chimney (diameter $\phi = 70\mm$) to allow passage of filling
pipes, calibration sources, temperature and pressure sensors.

The geode had the same shape of the target vessel, but a larger size;
the cylinder height was the same, whereas the outer radius was $160\cm$. The 
volume between the geode and the target was $19.6\m^3$. The geode surface
(a drawing of which is shown in Fig.~\ref{fig:geode}) had a total area
of $42\m^2$ segmented into 32 panels; each panel was equipped with 6 
$8''$ PMT's detecting the scintillation light produced in Regions I and II. 
The global PMT coverage was then $15\%$. 
Unlike the acrylic inner vessel, the geode was
opaque so as to optically shield the inner regions from 
the veto scintillation light. The external surface was 
white-coated in order to 
enhance the light collection in Region III and improve the Veto rejection 
efficiency, while the black inner surface reduced light reflections,
which could degrade the vertex resolution in the 
detector\cite{notaCdF}.
\subsection{The liquid scintillators}
\label{sec:liqscint}
About 5 tons of Gd-loaded scintillator and 107 tons of unloaded scintillator 
were used in the experiment. The main properties of the two scintillators 
are listed in Tab.~\ref{tab:scint}.
\begin{table}[hb]
  \caption{\small Main properties of the liquid scintillators
      used in the experiment.}
  \label{tab:scint}
  \begin{center}
    \begin{tabular}{|l|c|c|}
      \hline
      & Gd-loaded & unloaded \\
      \hline
      \hline
      Chemical content: & & \\
      \hspace{0.1cm} basic & Norpar-15 & Mineral oil \\
      & (50\% vol.) & (92.8\% vol.)\\
      \hspace{0.1cm} aromatics, alcohols & IPB+hexanol & IPB \\
      & (50\% vol.) & (7.2\% vol.) \\
      \hspace{0.1cm} wavelength shifters & p-PTP+bis-MSB & PPO + DPA \\
      & (1 g/l) & (1.5 g/l) \\
      \hline
      Atomic mass & & \\
      composition: & & \\
      \hspace{0.5cm} H & $12.2\%$ & $13.3\%$ \\
      \hspace{0.5cm} C & $84.4\%$ & $85.5\%$ \\
      \hspace{0.5cm} Gd & $0.1\%$ & \\
      \hspace{0.5cm} others & $3.3\%$ & $1.2\%$ \\
      \hline
      compatibility & \multicolumn{2}{c}{acrylic, Teflon} \vline \\
      \hline
      density ($20\celsi$) & 0.846 g/ml & 0.854 g/ml \\ 
      \hline
      Flash point & $69\celsi$ & $110\celsi$ \\
      \hline
      Scintillation yield&\multicolumn{2}{c}{5300 ph/MeV 
        (35\% of anthracene)} \vline \\
      \hline
      Optical attenuation & 4 m & 10 m \\
      length & & \\
      \hline
      Refractive index & 1.472 & 1.476 \\
      \hline
      Neutron capture & $30.5\usec$ & $180\usec$ \\
      time & & \\
      \hline
      Neutron capture & $\sim 6\cm$ & $\sim 40\cm$ \\
      path length & & \\
      \hline
      Capture fraction & $84.1\%$ & \\
      on Gd & & \\
      \hline
      \end{tabular}
  \end{center}
\end{table}

The solution of the Gadolinium salt Gd(NO$_3)_3$ in hexanol as well as the
mixing of the basic and aromatic compounds 
was performed in a dedicated hall close to 
the entrance of the underground tunnel.
The amount of Gadolinium (0.09 \% in weight) was chosen to optimize
neutron capture time and neutron detection efficiency. The measured
values for the average capture time, path length and capture efficiency are
listed in Tab.~\ref{tab:scint}.
A higher concentration would have required more alcohol, 
which could have lowered the high
flash point of the solution below the limit imposed by safety regulations. 
Moreover, as we shall see, the presence of the
nitrate ions in solution progressively deteriorated the optical
properties of the scintillator. A higher concentration would 
have further
compromised the chemical stability of the scintillator.

A fundamental quantity for normalizing the neutrino event rate is the
number of free protons in the target. An accurate
evaluation of this number relies on precise measurements of the density and of
the Hydrogen content of the Gd-loaded scintillator. The Hydrogen content was
determined by combustion of scintillator samples; the presence of volatile
elements (chiefly various types of alcohol) 
made this measurement particularly difficult. 
The value listed in Tab.~\ref{tab:scint} derives from averaging the dozen
measurements performed on scintillator samples at different 
laboratories~\cite{ETH,Lyon}. The overall relative uncertainty on the 
number of protons is $0.8\%$. 

\begin{figure}[htb]
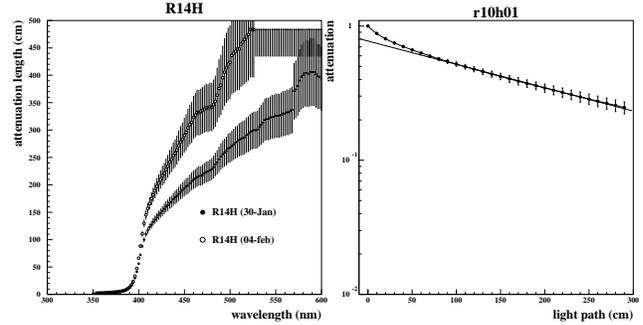

  \begin{center}
    \mbox{\includegraphics[bb=0 140 590 650,width=0.55\linewidth]{fig.14a}
          \hspace{-0.8cm}\includegraphics[bb=0 140 590 720,width=0.55\linewidth]{fig.14b}}
    \caption{\small Attenuation length vs. wavelength for the Gd-loaded 
      scintillator (of the same type of that used at CHOOZ) at different aging 
      stages (left) and scintillation light attenuation vs. path (right).} 
    \label{fig:photometer}
  \end{center}
\end{figure}

The scintillator transparency showed a degradation over time, 
resulting in a slight decrease of the photoelectron yield. The most probable 
cause is thought to be the oxidation by the nitrate ion. 
On one side, this forced us to repeatedly check the 
attenuation length in the detector by using calibration sources all along the 
acquisition period. The procedure followed is described in
\S\ref{sec:atten}. 
In parallel, we made a systematic laboratory study of the chemical 
and optical stability 
of liquid Gd-loaded scintillators, which is also of interest for future 
low-energy neutrino detectors. 
The main optical properties were determined with
both spectrophotometric 
and fluorimetric measurements. A 
Varian Cary 2200, UV-vis, 
$10\cm$ cell double-beam spectrophotometer was 
used to measure the absorbance of
scintillators as a function of the beam light 
wavelength in the $350\div 600$~nm
range, the absorbance being defined as
\begin{equation}
  A \equiv - \log_{10} \frac{I}{I_0}
  \label{abso}
\end{equation}
where $I_0,I$ are the intensity respectively of the incident and the emerging 
beams. Several paired cells, with Suprasil quartz windows were used for 
these measurements. The measurement of attenuation lengths of a few meters by
means of $10\cm$ cells required great care in the stability control of the 
apparatus as well as an accurate cleaning of the cell optical surfaces. 
Corrections were applied to take into account the multiple 
reflections inside the cell due to the different refraction indices of air, 
quartz and scintillator; the overall systematic uncertainty on the absorbance
(including the spectrometer stability and the planarity of the quartz windows)
was proved to be $0.5\%$. 
%
%
Examples of these measurements obtained with
a laboratory-blended scintillator sample (using the same composition as the
Region I scintillator) are shown in Fig.~\ref{fig:photometer} (left). 

An overall attenuation for the scintillation light was obtained by folding
the measured attenuation at different wavelengths with 
the fluorescence spectra 
of the phosphores and the PMT photocathode sensitivity. The light attenuation 
as a function of the path is shown in Fig.~\ref{fig:photometer} (right). 
Apart from the first $\sim 20\cm$,
needed for the absorption by the second shifter (bis-MSB), attenuation is
best described by an exponential decrease giving an ``effective'' length. 
Measurements performed on the laboratory samples at different aging 
stages reproduced the attenuation length values obtained for the target 
scintillator within $15\%$.

The time variation of the light attenuation length in the detector is best 
reproduced by the function
\begin{equation}
\label{eq:lambda}
  \lambda(t) = \frac{\lambda_0}{1+v t}
\end{equation}
which accounts for the observed exponential 
decrease of the number of photoelectrons with
time. Here $v$ is proportional to the velocity of the chemical reactions 
responsible for the scintillator deterioration. The reaction kinetics is known 
to depend on temperature according to an exponential law:
\begin{equation}
  v(T) = v_0 f(T) = v_0 a^{[(T-T_0)/10\celsi]}
  \label{vtemp}
\end{equation}
where the index 0 labels the values at room temperature. It is also known that
reactions occurring at time scales $\sim 1\s$ are accelerated by a factor 
$\sim 2$ if the temperature is raised by $10\celsi$ from room temperature.

\begin{figure}[htb]
  \begin{center}
    \mbox{\includegraphics[width=\linewidth]{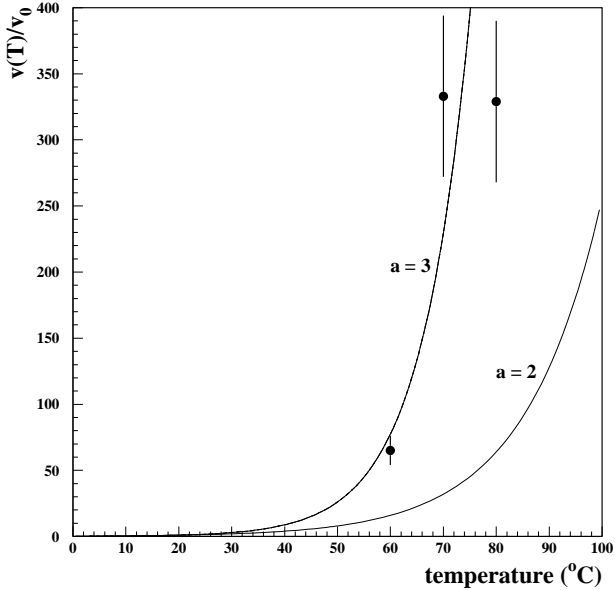}}
    \caption{\small Acceleration of the scintillator aging rate as a function of
 the temperature.}
    \label{fig:aging}
  \end{center}
\end{figure}

We attempted to accelerate the 
aging effects by heating different samples at 60, 70 
and $80\celsi$. As shown in
Fig.~\ref{fig:aging}, we found that an increase of $10$ degrees in the 
temperature corresponded to an acceleration factor $a\approx 3$ 
(instead of
$2$) in the aging rate. This discrepancy can be explained by 
simple thermodynamical arguments if (as in this case) such a reaction
develops at room temperature on a time scale of a few months.
The estimated value (at $T_0 = 20\celsi$) was 
$v_0 = (3.8 \pm 1.4)\cdot 10^{-3} \days^{-1}$, which was acceptable for a 1-year
data-taking duration experiment like CHOOZ. This prediction turned out to be 
close to the value ($(4.2 \pm 0.4)\cdot 10^{-3} \days^{-1}$) obtained by direct 
measurements in the detector.
\subsection{The photomultiplier system}
The PMTs provided both the charge and the 
time information needed to determine the event
position and the energy deposit. The average photoelectron (pe) 
yield associated with a typical positron signal ($E \approx 3\mev$) 
was $\approx 300$, which corresponded 
to an average $\approx 1.5$~pe at each PMT. The number 
of hit PMTs was $\approx 80\%$ of the total. As the PMTs had to work in a 
single pe regime, the most important feature 
for tube selection was the single 
pe resolution, \ie, the peak to valley ratio. Other important requirements 
were: low radioactivity and dark noise, high gain, large photocathode surface, 
low sensitivity to magnetic fields, good (but not ultra-fast) time properties. 
The EMI 9536KA B53~\cite{EMI} turned out to be the best PMT for our needs. The 
high amplification
($\approx 10^7$, corresponding to $\approx 30\mv$ for a 1-pe pulse) matched
the characteristics of the front-end electronics; the transit-time jitter
($\approx 8\nsec$ FWHM at 1-pe level) was 
comparable with the scintillator decay
time ($\approx 7\nsec$). The PMT envelope was made of an ultra 
low-radioactivity glass (B53) at our request.  
PMTs were power supplied by the CAEN SY527 
HV system (positive anode, grounded 
cathode), equipped with 
16-channel cards A734P, providing a maximum of 3 kV;
the HV channels were CAMAC controlled 
via the CAENET card C117B. The high voltage
was distributed to the dynodes through a ``tapered bleeder'' progressive
divider, with interdynode voltages progressively increasing in the last dynode 
stages. This divider was preferred to the linear one (with equal
interdynode voltages), since it was proved 
to provide enhanced linearity at a 
still acceptable high voltage for the required gain. The divider consisted in a
circular printed circuit board soldered to the PMT socket 
pins and glued to the 
back of the socket itself. The divider and the socket pins were enclosed in a
Plexiglass cylinder to prevent contact of liquid scintillator with
all the components.

A PMT test facility~\cite{nostro} was designed to determine the relevant 
parameters of all the PMTs. The measurement program allowed 
the quantities listed below to be determined and thus to compare the results 
with the ones contained in the PMT test card given by the producer:
\begin{description}
  \item[(i)] the operating voltage corresponding to 30 mV pulse (on $50\Omega$)
    for a single pe;
  \item[(ii)] the PMT noise level;
  \item[(iii)] the relative light sensitivity (proportional to the photocathode
    quantum efficiency $\times$ pe collection efficiency);
  \item[(iv)] the single pe pulse height spectrum and its peak to valley 
    ($P/V$) ratio;
  \item[(v)] the PMT time characteristics (time jitter and ``walk'' effects).
\end{description}
The results were used to define the optimal working conditions and decide a
proper geometrical arrangement of all the PMT's in the detector. 
Since the light level of neutrino events in CHOOZ corresponded 
to a few photoelectrons at
each PMT, it is at this regime that the listed parameters were determined.
The best suited source was the Hamamatsu light pulser PLP-02 with a laser diode
head SLD-041 emitting at $410\nm$ a wavelength close to the maximum 
photocathode sensitivity and to the maximum bis-MSB emission wavelength. 
Details of the features of this source, the coupling 
to the optical bench and
to PMT housing structure can be found in \cite{nostro}. 
The determination of the PMT relative 
sensitivity was cross-checked by using two
additional ``passive'' light sources: 
a disk of NE110 scintillator activated by 
a low intensity $\nucl{60}{Co}$ and a $\nucl{241}{Am}$ $\alpha$-source coupled
to a NaI crystal. All our sensitivity measurements were found to be 
consistent with each other and with the quantum efficiency as measured by EMI.
\subsection{Detector simulation}
\label{sec:MC}
The Monte Carlo (MC) simulation of the detector 
is based on a Fortran code linked
with the CERN GEANT 3.21 package~\cite{geant}. The program allowed us to 
simulate
the detector response for different particles, in particular for the positron 
and the neutron following an $\Pagne$ interaction. The use of GEANT routines 
had a two-fold objective: 
\begin{description}
\item[-] define the detector geometry and the physical parameters of  
the different materials (scintillators, acrylic, steel) in the detector and
\item[-] track positrons, electrons and 
photons in the detector and evaluate the
energy deposit in the scintillator. 
\end{description}
A specially written routine was used to track secondary particles below 
$10\kev$, the minimum cut-off energy allowed by GEANT, 
down to $1\kev$, in order to
better account for the scintillation saturation effects arising at 
low energies. The routine calculates the range 
corresponding to the particle energy and divides this range into 10 steps; the 
ionization loss is evaluated at each step, and the particle is tracked until 
its residual energy reaches $1\kev$.
Each energy deposit is then converted into the scintillation photon yield; 
ionization quenching effects are taken into account  
according to Birk's law, with $\beta = 1.3\cdot 10^{-2} \cm/\mev$ derived from 
the measurements performed on the similar MACRO 
liquid scintillator~\cite{rliu}.

Scintillation photons are isotropically generated and tracked up to the PMT
surface. Effects due to light attenuation by the scintillators, due to 
reflection on the PMT glass, to the light refraction at
the target boundary and to absorption on opaque surfaces (such as the Geode
and the calibration pipes) are also simulated. The probability of detecting a
photon hitting the PMT surface is weighted according to the quantum efficiency.

The MC code also includes the charge response of the digitizing 
electronics; the charge output for each event is formatted as
for real data, which allowed us to reconstruct MC and real events by means
of the same algorithm. The evaluation of the neutrino detection 
efficiencies strongly relies on MC evaluations; it was therefore crucial that
the MC output was as close as possible to the data. The most important 
difference
between the real and simulated geometry concerns the 
shape of the PMT's, assumed
to be flat and lying on the geode surface (whereas the PMT glass is 
hemispherical and protrudes towards the centre by a few cm). As we 
will see, this 
approximation (needed to avoid a large waste of computing
time) is also responsible for the fragility of the reconstruction algorithm in 
the case of events close to the PMTs.

User-defined routines~\cite{bernard} were introduced in the main MC code to 
follow the neutron moderation and capture processes. Tabulated cross section 
values were used for both elastic scattering and capture on Gd, H, C, O, 
Fe~\cite{BNL}. At 
each step (whose length depends on the total cross section at the neutron 
velocity) a decision is taken if either scattering or capture processes (or 
neither) arise. In the case of captures, $\gamma$-rays are emitted in cascade 
according to the de-excitation scheme of the nucleus involved; each photon is 
then tracked by GEANT routines. Captures on Hydrogen are followed by the 
emission of one $2.2\mev$ $\gamma$-ray. The case of Gadolinium is harder to
manage; the excited nucleus can decay to
the ground state through a series of intermediate excited levels, both 
discrete and continuously distributed. Each capture event typically releases 
about three $\gamma$-rays with a total energy depending on which Gd isotope 
captures the neutron. The information used about the energy
levels involved and the relative decay branches are taken from~\cite{groshev}. 
%
\section{The trigger}
The electronic system for the CHOOZ experiment collected and processed
PMT output signals to select events due to neutrino interactions
and to separate them from the background. The charge and time information 
provided by the PMT's were used to reconstruct the energy 
and position of each event. 

The signature of a neutrino interaction relied 
on the delayed coincidence of two signals within a $100\usec$ time window: the 
former, due to the positron, has an average energy of about $3\mev$; the 
latter, due to the delayed neutron capture on Gadolinium, has a total 
$\gamma$-ray energy $\approx 8\mev$. The identification of this event 
combination was
therefore based on a two-level trigger system; the first level selected events 
with an energy release $> 1.3\mev$ (L1lo) or $>3\mev$ (L1hi); the occurrence of
the sequence (L1lo,L1hi), within the $100\usec$ time window, generated the 
second level which triggered the data digitization 
and recording by the on-line system.

\subsection{The first-level trigger}
\begin{figure}[htb]
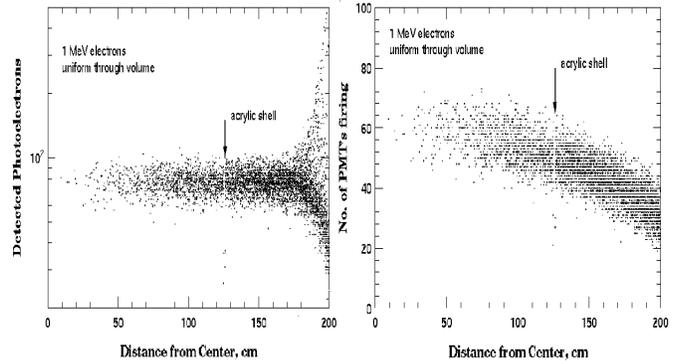

\begin{center}
\mbox{
\hspace{-0.3cm}\includegraphics[height=0.543\linewidth,width=0.493\linewidth]{fig.16a}
\hspace{-0.1cm}\includegraphics[height=0.55\linewidth,width=0.5\linewidth]{fig.16b}}
\end{center}
\caption{\small Number of detected photoelectrons QSUM (left) and number of hit
PMT's NSUM (right) for $1\mev$ electron events, as a function of the distance 
from the detector centre.}
\label{fig:Nphe_NPMT}
\end{figure}

The electronic sum of the signals corresponding 
to photoelectrons detected by all
PMT's gave a first rough measurement of the energy released by any event. A 
Monte Carlo simulation of electrons uniformly generated in the detector
(whose results are summarized in Fig.~\ref{fig:Nphe_NPMT}) showed that the 
collected charge is linearly dependent on the energy deposit and almost 
independent of the position within Region I.
In Region II this changes and the total number of photoelectrons for events
within $30\cm$ from the PMT's can be as much as 10 times larger than for events
at the centre. So, a trigger exclusively based on the 
charge information would not have rejected 
lower energy events generated close to the 
PMT surface.

A ``topological'' cut was then applied by also requiring a minimum number of 
hit PMT's. Fig.~\ref{fig:Nphe_NPMT} clearly shows that this extra condition 
preferentially rejected events close to the PMT's. 

\begin{figure}[htb]
\begin{center}
\mbox{\includegraphics[width=0.9\linewidth]{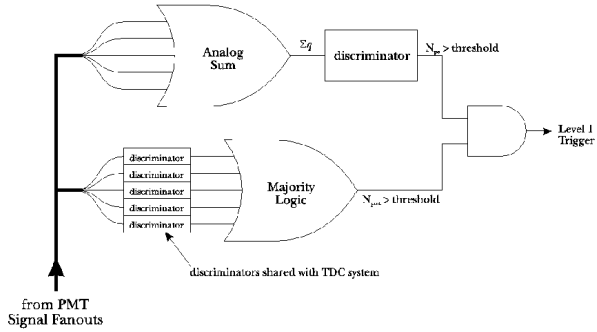}}
\end{center}
\caption{\small First-level trigger scheme. Both the number of photoelectrons 
(QSUM) and the number of hit PMT's (NSUM) are required to fulfil a certain 
threshold condition.}
\label{fig:Level1}
\end{figure}

The first-level (L1) trigger was generated when both the charge and the 
PMT multiplicity fulfilled the above criteria. A simple schematic of the 
trigger circuit is presented in Fig.~\ref{fig:Level1}. The signals 
from each PMT were fed into the front-end electronics made 
of fan-in/out modules developed for the purpose. 
These modules provided a linear sum of the input signals (QSUM) whose 
amplitude was proportional to the total number of photoelectrons. The PMT 
multiplicity signal (NSUM) was obtained in a similar way. A copy of the PMT 
signals was fed into Lecroy LRS 4413 discriminators which had their threshold 
adjusted to $15\mv$ ($\approx$~one half of the single photoelectron amplitude).
The linear sum of these discriminator outputs, whose amplitude was proportional
to the number of channels over threshold on each discriminator board, formed 
the NSUM signal. The L1 trigger was finally asserted when both the 
resulting QSUM and NSUM signals exceeded the preset thresholds.
\subsection{The second level trigger}
The second-level (L2) condition (the one identifying 
the neutrino interactions) triggered on the occurrence of two events satisfying
the L1 condition within a $100\usec$ wide time window. 
%
This width represented a trade-off between a high neutrino detection 
efficiency and a high rejection of the accidental background.
The positron--neutron delay followed an exponential distribution 
with an average $\tau \simeq 30 \usec$, therefore only $5\%$ of the neutrino 
events were missed by the delayed coincidence window. 

The L2 trigger logic is somewhat more complicated than the
simple scheme just described. There were two different
conditions for the fulfilment of the L1 trigger, corresponding to two 
different energy thresholds: the low 
condition (L1lo) roughly corresponded to $1.3\mev$ and had a rate
$\approx 130 \s^{-1}$, the high condition (L1hi) corresponded to $\approx 
3\mev$ and had a rate $\approx 30 \s^{-1}$. The higher threshold rejected 
the large background due to 
the natural $\gamma$-radioactivity (which practically ends with 
the $2.6\mev$ $\nucl{208}{Tl}$ line) and was 
sufficiently low to detect signals 
due to the neutron capture on Gd with almost full efficiency.
%
%

The L2 trigger was vetoed if some activity was present in Region III; this 
selection rejected events associated with cosmic rays. The VETO 
condition was satisfied whenever VSUM, the sum of the signals of the 48 PMT's 
located in Region III, was over a preset threshold. If no VETO 
condition occurred during a $1\msec$ time interval preceeding L2, the 
L2 triggered: the acquisition of further secondary particles went on for 
an extra-time of $100\usec$ and then stopped. The on-line processor took 
$\approx 80\msec$ to read all the electronics and to record the data on disk.
During this time the computer {\it busy} condition was on and the
acquisition of any further event was disabled. 

Although a neutrino interaction corresponded to an L1lo-L1hi or 
L1hi-L1hi time sequence, the trigger logic enabled the acquisition of 
events with an L1hi-L1lo sequence as well, allowing a systematic study of
the accidental background.

Finally, although the acquisition rate ($\approx 0.15 \s^{-1}$)
was rather low, it was large compared to
the expected neutrino rate at full reactor power
($\approx 30$ events $\days^{-1}$). The typical event size 
was roughly 30 Kbytes;
the acquisition system could handle a 
daily amount of data in the
order of 0.5 Gbytes. This capability was achieved through fast readout 
electronics and large storage devices.
\subsection{Trigger for calibration runs}
Acquisition schemes other than
the normal neutrino trigger were required for 
calibrations or particular test runs. The trigger logic 
could be changed at the 
beginning of each run by loading an appropriate {\it trigger table}, a 
combination of the level trigger, veto and busy 
signals assembled by a dedicated
logic unit Caen C542.
\subsection*{Neutron sources}
Neutron sources were extensively used for detector calibration, both to 
define an absolute energy scale and 
to measure the neutron detection efficiency. A $\nucl{252}{Cf}$ source was the 
most frequently used. This nucleus
undergoes spontaneous fission, simultaneously emitting prompt neutrons and
$\gamma$-rays below $10\mev$. The
neutron multiplicity obeys Poissonian statistics with an average value 
$3.787$~\cite{Spencer} and the kinetic energy is Maxwell-distributed, with 
an average energy equal to $1.3\mev$. 

We also used a Am/Be source. This emits neutrons, part of which follow the 
emission of a $4.4\mev$ prompt $\gamma$-ray (see \cite{ambe} and references
therein for details).

With the source inserted in the centre of Region I, almost all the neutrons 
were contained and captured: some ($\approx 15\%$) were captured on Hydrogen, 
the majority on Gadolinium. The neutron binding energy was released in the form
of $\gamma$-rays (one $2.2\mev$ $\gamma$ for H, an average of 3 $\gamma$'s and
$\sum_\gamma E_\gamma \approx 8\mev$ for Gd). In order to record events due to 
neutron capture on H as well, we defined a different 
trigger logic, by-passing the
L1hi threshold condition; the L2 condition relied on the occurrence
of two signals both satisfying the L1lo threshold. 
The ``positron'' role was, in 
this case, played by prompt $\gamma$'s, 
while the second L1 trigger was provided 
by the first neutron capture (if within the $100\usec$
coincidence time-window). 

\begin{figure}[htb]
  \begin{center}
    \mbox{\includegraphics[width=0.8\linewidth]{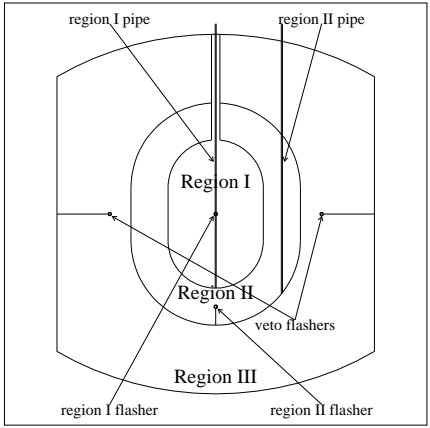}}
    \caption{\small Location of laser flashers and calibration pipes in the 
             detector.}
    \label{fig:calisou}
  \end{center}
\end{figure}

The source could be inserted in the detector 
through two vertical calibration pipes: a
central pipe, along the symmetry axis of the detector, and a parallel one,
in Region II, just at half distance between the geode and the target boundary, 
as shown in Fig.~\ref{fig:calisou}. It 
was therefore possible to study the detector response anywhere along the
$z$-axis in both regions. In particular, as we will see in the next section, 
calibration runs at different, known, positions allowed 
accurate tuning of the
reconstruction algorithm and comparison with Monte Carlo forecasts.
\subsection*{$\gamma$ sources}
Daily calibration runs were performed by means of a $\nucl{60}{Co}$ 
source, a well-known $\gamma$-emitter providing a low-energy calibration point 
at $2.5\mev$ (the sum of $1.17$ and $1.33\mev$ simultaneous $\gamma$-lines). As
we will see in the following sections, these runs were particularly useful to 
follow the detector evolution (a critical point because of the aging of
the Gd-loaded scintillator), and to tune and check the trigger efficiency.

The neutrino trigger was not suited for acquiring these events, since the 
delayed neutron signal was missing. In this case, the L2 trigger was made to 
coincide with the L1lo trigger. With such a trigger we also recorded 
a great number of events due to $\gamma$-radioactivity natural background. 
\subsection*{Laser}
A laser based calibration system was also available. It consisted 
of a nitrogen UV Laser emitter ($\lambda = 337\nm$) 
with a repetition rate $\approx 10\hz$ and
a time jitter of less than $1\nsec$ between the trigger and the light emission.
The light intensity could be selected by 
using two variable neutral density filter
wheels rotated by remote-controlled stepping motors. 
The attenuated light was fed to any one of six locations (remotely selectable),
using UV quartz fibers (only two out of the four Veto flasher positions are
shown in Fig.~\ref{fig:calisou}).
At each position the bare fiber was terminated in a 
quartz vial filled with a dilute p-terphenyl solution, comprising the 
{\it flasher}. The fluorescence process shifted the laser wavelength to the 
same emission spectrum produced by scintillation, resulting in an isotropic 
point source having the same attenuation properties, time characteristics and
PMT quantum efficiency as light generated by energetic particles.  Time and
pulse height calibrations of the detector could be carried out and monitored ona pulse-to-pulse basis. The monitors consisted of a photodiode, used to monitor
laser intensity, and a separate fiber that directed light passing through the 
attenuator into a monitor PMT located next to the laser. Each of the monitors 
could be calibrated relative to an absolute standard.

In this case the L2 was given by the internal laser trigger after a delayed of 
$\approx 200\nsec$ relative to the light emission. 
\subsection{The neural network trigger}
This trigger~\cite{NNtrigger} was intended to work as
a fast acquisition filter for selecting events inside
a fiducial volume, and secondly, it could be used for 
triggering upon coincidence of two L1lo triggers, therefore acquiring
those neutrino events followed by neutron capture on Hydrogen (instead of
Gadolinium).

This technique is not 
essential in the case of CHOOZ, since the amount of H-captures is 
$\approx 15\%$ 
of the total neutrino interactions in the target, but it 
could be important in on--going reactor experiments (namely KamLAND), where 
only a high-flash liquid scintillator (not Gd-loaded) is used.

The neural network algorithm was implemented on a VME parallel processor
which reconstructed in $\approx 180\usec$ 
the energy and position of each event from the
information provided by the NNADC digitizing system (see next section).

Due to the progressive deterioration of the Gd scintillator
the potentialities of this trigger could not be fully exploited in
this experiment.
\section{Data acquisition}
The data acquisition system consisted of various and distinct systems 
controlled by a VME-based processor:
\begin{center}
\begin{tabular}{lll}
NNADC &$\rightarrow$ & ADC units (VME),\\
SWFD & $\rightarrow$ & Slow ($20\Mhz$) WFD units (VME),\\
FWFD & $\rightarrow$ & Fast ($150\Mhz$) WFD units (FastBus),\\
FBADC & $\rightarrow$ & ADC units (FastBus),\\
FBTDC & $\rightarrow$ & TDC units (FastBus).
\end{tabular}
\end{center}
The PMT signals were fanned out and fed into each digitizing 
system, according to the diagram shown in Fig.~\ref{fig:eledia}. 
Since the PMT's were used in the grounded cathode configuration, their signals
were fed to the circuits via AC couplings.
For the
first three systems, the signals from groups of PMTs were fanned in;
for the NNADC's and SWFD's the PMT signals were linearly added by
groups of 8, a PMT {\it patch}. This reduced the 
number of channels to 24. For the FWFD's the PMT groups varied from 4 to 
8 during the experiment.
\begin{figure}[htb]
  \begin{center}
    \mbox{\includegraphics[width=\linewidth]{fig.19}}
    \caption{\small Electronics layout for the CHOOZ experiment, including
             front-end, trigger and digitizing modules.}
    \label{fig:eledia}
  \end{center}
\end{figure}
\subsection{The on-line system}
The acquisition system combined different
bus standards (see Fig.~\ref{fig:online}); the digitizers
listed above used both the VME and the FastBus standard, while the trigger 
electronics (including discriminators and logic units) was CAMAC based. The 
data were managed by a VME, OS-9 operating processor, 
whose central unit consisted of a Motorola 68040 microprocessor mounted
in a CES FIC 8234 board. After completing the event readout, the data
were sent through Ethernet to a dedicated SUN/Unix station and written on disk.
A LabVIEW run controller was the interface between the user and the SUN; the
controller also provided real-time information about the run and data quality.
\begin{figure}[htb]
  \begin{center}
    \mbox{\includegraphics[width=0.9\linewidth]{fig.20}}
    \caption{\small On-line system architecture, with special reference to the
             different bus standards (VME, FastBus, CAMAC) and their 
             interconnection (VIC,VMV).}
    \label{fig:online}
  \end{center}
\end{figure}
\subsection{The digitizing electronics}
\subsection*{The NNADC's}
%
The signals from the 24 PMT patches were sent to two VME 12-bit ADC banks, 
each composed of three 8-channel Caen V465 boards. Two banks were necessary 
to avoid any dead time between 
positron and neutron pulses, the time needed for 
the charge integration being $\approx 14\usec$. The charge was integrated
within a fast $200\nsec$ gate. An external logic 
alternately switched the gate to the banks on the occurrence of an L1lo 
trigger. The digitized ADC values were arranged in a FIFO, thus allowing an
internal multi-hit buffering in each bank. The ADC readout 
was performed by a Motorola 68030-based Themis TSVME133 VME processor board 
(referred to as 
``Themis'' in Fig.~\ref{fig:online}). 
The TSVME133 was usually in a {\it wait} status until it 
received a VME interrupt generated when one of the two ADC banks was {\it not
empty}. 

The NNADC circuitry also included:
\begin{itemize}
\item[-] 1 scaler Caen V560N:
\item[-] 1 flash-ADC Caen V534:
\item[-] 1 input/output register Caen V513:
\item[-] 1 Adaptive Solutions CNAPS parallel processor board.
\end{itemize}
The scaler and flash-ADC provided redundant information on the 
relative timing of events. 
The flash-ADC board was also used to 
sample the trigger signals (L1lo, L1hi, L2 and VETO).
The other two boards were used to produce the Neural Network 
trigger.

\subsection*{The SWFD's}
The Slow WaveForm Digitizer system used a set of 5 flash-ADC Caen V534 boards,
with $20\Mhz$ internal clock speed. These digitizers recorded the 30 Geode
plus Veto patch signals from $100\usec$ before to $100\usec$ after the L2 
trigger. All the input signals needed a 
proper stretching in order to fit to the 
inherent $50\nsec$ resolution. Therefore these digitizers were 
essentially peak 
sensing devices and could not be used for pulse shape analysis. Extra
channels recorded all the detector triggers.
\subsection*{The FWFD's}
With their $155\Mhz$ sample clock, these digitizers were 
the only equipment which could permit the study of pulse shapes. 
These units were equipped with an 8-page memory, the depth of one page being 
$206\nsec$, in order to record up to 8 events before the stop. 
\subsection*{The FBADC's}
The FastBus ADC system had four 96-channel Lecroy 1885F boards to separately
measure the anode charge of each geode PMT. Even in this case, the boards were 
arranged in two alternating banks in order to 
record both events associated with the L2 trigger.
\subsection*{The FBTDC's}
The FastBus TDC's recorded the time information of each geode PMT. Referring to
Fig.~\ref{fig:eledia}, the input signals consisted of the logical (NIM) pulses
generated by the trigger discriminators at minimum threshold. This system 
consisted of two 96-channel Lecroy 1877 multi-hit units able to record up to 16
hits in one single channel with $2\nsec$ resolution. 

\section{Detector calibration}
The reconstruction methods were based, for each event, on the
VME ADC's digitization of the signals from the 24 patches which grouped  
the 192 PMTs looking at Region I. It was checked that this 
grouping does not significantly affect the quality of the reconstruction.

A good determination of the position and energy of each event and of the 
detection efficiency depends to a large extent on the knowledge of the main 
parameters of the detector (scintillator light yield and attenuation length,
PMT and electronics gains). 
\subsection{PMT gain}
\begin{figure}[htb]
  \begin{center}
    \mbox{\includegraphics[width=0.8\linewidth]{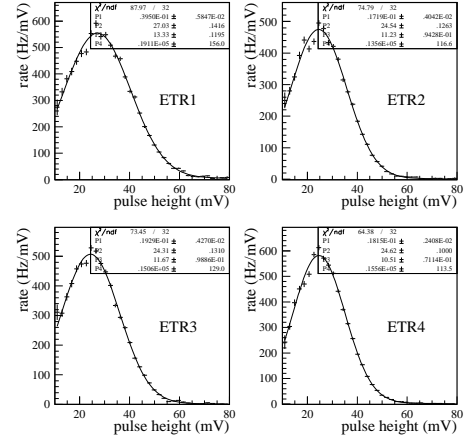}}
    \caption{\small Pulse height spectra for a sample of four PMTs.}
    \label{fig:PMTrate}
  \end{center}
\end{figure}

\begin{figure}[htb]
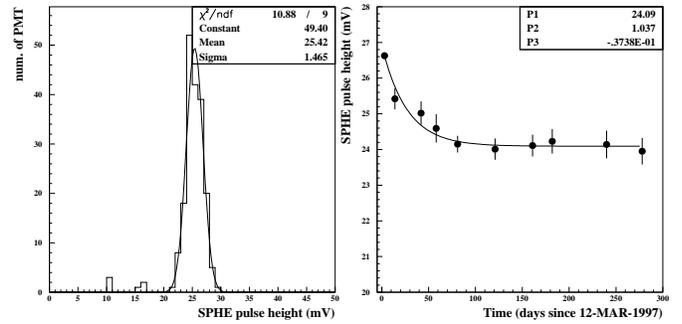

  \begin{center}
    \mbox{\includegraphics[width=0.54\linewidth]{fig.22a}
          \hspace{-0.5cm}\includegraphics[width=0.54\linewidth]{fig.22b}}
    \caption{\small Distribution of the single pe peak for all PMTs 
             (left) and its time evolution since the start of data taking 
             (right).}
    \label{fig:PMTgain}
  \end{center}
\end{figure}

The stability of the PMT gain was checked by periodic measurements of the
single photoelectron pulse height spectrum of each PMT. Each spectrum was 
obtained by recording the PMT counting rate, out of a discriminator, as a 
function of its threshold. The data were taken by a LabVIEW code running in parallel with the main acquisition.
%
%
No artificial light source was needed, since the single counting rate, at the 
one-pe level, due to the $\gamma$-radioactivity and to the dark noise, was 
$\approx 1 \khz$. Examples of these curves are shown in Fig.~\ref{fig:PMTrate}.
The single pe spectra, obtained by differentiating the counting rate curves, 
were then fitted to get the PMT gains; values around 30 mV for the geode PMTs 
and 5 mV for the Veto ones were needed to match our front-end electronics. 
Fig.~\ref{fig:PMTgain} 
shows the distribution of the average single pe pulse 
height for the whole of the geode PMTs and its evolution since the 
beginning of the data taking. One concludes that the PMTs stabilized 
about one month after burn-in and that the asymptotic gain was $\approx
10\%$ lower than the starting value.
\subsection{Determination of the photoelectron yield}
A similar method was used to determine the photoelectron yield for events at the
detector centre. We used the laser flasher at the detector centre as a 
calibrated light source. The laser  provides its own trigger, so the counting rate is not affected by radioactivity.

An estimate of the number of photoelectrons $N^i$ detected by each PMT
comes from the probability of having zero photoelectrons 
when the average number is 
$N^i$. From Poisson statistics, we obtained for the i-th PMT
\begin{equation}
  N^i = -\log(1-\frac{N_{hit}^i}{\varepsilon^i N_{shot}})
  \label{pheyie}
\end{equation}
where $N_{hit}^i,N_{shot}$ are respectively the counts over threshold for that 
PMT and the number of laser shots. (\ref{pheyie}) includes also a detection 
efficiency $\varepsilon^i$, equal to the probability for a single pe pulse 
height to be over threshold, which can be evaluated for each PMT from the pulse
height spectra shown in Fig.~\ref{fig:PMTrate}. 
The light intensity was kept at a level of about $0.5$~pe/PMT so as to increase
the sensitivity of the method. 
This number was divided by the energy deposit corresponding to the
light intensity in order to determine the absolute light yield. As an energy 
reference we used the $2.5\mev$ $\nucl{60}{Co}$ ``sum'' line. By averaging the 
number of photoelectrons all over the PMTs, we obtained 
$(0.65 \pm 0.03)$~pe/MeV/PMT and a yield of $(125\pm 5)$~pe/MeV.
\subsection{ADC calibration in a single photoelectron regime}
\begin{figure}[htb]
  \begin{center}
    \mbox{\includegraphics[width=\linewidth]{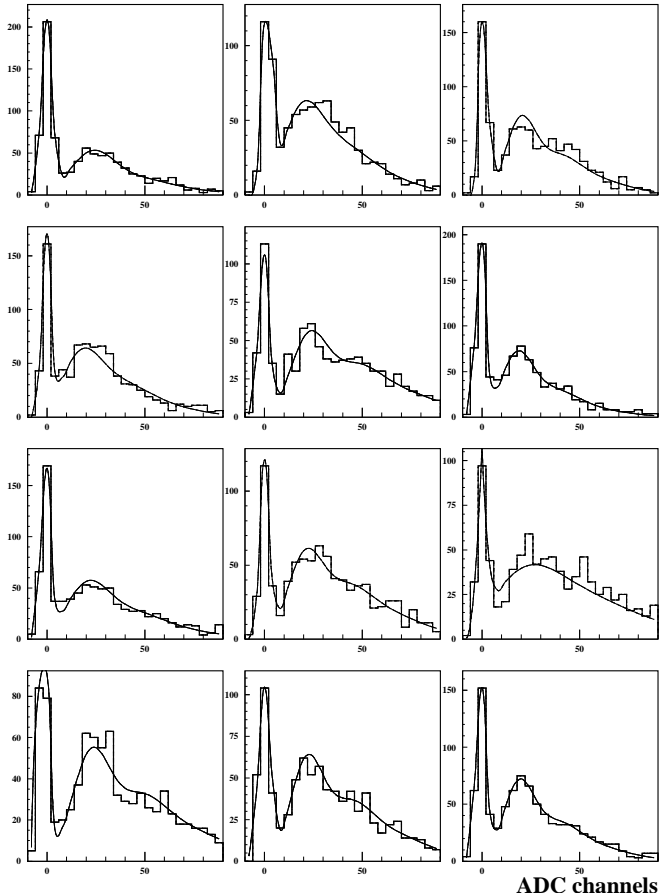}}
    \caption{\small ADC spectra obtained with a laser calibration, where the 
             single pe peak is clearly visible. The fitting function results
             from the sum of the pedestal and the first three photoelectrons.}
    \label{fig:patchgain}
  \end{center}
\end{figure}

Calibration runs using the laser at a low intensity were periodically 
performed to test the single photoelectron gain and to calibrate the VME ADCs. 
The good ADC resolution and the high PMT gain allow us to distinguish the 
contributions to the ADC spectra due to different numbers of photoelectrons. 
A sketch of these spectra is presented in Fig.~\ref{fig:patchgain}. The fit 
function results from the sum of a pedestal and a number of pe-distributions; 
each of these terms is represented
by a Gaussian function (weighted according to Poisson statistics) whose 
peak position is linear with the number of photoelectrons.

Apart from the ADC calibration, the fit also provides an independent 
determination of the number of photoelectrons 
collected by each PMT patch which is 
consistent with the measured photoelectron yield.
\subsection{Light attenuation in the Gd-loaded scintillator}
\label{sec:atten}
\begin{figure}[htb]
  \begin{center}
    \mbox{\includegraphics*[bb=0 130 600 700,width=0.85\linewidth]{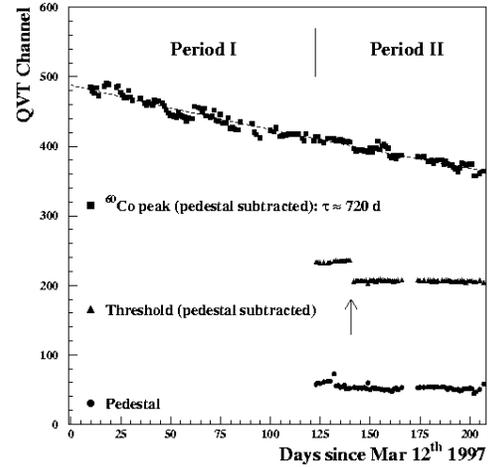}}
    \caption{\small Peak associated with the $\nucl{60}{Co}$ $2.5\mev$ line, as
             a function of time, as measured by means of a Lecroy QVT. The 
             detected charge follows an exponential decrease, with decay time 
             $\approx 720\days$.}
    \label{fig:CobQVT}
  \end{center}
\end{figure}

The degradation of light transparency in Region I (see \S\ref{sec:liqscint}) 
was regularly monitored in the apparatus by recording the number of 
photoelectrons associated with the $\nucl{60}{Co}$ $\gamma$-line, as shown in 
Fig.~\ref{fig:CobQVT}.

The light absorption of Region I scintillator was periodically measured  
throughout the acquisition period. 
The method we used consisted in displacing a light source along the 
calibration pipe and in recording the charge detected by the top and bottom PMT
patches. If we assume an exponential light attenuation (which is a good 
approximation for a light path longer than $10\cm$, as already remarked), 
these charge values are related by the simple expression
\begin{equation}
  \frac{Q_T}{Q_B} = \frac{\Omega_T}{\Omega_B} \exp(\frac{d_T-d_B}{\lambda_{Gd}})
  \label{QTQB}
\end{equation}
$\Omega_{T,B}$ being the solid angle subtended by the top (bottom) patch PMTs
and $d_{T,B}$ the average distance between these PMTs and the source position.

For this measurement we used a radioactive
source, namely $\nucl{252}{Cf}$. This source was preferred to a $\gamma$
source (such as $\nucl{60}{Co}$) since the double signature, provided by the
prompt $\gamma$'s and the neutron captures, makes the identification of source 
events much easier. The $\nucl{252}{Cf}$ neutron emission and capture, the 
involved light paths and solid angles were simulated by the MC method.

\begin{figure}[htb]
  \begin{center}
    \mbox{\includegraphics[width=0.7\linewidth]{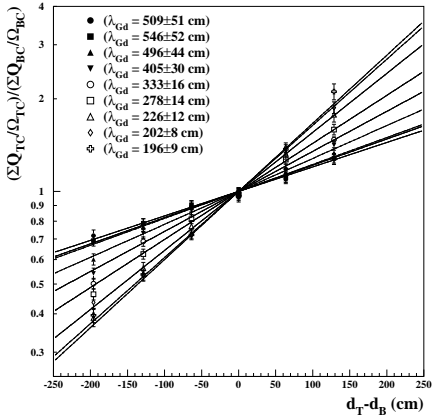}}
    \caption{\small Attenuation length measurements at different stages of the 
             data taking period.}
    \label{fig:topbot}
  \end{center}
\end{figure}

\begin{figure}[htb]
  \begin{center}
    \mbox{\includegraphics[width=0.7\linewidth]{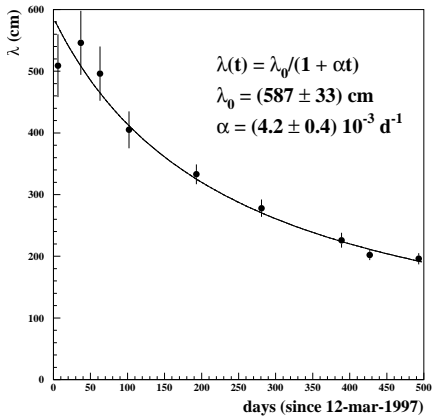}}
    \caption{\small $\lambda_{Gd}$ versus time with best-fit function
             superimposed.}
    \label{fig:tlam}
  \end{center}
\end{figure}

The results obtained from different measurements are displayed in
Fig.~\ref{fig:topbot}, superimposed on the exponential best-fit curves 
according to (\ref{QTQB}). The charge values in use are those corresponding
to the $2.2\mev$ $\gamma$-line due to the neutron capture on Hydrogen. We did 
not use the $8\mev$ line in order to reduce systematics due to ADC 
saturation effect which might have arisen when 
the source approached the edges of Region I.
The fitted $\lambda_{Gd}$ values are plotted versus time in 
Fig.~\ref{fig:tlam}. The time evolution of the absorption length was fitted by 
the empirical function (\ref{eq:lambda}) (already used in the laboratory test).
%
%
By taking the $\lambda_{Gd}$ values, at the 
beginning and at the end of the experiment, one can estimate
a $35\%$ reduction in the 
photoelectron yield, in agreement with the measurement of the 
QSUM for events at the detector centre (Fig.~\ref{fig:CobQVT}).
\subsection{Electronics amplification balancing}
The electronic gain may differ from patch to patch and slightly vary with
time because of the behaviour of the active electronic 
components in the circuits treating the signals from 
the PMTs. Amplification balancing 
factors were obtained three times a week using the ADC charge values 
corresponding to the $8\mev$ line peak generated 
with the $\nucl{252}{Cf}$ source at the detector centre.
%
\section{Event reconstruction techniques}
\label{sec:reco}
\subsection{The standard minimization algorithm}
The standard algorithm uses a maximum likelihood method to reconstruct the
energy $E$ and the vertex $\vect{x}$ of an event. The likelihood is defined as 
the joint Poissonian probability of observing a measured distribution of 
photoelectrons over the 24 patches for given $(E,\vect{x})$ coordinates in the 
detector. 
So, for an event occurring at time $t$ after the start of data taking, we can 
build a likelihood function as follows:
\begin{equation}
  {\cal L}(N;\overline{N}) = 
             \prod_{j=1}^{24} P(N_j;\overline{N}_j(E,\vect{x},t)) =
             \prod_{j=1}^{24} \frac{\overline{N}_j^{N_j}}{\fact{N_j}}
             e^{-\overline{N}_j}
  \label{nphelike}
\end{equation}
where $N_j$ is the observed number of photoelectrons and $\overline{N}_j$ the
expected one for the j-th patch, given an event $(E,\vect{x},t)$. The reason 
for using a Poissonian instead of Gaussian statistics is due to the frequent 
occurrence of low energy events, with low number of photoelectrons detected by 
some PMT patches.

The values $N_j$ are obtained from the data recorded by the VME ADC's by 
applying the reference ADC gain $g_0$ and the balancing factors $f_j$: so
\begin{equation}
  N_j = \frac{Q_j}{g_0 f_j(t)}
  \label{nobs}
\end{equation}
The predicted number of photoelectrons for patch $j$ is computed by considering
a local deposit of energy, resulting in a number of visible photons which are
tracked to each PMT through the different attenuating Region 1 and 2
scintillators. Therefore
\begin{equation}
  \overline{N}_j = \alpha E \eta \sum_{k=1}^8 \frac{\Omega_{jk}(\vect{x})}{4\pi}
    \exp\left(-\frac{d_{1jk}(\vect{x})}{\lambda_{Gd}(t)}
         -\frac{d_{2jk}(\vect{x})}{\lambda_{Hi}}\right)
  \label{nthe}
\end{equation}
where
\begin{center}
\begin{tabular}{cl}
$E$ & is the energy deposited in the scintillators,\\
$\alpha$ & is the light yield of the scintillator,\\ 
$\eta$ & is the average PMT quantum efficiency,\\
$\Omega_{jk}$ & is the solid angle subtended by the k-th PMT \\
& from the position,\\
$d_{1jk}$ & is the path length in Region I,\\
$d_{2jk}$ & is the path length in Region II,\\
$\lambda_{Gd}$ & is the attenuation length in Region I scintillator,\\
$\lambda_{Hi}$ & is the attenuation length in Region II scintillator.
\end{tabular}
\end{center}
To reduce computing time PMTs are considered to be flat and the solid
angle is approximated by the following expression
\begin{equation}
  \Omega_{jk} = 2\pi \left(1 - \frac{d_{jk}}{\sqrt{d_{jk}^2+r_{PMT}^2  
\cos\theta}}\right)\,
  \label{solid}
\end{equation}
$r_{PMT}$ being the PMT photocathode radius, $\theta$ the angle between the 
event-PMT direction and the inward unit vector normal to the PMT surface and 
$d_{jk} = d_{1jk} + d_{2jk}$. 

Instead of directly using Eq.~(\ref{nphelike}) it is more convenient to  
exploit the ``likelihood ratio test'' theorem to convert the likelihood
function into a form which obeys the $\chi^2$ distribution\cite{Eadie}. We let 
${N_j}$ be the best estimate of the true (unknown) photoelectron distribution
and form the likelihood ratio $\lambda$ defined by
\begin{equation}
  \lambda = \frac{{\cal L}(N;\overline{N})}{{\cal L}(N;N)}
  \label{likerat}
\end{equation}
The ``likelihood ratio test'' theorem states that the ``Poissonian'' $\chi^2$,
defined by 
\begin{equation}
  \chi^2 = -2 \log \lambda = 2 \sum_{j=1}^{24} [ \overline{N}_j - N_j + N_j 
           \log(\frac{N_j}{\overline{N}_j})],
  \label{chipois}
\end{equation}
asymptotically obeys a chi-square distribution\cite{Cousins}. 
It is easy to prove that the minimization of $\chi^2$ is equivalent to 
maximization of the likelihood function, so that the $\chi^2$ statistic may be 
useful both for estimating the event characteristics and for goodness-of-fit 
testing.

\begin{figure}[htb]
  \begin{center}
    \mbox{\includegraphics[width=0.8\linewidth]{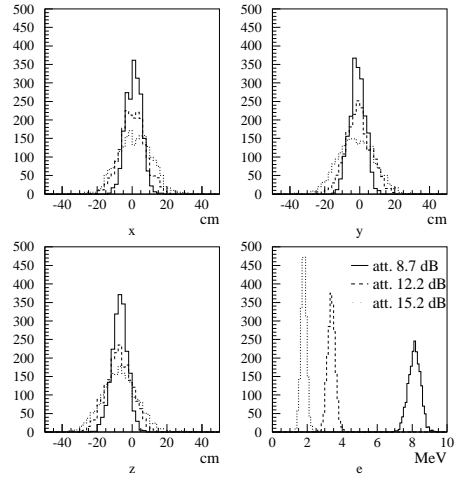}}
    \caption{\small Comparison of position and energy distributions for runs 
             with the laser flasher at the detector centre, corresponding to 
             three different light intensities.}
     \label{fig:laser}
  \end{center}
\end{figure}

\begin{figure}[htb]
  \begin{center}
    \mbox{\includegraphics[width=\linewidth]{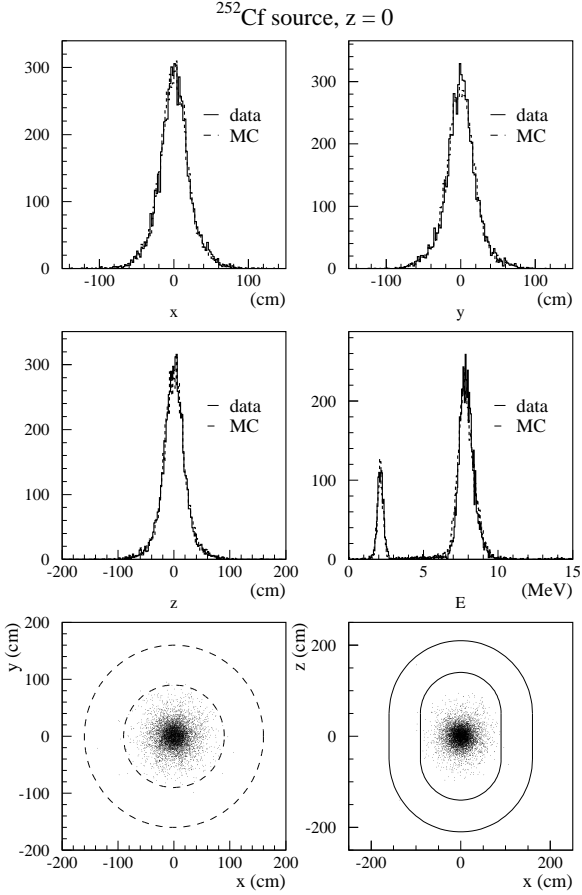}}
    \caption{\small Data and Monte Carlo distributions of neutron events with 
             the $\nucl{252}{Cf}$ source at the detector centre. Dashed and 
	     solid lines in the bottom figures represent respectively the top
	     and the side view of the edge of the target and the geode.}
    \label{fig:cf0}
  \end{center}
\end{figure}

\begin{figure}[htb]
  \begin{center}
    \mbox{\includegraphics[width=\linewidth]{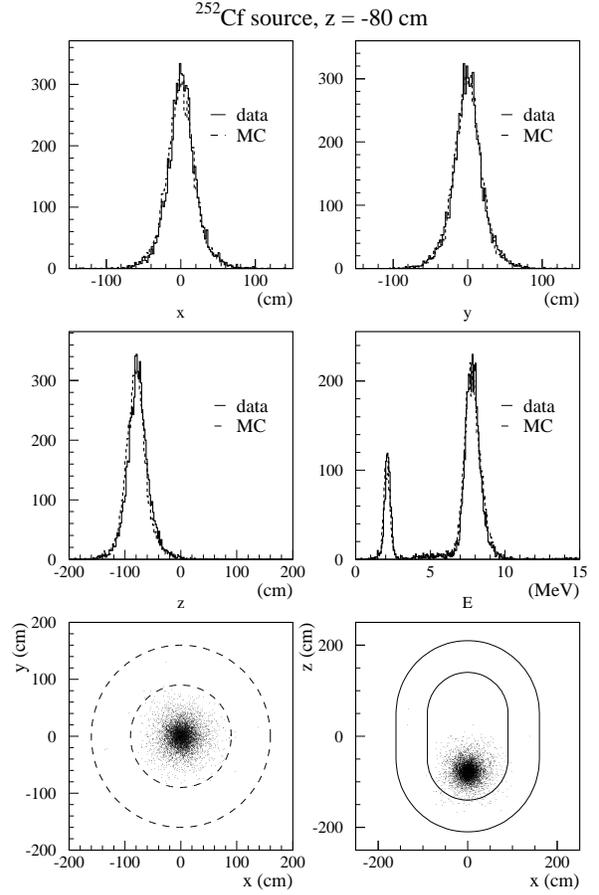}}
    \caption{\small Same as before, with the source at $z=-80\cm$.}
    \label{fig:cf80}
  \end{center}
\end{figure}

We used the MIGRAD minimizer provided by the \\ 
CERN MINUIT package~\cite{MINUIT} 
to minimize (\ref{chipois}). The search for the minimum
$\chi^2$ proceeds through the computation of the first derivatives of 
(\ref{chipois}). This routine proved very powerful, provided 
the starting values for the fit parameter are accurate. We studied the $\chi^2$
profile by reconstructing Monte Carlo generated events. Several relative minima
were found, most of them differing by more than $1\sigma$ from the generated 
$(E,\vect{x})$ coordinates. This is the reason why a 
suitable choice of these starting values is crucial
in event reconstruction. In our case, an indication of the event position comes
from the asymmetry of the charge distribution: no significant 
asymmetry will be visible for events at the centre while the charge 
distribution will be more and more asymmetric for events approaching the 
detector boundary. We subdivided the PMT patches into 6 ``superpatches'',
two patches for each coordinate, and built
%
%
the starting point 
for the i-th coordinate according to the following equation:
\begin{equation}
  x_{i0} =\frac{\sqrt{Q^i_+}-\sqrt{Q^i_-}}{\sqrt{Q^i_+}+\sqrt{Q^i_-}} D^i,
  \,\,\,i=1,2,3,
  \label{start}
\end{equation}
where the indices $+-$ refer to the opposite superpatches of the i-th axis and 
$D^i$ is the half size of the detector along that axis. Once the $x_{i0}$
corresponding to the starting position is known, the starting energy value is 
obtained from (\ref{nthe}) after replacing $\vect{x}$ with $\vect{x}_0$ and 
${\overline{N}_j}$ with ${N_j}$.
%
\label{sec:perfo}
The event reconstruction was tested by analysing events generated with
calibration sources in various positions inside the detector. Let us review the
results.
%
Distributions of reconstructed events generated by the laser flasher at the 
detector centre are presented in Fig.~\ref{fig:laser}. The standard deviation 
of the distributions shown gives an indication of the resolution, both in 
energy and position. The fit yields $\sigma_x \approx 4\cm$ for each
coordinate and an energy resolution $\sigma_E \simeq 0.33\mev$ at an equivalent
energy of $8.1\mev$ in which the statistical fluctuations in the
number of photoelectrons prevail. Photoelectron statistics also affect the 
position resolution. The 
effect is clearly visible in the data taken with the same laser flasher for 
three different light intensities (in Fig.~\ref{fig:laser} a larger filter 
attenuation corresponds to a lower intensity).
As can be seen in Fig.~\ref{fig:laser}, 
the average $z$ coordinate is displaced by $7\cm$ from
the nominal position; this is due to the effect of the shadow of the 
flasher already mentioned; in this case, the flasher points downward, thus 
producing a shadow in the upward direction and displacing the event below the 
true source position. 
%
In Figs.~\ref{fig:cf0}, \ref{fig:cf80} we report 
the results of the calibration runs with the $\nucl{252}{Cf}$ source at two 
different positions
($z=0,-80\cm$) along the calibration pipe. The position distribution is 
Gaussian for all the coordinates, with $\sigma_x \approx 19\cm$.
An equivalent number of neutrons was generated, at each position, by our Monte
Carlo code and similarly analysed. The figures also display the Monte 
Carlo data to emphasize the agreement between data and expectations.
%

\begin{figure}[htb]
  \begin{center}
    \mbox{\includegraphics[width=0.8\linewidth]{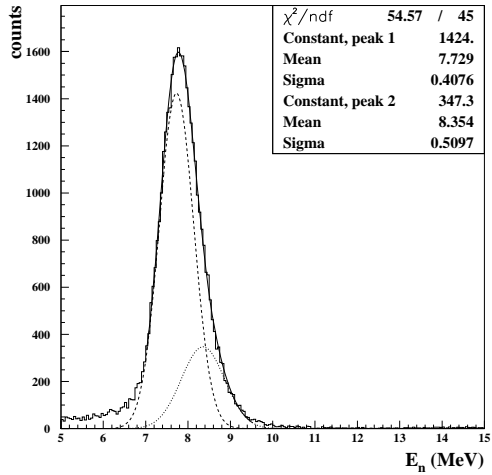}}
    \caption{\small Reconstructed energy spectrum for events associated with 
             neutron capture on Gd. Contributions from $\gamma$-lines at
             $7.94\mev$ (capture on $\nucl{157}{Gd}$) and $8.54\mev$ (capture 
	     on $\nucl{155}{Gd}$) are singled out. 
             The double-Gaussian fit parameters are also shown.}
    \label{fig:Gdiso}
  \end{center}
\end{figure}

The energy spectra show the energy lines due to neutron capture on Gd 
and H. The shape of the Gd-capture energy spectrum, as pointed out in
Fig.~\ref{fig:Gdiso}, results from the superimposition of $\gamma$ lines due to
neutron capture on $\nucl{157}{Gd}$ and $\nucl{155}{Gd}$ (see 
Tab.~\ref{tab:Gdiso} for a reference). 
Due to the scintillator saturation (see \S\ref{sec:MC}) the fitted peak 
values are $\approx 2.5\%$ lower than the nominal ones.
%
%
\subsection{Reconstruction problems}
We have just seen that event reconstruction gives good results, for 
calibration runs, as long as the source position is inside Region I. 
Unfortunately this is not always true and
more and more problems arise for events closer and closer to the geode surface.

\begin{figure}[htb]
  \begin{center}
    \mbox{\includegraphics[width=\linewidth]{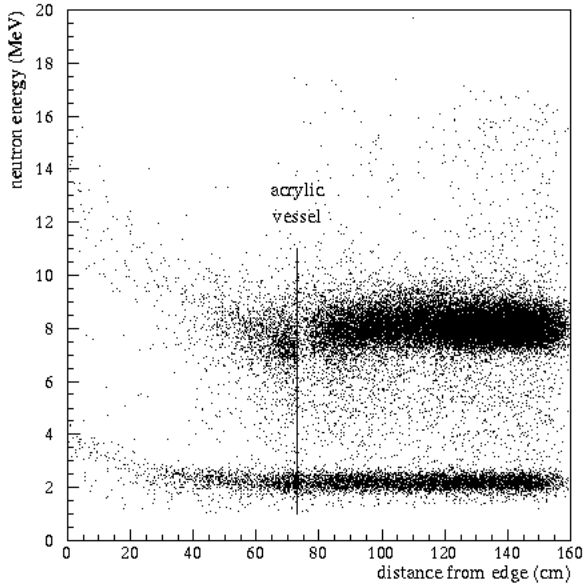}}
    \caption{\small Energy versus distance from edge for neutron events 
             generated by the $\nucl{252}{Cf}$ source along the 
             calibration pipe. The energy is flat whenever the reconstructed 
             position is more than $30\cm$ from PMTs.}
    \label{fig:cfevsz}
  \end{center}
\end{figure}

The main problem concerns the $1/r^2$ divergence of the light collected by 
one PMT; the exponential light attenuation, entering formulae
(\ref{QTQB},\ref{nthe}), becomes inaccurate in the vicinity of the PMTs.
In such a case, the approximation of a flat PMT surface is also no longer 
adequate to evaluate the solid angle by means of (\ref{solid}). These effects 
are shown in Fig.~\ref{fig:cfevsz}. The neutron capture energy (both for
Gadolinium and Hydrogen) is overestimated at reconstructed distances smaller
than $\sim 30\cm$ from the geode surface.

Fig.~\ref{fig:cfevsz} also shows one further weakness inherent in the 
minimization procedure. As remarked above, the MIGRAD minimizer heavily depends
on the knowledge of the first derivatives of (\ref{chipois}) with respect to 
the fit parameters and fails if this computation is not accurate enough. This 
situation arises for events near the acrylic vessel, where the first 
derivatives of (\ref{chipois}) exhibit a discontinuity due to the different 
light attenuation of the Region I and Region II scintillators; as a result, a 
discontinuity in the energy vs. distance profile is found around the vessel 
surface position. A slight improvement was obtained by calling the CERN SIMPLEX
routine when MIGRAD failed. This minimizer, although much slower and
in general less reliable than MIGRAD, is more suitable in this case since it
does not use first derivatives and it is not so sensitive to 
large fluctuations in the function value.

\begin{figure}[htb]
  \begin{center}
    \mbox{\includegraphics[width=\linewidth]{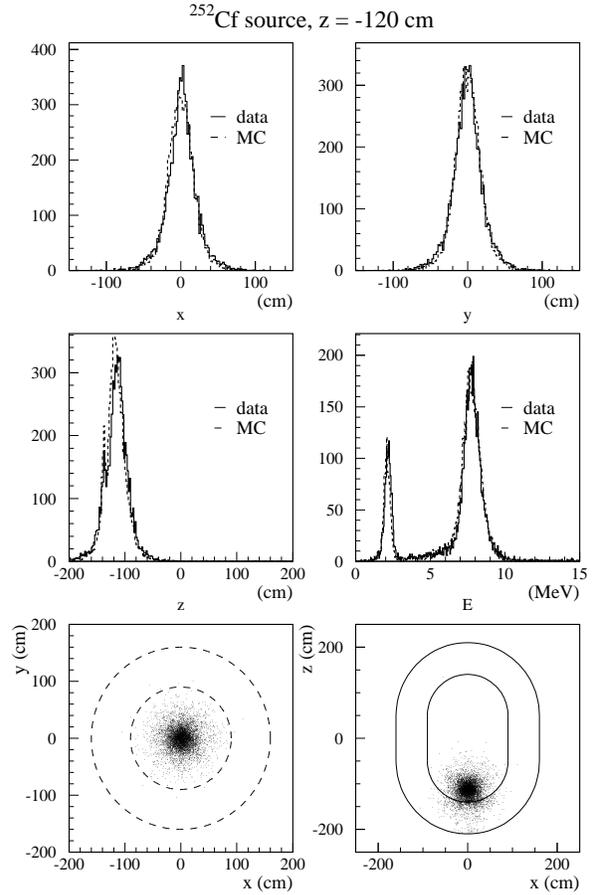}}
    \caption{\small Distributions of neutron events with the $\nucl{252}{Cf}$
             source at $z=$ -120 cm. The discontinuity in the $z$ 
             distribution at the vessel surface is visible also in Monte Carlo 
             generated events.}
    \label{fig:cf120}
  \end{center}
\end{figure}

Fig.~\ref{fig:cf120} shows 
the distribution of $\nucl{252}{Cf}$ neutron events
generated at $z=-120\cm$. The discontinuity in the $z$ distribution is also
present for Monte Carlo generated events, thus proving that this is an inherent
feature of the minimization procedure, and was properly taken into account in
evaluating the neutrino detection efficiency.
%
\section{Neutrino event selection and background rejection}
Neutrino events must be identified in a diverse and much more numerous
background. Off-line selections, based on 
the reconstructed event variables, were chosen to enhance the signal/background
ratio; at full reactor power this ratio is $>20$. We performed a careful
evaluation of the efficiencies associated with the selection criteria. The 
analysis procedure is reviewed and the characteristics of the neutrino event 
sample are presented. 
%
\subsection{The data sample}
The data acquisition period extended over 450 days, from 12 March 1997 
till 20 July 1998; about 2000 runs were taken, including standard neutrino runs
(about 1/4 of the total) and daily calibrations. The experiment stopped taking 
data five months after the 
last reactor shut-down (8 February 1998) when it became clear 
that, due to problems related with the cooling system, neither reactor would 
resume its normal operating conditions for at least one year. 

\begin{figure}[htb]
  \begin{center}
    \mbox{\includegraphics[width=0.7\linewidth]{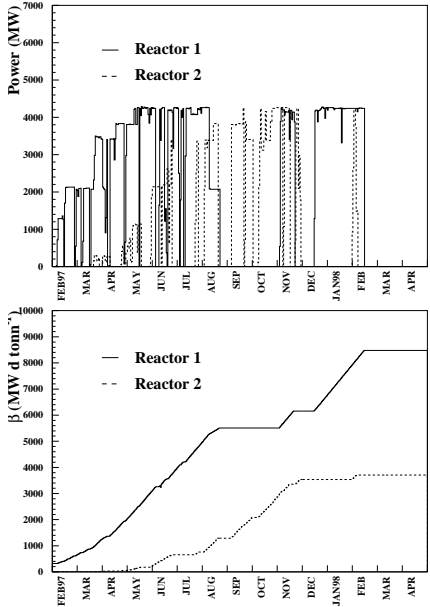}}
    \caption{\small Power (top) and burn-up (bottom) evolution for CHOOZ
             reactors. Both have been off since February 1998.}
    \label{fig:powburn}
  \end{center}
\end{figure}

Data taking is summarized in Tab.~\ref{tab:dataacq}; the total thermal
energy released (which can be intended as a neutrino integrated luminosity) 
is also listed. The total live time amounts to $\approx 340\days$, $40\%$
of which was with both reactors off. 
The power evolution of the CHOOZ reactors is illustrated in 
Fig.~\ref{fig:powburn}. The set of power values for both reactors almost 
continuously covers the entire range up to full power.  
It is  worth noting that this is a very unique feature
of the CHOOZ experiment.

Only one of the two CHOOZ reactors was on for at 
least $80\%$ of the total live time of the experiment. This allowed us to 
extract the separate contribution of each reactor to the neutrino signal and to
perform, thanks to their different distances from the detector, a two-distance 
oscillation test. 
\subsection{Candidate event selection}
Radioactivity background events are greatly reduced by
applying loose energy cuts to the neutron--like signals; events are selected if
QSUM$_\Pn > 13000$ ADC counts, which roughly corresponds to a $4\mev$ energy
deposit at the detector centre. The residual events ($\approx 7.2\cdot 10^5$ 
over a total number of $\approx 1.2\cdot 10^7$ L2 triggers) are then 
reconstructed by the standard minimization procedure described above. 

\begin{figure}[htb]
  \begin{center}
    \mbox{\includegraphics[width=0.8\linewidth]{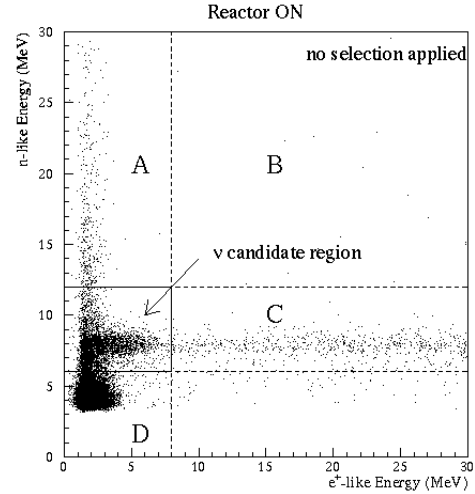}}
    \caption{\small Neutron versus positron energy for neutrino-like events
             collected during the reactor-on period. A
             preliminary cut to the neutron QSUM is applied to reject most of
             the radioactivity background.}
    \label{fig:esvsepon}
  \end{center}
\end{figure}

\begin{figure}[htb]
  \begin{center}
    \mbox{\includegraphics[width=0.8\linewidth]{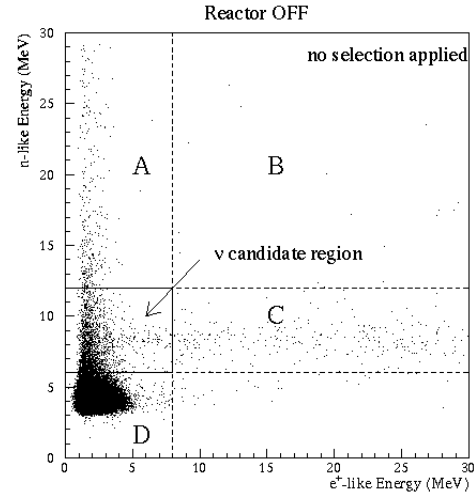}}
    \caption{\small Same as before, with neutrino-like events collected during
             the shut-down of both reactors.}
    \label{fig:esvsepoff}
  \end{center}
\end{figure}

\begin{figure}[htb]
  \begin{center}
    \mbox{\includegraphics[bb=60 160 530 680,width=0.8\linewidth]{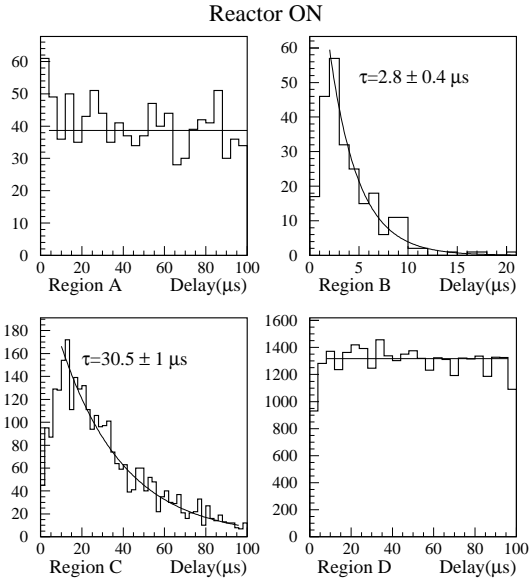}}
    \caption{\small Distribution of positron--neutron delay for the different 
             event categories. The best fit curves are also drawn and the
             relative parameters indicated.}
    \label{fig:cordelay}
  \end{center}
\end{figure}

An analysis of this preliminary sample gives a first illustration of the 
properties of the neutrino signal (reactor--on data) and the associated 
background (reactor--off data). Figures \ref{fig:esvsepon}, 
\ref{fig:esvsepoff} 
show the correlation of neutron--like vs. positron--like 
energy for this sample. 
Neutrino events, as indicated in Fig.~\ref{fig:esvsepon}, populate a region in 
the $(E_\Pep,E_\Pn)$ plot delimited by $E_\Pep < 8\mev$ and $6< E_\Pn<12\mev$.
Background events, depending on their position in these scatter plots,
are classified in the following categories:
\begin{description}
\item A) events with $E_\Pep < 8\mev$ and $E_\Pn >12\mev$:
\item B) events with $E_\Pep > 8\mev$ and $E_\Pn >12\mev$:
\item C) events with $E_\Pep > 8\mev$ and $6<E_\Pn<12\mev$: 
\item D) events with $E_\Pep < 8\mev$ and $E_\Pn<6\mev$:
\end{description}
The neutron energy distribution of C) events presents a clear $8\mev$ peak, 
typical of the neutron capture on Gadolinium, still persisting
in the reactor-off data. These events can then be interpreted as a 
correlated background associated with high energy spallation neutrons 
from cosmic ray interactions in the rock surrounding the detector; the 
neutrons entering the detector are slowed down 
to thermal velocities via elastic
scattering on protons and then captured; 
the proton recoil signal, whose energy 
spectrum is basically flat and extends to high energies, 
mimics the positron signal. 
This interpretation is confirmed by the neutron 
delay distribution. An example is shown in Fig.~\ref{fig:cordelay}, where
events in sample C) follow an exponential decay 
distribution whose life time $\tau= (30.5\pm 1.0)\usec$ is correct for
neutron capture in the Gd-doped scintillator.

Events in categories A) and D) show a flat delay distribution, therefore 
implying an accidental coincidence of uncorrelated signals. In both cases, the 
positron signal is faked by a low energy ($E_\gamma \leq 3\mev$) radioactivity 
event. The neutron signal may be associated with either another
radioactivity event, 
as in category D), or a high activity signal (most likely due to a 
proton recoil) as in category A).

Finally, the delay of B) events is exponentially distributed with a lifetime 
$\tau = (2.8 \pm 0.4)\usec$ which is compatible 
with that of a muon decay at
rest. These events can be associated 
with residual cosmic muons stopping in the 
detector and then decaying.
Both the muon energy loss and the Michel electron energy are 
much higher than the typical energy of a reactor neutrino interaction; these 
events can be rejected by applying an energy selection both to the 
positron--like and the neutron--like signals.

\begin{figure}[htb]
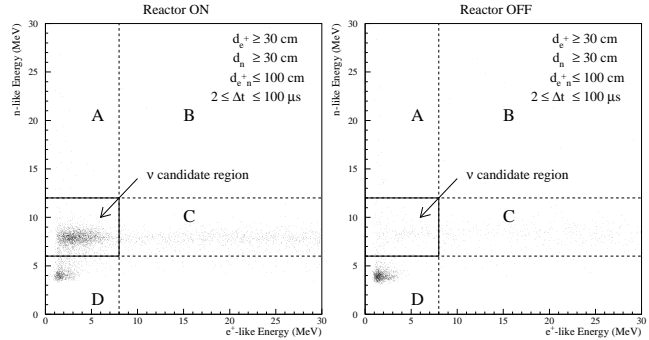

  \begin{center}
    \mbox{\hspace{-0.2cm}\includegraphics[width=0.55\linewidth]{fig.37a}
          \hspace{-0.7cm}\includegraphics[width=0.55\linewidth]{fig.37b}}
    \caption{\small Neutron versus positron energy for neutrino-like events
             selected from the preliminary sample by applying the 
             ``topological'' cuts here indicated.}
    \label{fig:esvsepcut}
  \end{center}
\end{figure}

\begin{figure}[htb]
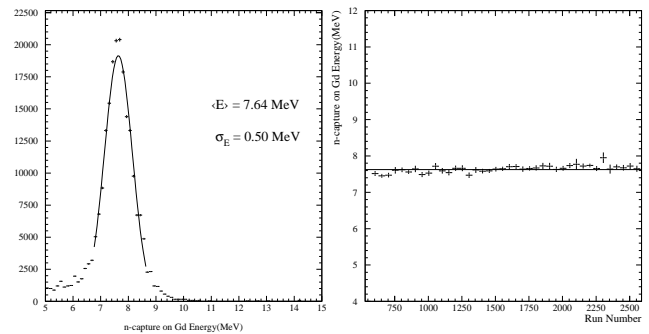

  \begin{center}
    \mbox{\hspace{-0.2cm}\includegraphics[width=0.55\linewidth]{fig.38a}
          \hspace{-0.7cm}\includegraphics[width=0.55\linewidth]{fig.38b}}
    \caption{\small Neutron energy distribution for correlated background events
             (left) and average $E_\Pn$ vs. run number (right).} 
    \label{fig:enestab}
  \end{center}
\end{figure}

The accidental background can be significantly reduced by applying fiducial
volume cuts. Less than $10\%$ of the events in regions A) and D), as shown
in Fig.~\ref{fig:esvsepcut}, survive the 
selection cuts (distances from the geode $d > 30\cm$ for both positron and 
neutron and relative distance $<100\cm$). 
Conversely the correlated background is 
more difficult to 
eliminate since these events exhibit the same 
features as neutrino interactions. 
This is the reason why the final background rate is dominated by the correlated
component, in spite of the shielding provided by the rock overburden. 
However, the neutron signals associated with this background component
are an important tool to follow the energy 
calibration stability throughout the experiment. The results are shown in 
Fig.~\ref{fig:enestab}, where the average neutron energy, obtained by a 
Gaussian fit of the Gadolinium capture peak, is plotted vs. the run number. The
relative stability throughout the data taking period (drift at a 0.8\% level) 
is an independent verification of the reliability of the adopted reconstruction
technique.
\subsection{Final selection}
Both the energy and the topological cuts 
were studied and optimized by relying on the Monte 
Carlo simulation of neutrino events. 
These predictions were cross-checked with 
the calibration data to gain confidence in the final efficiencies. 
We adopted the following criteria for selecting neutrino events:
\begin{description}
\item 1) Positron energy: $E_\Pep < 8\mev$:
\item 2) Neutron energy: $6<E_\Pn < 12\mev$:
\item 3) Distance from geode boundary: $d_\Pep > 30\cm$, $d_\Pn > 30\cm$:
\item 4) Relative positron--neutron distance: $d_{\Pep \Pn} < 100\cm$:
\item 5) Neutron delay: $2 < \Delta t_{\Pep \Pn} <100 \usec$:
\item 6) Neutron multiplicity: $N_\Pn=1$.
\end{description}

The cut on the positron--like energy is chosen to accept all possible positron 
triggers. The L1lo trigger threshold limits the lower end while the upper
limit was set to $8\mev$ since the probability of having a larger positron 
energy is negligible ($<0.05\%$).

Cut 2) selects the neutrino events associated with the neutron captures on
Gadolinium, which are more 
than $80\%$ of the total. The lower cut introduces an
additional inefficiency due to $\gamma$ 
rays escaping the containment region. As
illustrated in Fig.~\ref{fig:esvsepcut}, 
$6\mev$ is a suitable choice to 
separate neutrino events from the residual low energy uncorrelated background, 
thus optimizing the signal to noise ratio. 

The cut on the neutron delay covers about three capture times for neutrons in 
Region I. The $2\usec$ lower cut was introduced to reduce the effects
of the signal overshoot inherent the AC coupling of PMT bases and front-end 
electronics. These effects are particularly troublesome in the case of NNADC's, 
which integrate current signals of both polarities. 

The correlated background can be reduced by applying a cut on the 
secondary particle multiplicity\footnote{
The positron (or primary) signal is associated with the L1 trigger preceding
the L2. Every signal associated with or occurring after the L2 is 
referred to as a secondary event.}.
Muon spallation processes usually generate several neutrons, 
therefore more than one 
particle is likely to enter the detector and give a detectable signal.

Events satisfying the selection criteria will be referred to as {\it neutrino 
event candidates} from now on. The next section reviews 
the efficiency for selecting neutrino interactions and the
background rejection resulting from individual cuts.
\subsection{Positron efficiency}
\subsection*{Energy cut}
\begin{figure}[htb]   
   \begin{center}
     \mbox{\includegraphics[width=0.95\linewidth]{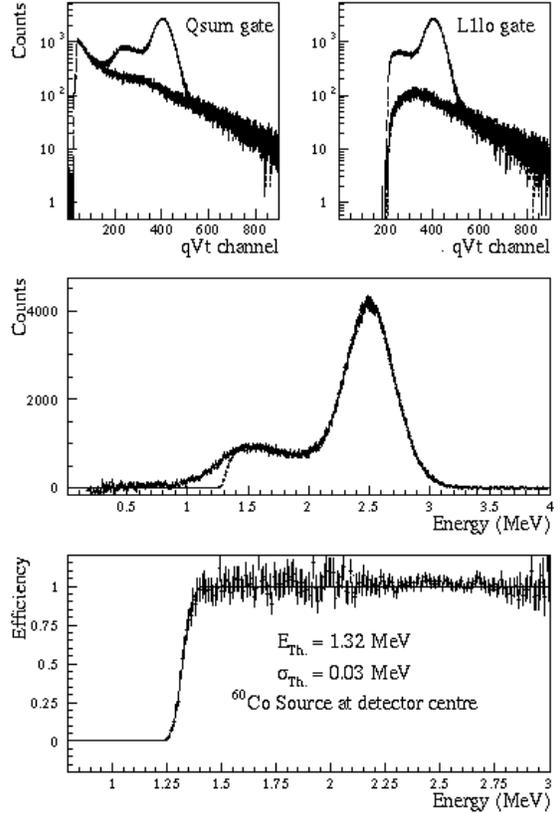}}
     \caption{\small Determination of the L1lo equivalent energy threshold. The
              top figures show the QSUM spectra measured with a $\nucl{60}{Co}$
              at the detector centre by means of an internal and external 
              triggered qVt (the corresponding background is superimposed). The
              central plot shows the background subtracted spectra. The
              bottom histogram, displaying the efficiency curve, follows an
              integral Gaussian function whose parameters are indicated.}
     \label{fig:threffi}
   \end{center} 
\end{figure}

The expected positron energy spectrum is obtained by 
folding the kinetic energy 
spectrum coming from (\ref{pos:energy}) with the detector response function. 
Two problems then arise in evaluating the positron efficiency. Firstly, the 
neutrino spectral shape slightly varies along a reactor cycle as a consequence 
of the fuel burn-up; however, the information daily provided by E.D.F. allows 
us to accurately follow this variation. Secondly, the energy cut operated by 
the L1lo trigger increases with time because of the aging of Region I 
scintillator. A daily check of the equivalent energy threshold is required 
to account for this effect. The procedure we followed is shown in 
Fig.~\ref{fig:threffi}. A Lecroy qVt
recorded the energy spectra due to a $\nucl{60}{Co}$ source placed at the 
detector centre. Two spectra were taken, the first one with an external trigger
provided by the L1lo signal and the second with an internal trigger set
at a lower threshold. In addition to the $2.5\mev$ ``sum'' line, each spectrum
exhibits a low energy tail due to the energy loss in the calibration pipe.
The ratio of the two spectra (after background 
subtraction) yields the L1lo trigger efficiency curve. The equivalent energy 
threshold results from fitting this curve by an integral Gaussian 
function. 

\begin{figure}[htb]
   \begin{center}
     \mbox{\includegraphics[width=0.7\linewidth]{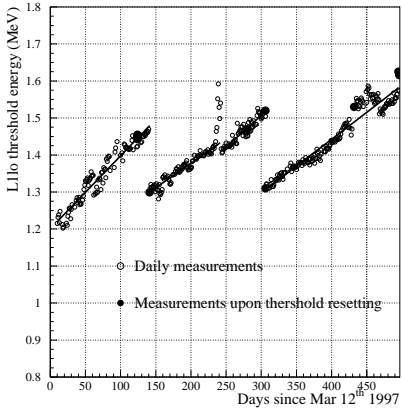}}
     \caption{\small Equivalent energy threshold at detector centre as a 
              function of time. The jumps visible here are due to the threshold 
              setting.}
     \label{fig:thrtime}
   \end{center} 
\end{figure}

The energy threshold varies with time as illustrated in Fig.~\ref{fig:thrtime}.
Its behaviour can be described by a linear function. Discontinuities 
arise in coincidence with the threshold resettings, which were needed
to bring the L1lo threshold back to the optimal value.

\begin{figure}[htb]
   \begin{center}
     \mbox{\includegraphics[width=\linewidth]{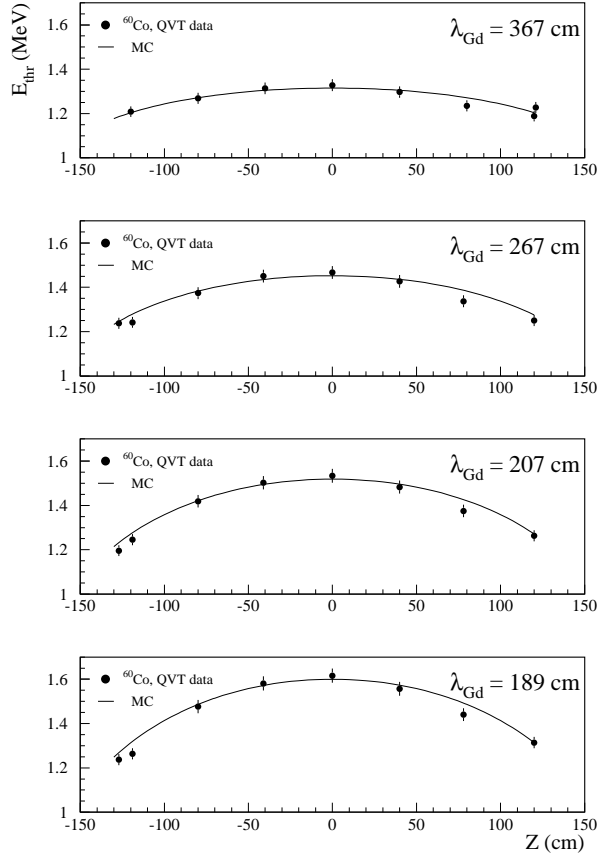}}
     \caption{\small Energy threshold as a function of $z$ for different values
               of the attenuation length for the Gd-doped scintillator. The
               measurements obtained with the $\nucl{60}{Co}$ source
               follow the expected behaviour.}
     \label{fig:zeffi}
   \end{center} 
\end{figure}

More extensive $\nucl{60}{Co}$ calibrations were periodically performed at
different positions of the source along the central pipe, in order to 
measure the $z$
dependence of the energy threshold. The results are shown in
Fig.~\ref{fig:zeffi} at four different stages of data acquisition. The overall
dependence on the position of the energy threshold 
can also be predicted by a Monte
Carlo simulation of the QSUM and NSUM behaviour. The
attenuation length values are the same as those 
used in the event reconstruction. The threshold measurement always fits well 
with expectations, as evident from the
same figure. We can therefore rely on Monte Carlo predictions to
determine the energy threshold at each position in the detector, where the 
source cannot be placed. An expression for the positron threshold as a
function of time and position can be derived by combining the daily
threshold measurements at the detector centre with Monte Carlo evaluations.  
An explicit form of this threshold is given in cylindrical coordinates 
$(\rho,z,\phi)$ by the following equation: 
\begin{equation} 
E_{thr}(t,\vect{x}) = E_{thr}(t,\vect{x}=0) \times (1-\alpha(t)\rho^2)
                      \times (1-\beta(t)z^2) 
\label{thrtime} 
\end{equation} 
where the energy threshold at centre $E_{thr}(t,\vect{x}=0)$ is extracted by 
interpolating the daily measurements shown in Fig.~\ref{fig:thrtime} and 
$\alpha(t), \beta(t)$ are linear functions obtained by Monte Carlo. 
\subsection*{Distance from the geode cut}
The positron loss due to the selection 3) (distance from the geode boundary) was
evaluated by a Monte Carlo simulation of neutrino interactions. 10000 events 
were uniformly generated in a volume including a $10\cm$ wide 
shell surrounding the target, so as to take into account spill-in/spill-out 
effects. The estimated efficiency proves to be almost independent of the 
scintillator degradation; we can then consider it as a constant and get the 
average value
\begin{equation}
\varepsilon_{d_\Pep} = (99.86 \pm 0.1) \%
\label{effi:dpos}
\end{equation}
\subsection{Neutron efficiency}
\subsection*{Neutron capture cut}
The neutron capture efficiency is defined as the ratio of neutrons captured by
Gadolinium nuclei to the total number of captures. This ratio enters the global
neutron efficiency since the neutron energy selection excludes events
associated with neutron captures on Hydrogen. This capture efficiency was
studied by means of a normal and a 
special tagged $\nucl{252}{Cf}$ source, providing a fission tagging signal. The
calibration pipe was removed for these particular studies in order to
avoid effects due to the iron content in the pipe. The source was at the end of
a plumb-line long enough to reach the bottom of the target vessel.
Fission neutrons have a longer mean pathlength than neutrons from
neutrino interactions ($\approx 20\cm$ in scintillator instead of
$\approx 6\cm$), so calibration data must be coupled with Monte Carlo 
predictions to correctly take into account the effects associated with the
neutron
spill-out (which are relevant for events approaching the boundary of Region I).

Neutron events are associated with Gd captures if $4<E_\Pn<12\mev$. The ratio 
obtained includes a $1\%$ correction due to Gd-capture events with visible
energy below $4\mev$. The combination of data and MC studies results in an 
average efficiency over the entire detector of
\begin{equation}
\varepsilon_{Gd} = (84.6 \pm 0.85)\%
\label{effi:Gd}
\end{equation}
\subsection*{Energy containment cut}
The $6\mev$ lower energy cut introduces an inefficiency in neutron detection 
due to the $\gamma$ rays from Gd-capture (by definition inside Region I) 
escaping the visible volume. Since this efficiency 
is almost independent of the 
neutron pathlength we can rely on the calibration data to evaluate this
number; Monte Carlo predictions are used to cross-check the data. 
Since this effect
is expected to be more relevant for  events near the edge of Region I, a
fine source scanning was made in this region. A special set of 
runs, was made with  $2\cm$ steps, 
between $1.5\cm$ and $11.5\cm$, from the bottom edge 
of the acrylic vessel. 

The escape fraction was defined as the ratio of the number of events 
with $4<E_\Pn<6\mev$ over those with $4<E_\Pn<12\mev$. 
The values turn out to
range from $2.9\%$ at the centre to up to $6.6\%$ at the bottom edge.
By averaging over the target volume we get
\begin{equation}
\varepsilon_{E_\Pn} = (94.6 \pm 0.4)\%
\label{effi:eneneut}
\end{equation}
\subsection*{Distance from the geode cut}
We followed the same procedure used to evaluate the corresponding positron 
efficiency. We obtain
\begin{equation}
\varepsilon_{d_\Pn} = (99.5 \pm 0.1)\%
\label{effi:dene}
\end{equation}
This value is not significantly affected by the scintillator degradation.
\subsection*{Delay cut}
\begin{figure}[htb]
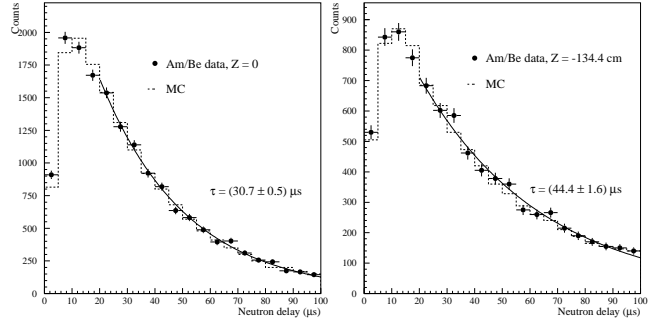

  \begin{center}
    \mbox{\hspace{-0.2cm}\includegraphics[width=0.55\linewidth]{fig.42a}
          \hspace{-0.7cm}\includegraphics[width=0.55\linewidth]{fig.42b}}
    \caption{\small Neutron delay distribution measured with the Am/Be source 
	     at
             the detector centre (left) and at the bottom edge of the acrylic 
             vessel (right). The time origin is defined by the $4.4\mev$ 
             $\gamma$-ray. No time analysis cut is applied yet.}
    \label{fig:Amdel}
  \end{center}
\end{figure}

The neutron delay cut was studied 
by using both the $\nucl{252}{Cf}$ and an Am/Be 
source; the latter emits single neutrons only, 
preventing biases due to the ADC integration dead time arising at high neutron
multiplicities. Fig.~\ref{fig:Amdel} shows the delay distributions obtained 
with the Am/Be source placed at the centre and at the 
bottom edge of the target.
The neutron capture time at the centre can be used to determine the Gd content 
in the Region I scintillator (to be used in the Monte Carlo code). 
The fitted decay time $\tau = (30.7 \pm 0.5)\usec$ corresponds to a mass 
fraction of $(0.0940\pm 0.0015)\%$. With reference to 
Fig.~\ref{fig:Amdel}(right), at the bottom of the detector the neutron capture 
on Hydrogen becomes more important and the fitted decay time increases 
accordingly. Both spectra show a low first bin content. This is due to neutron 
moderation time which corresponds to the relatively fast neutrons emitted by 
the Am/Be source. 

Given the chemical composition of the 
scintillator, it is possible to rely on the Monte Carlo simulation to predict 
the efficiency related to the neutron delay cut in the case of neutrino induced
events. The
estimated loss due to the $2\usec$ analysis cut amounts to $1.6\pm 0.2\%$; the 
fraction of neutrons with $\Delta t>100 \usec$ is $4.7\pm 0.3\%$. We 
therefore end up with:
\begin{equation}
\varepsilon_{\Delta t} = (93.7 \pm 0.4)\%
\label{effi:delt}
\end{equation}
\subsection{Relative distance cut efficiency}
Again we used the reconstruction of 10000 Monte Carlo generated 
events to predict the ef\-ficien\-cy due to the positron--neutron distance cut. We 
obtained
\begin{equation}
\varepsilon_{d_{\Pep \Pn}} = (98.4 \pm 0.3)\%
\label{effi:drel}
\end{equation}
Again this efficiency is nearly independent 
of the time evolution of the Gd-doped
scintillator.
\subsection{Neutron multiplicity}
The neutron multiplicity cut rejects neutrino events if a ``spurious'' L1lo 
trigger occurs along with the positron--neutron pair. L1lo triggers are mainly 
associated with $\gamma$-background events, $\approx 97\%$ of which have 
energies lower than the high threshold; so, the background above the L1hi 
threshold can be neglected in what follows. The error associated with this 
approximation is negligible.

Let us consider all possible sequences:
\begin{description}
\item 1) $\Pep-\Pn-\gamma$ with $t_\gamma-t_\Pn<100\usec$:
\item 2) $\Pep-\gamma-\Pn$:
\item 3) $\gamma-\Pep-\Pn$ with $t_\Pep-t_\gamma<100\usec$:
\end{description}

\fbox{\sf Case 1}\\
The L2 triggers on the $(\Pep,\Pn)$ pair and the $\gamma$ signal is mistaken 
for a second neutron; the neutrino event is then rejected because of cut 6). 
Given the $\gamma$ rate $R_\gamma$, the probability of such a sequence is
\begin{equation} 
  1 - \varepsilon_1 = R_\gamma \Delta t_\nu
  \label{effi1}
\end{equation} 
(where $\Delta t_\nu$ is the $98 \usec$ acceptance window)
from which $\varepsilon_1$ is extracted. The average efficiency is
$(98.6\pm 0.3)\%$.

\fbox{\sf Case 2}\\
If the positron energy exceeds the high threshold, the L2 triggers on the
$(\Pep,\gamma)$ pair; then two neutron--like signals are detected (the ``true''
neutron plus the $\gamma$) and the
neutrino event is rejected by cut 6). If the positron energy is lower then
the L1hi threshold, the L2 triggers on the 
$(\gamma,\Pn)$ and the $\gamma$ is mistaken for a positron--like event. 
Since the $\gamma$ signal is not correlated with the neutron, the probability 
of surviving the topological cuts 3) and 4)
is quite low ($<2\%$) and can be neglected. As a conclusion, neutrino events 
occurring in such a sequence are always rejected, whatever the positron energy 
is.

Let $P_\Pn(t_c)$ be the neutron delay distribution shown
in Fig.~\ref{fig:Amdel} and $t$ the time between the positron and the $\gamma$
signals. The neutrino efficiency can be written as follows:
\begin{equation}
  \varepsilon_2 = 1 - \int_0^{100} P_\Pn(t_c) \diff t_c \int_0^{t_c} 
                  R_\gamma \diff t = 1 - R_\gamma \overline{t_c}
  \label{effi2}
\end{equation}
where $\overline{t_c} = (30.5\pm 0.5)\usec$ is the average neutron delay.

\fbox{\sf Case 3}\\
Like the above case, if the positron signal met the high trigger 
condition, L2 would trigger on the $(\gamma,\Pep)$ pair and the selection 
criterium 6) would reject the event; otherwise the L2 would trigger on the 
$(\Pep,\Pn)$ pair and the neutrino events would be accepted. So
\begin{equation}
  \varepsilon_3 = 1 - R_\gamma f_H \Delta t_\nu
  \label{effi3}
\end{equation}
$f_H = (0.45 \pm 0.05)$ being the positron fraction above the high energy 
threshold.

\fbox{\sf Combined efficiency}\\
The neutron multiplicity cut efficiency is obtained by multiplying the three
above efficiencies. 
We evaluated it on a run by run basis since $R_\gamma$
depends on the L1lo threshold. The average value is
$\varepsilon_{2n} = (97.4\pm 0.5)\%$.

A summary of all the efficiencies is presented in Tab.~\ref{tab:effi}.
\begin{table}[htb]
  \begin{center}
    \caption{\small Summary of the neutrino detection efficiencies.}
    \label{tab:effi}
    \begin{tabular}{||l||c||c||}
      \hline
      \hline
      selection & $\epsilon (\%)$ & rel. error $(\%)$ \\
      \hline
      \hline
      positron energy$^\ast$ & $97.8$ & $0.8$ \\
      \hline
      positron-geode distance & $99.9$ & $0.1$ \\
      \hline
      neutron capture & $84.6$ & $1.0$ \\
      \hline
      capture energy containment & $94.6$ & $0.4$ \\
      \hline
      neutron-geode distance & $99.5$ & $0.1$ \\
      \hline
      neutron delay & $93.7$ & $0.4$ \\
      \hline
      positron-neutron distance & $98.4$ & $0.3$ \\
      \hline
      neutron multiplicity$^\ast$ & $97.4$ & $0.5$ \\
      \hline
      \hline
      combined$^\ast$ & $69.8$ & $1.5$ \\
      \hline
      \hline
      \multicolumn{3}{l}{$^\ast${\small average values}} 
    \end{tabular}
  \end{center}
\end{table}
\section{The neutrino signal}
\begin{figure}[htb]
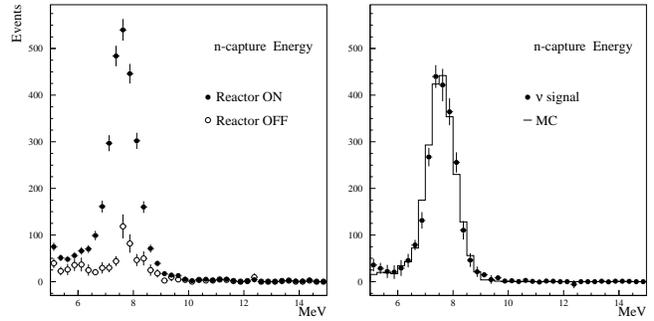

  \begin{center}
    \mbox{\hspace{-0.2cm}\includegraphics[width=0.55\linewidth]{fig.43a}
          \hspace{-0.7cm}\includegraphics[width=0.55\linewidth]{fig.43b}}
    \caption{\small Neutron energy spectra for reactor-on and reactor-off 
             periods (left) and background subtracted spectrum compared to
             Monte Carlo expectations (right).}
    \label{fig:eneuca}
  \end{center}
\end{figure}

\begin{figure}[htb]
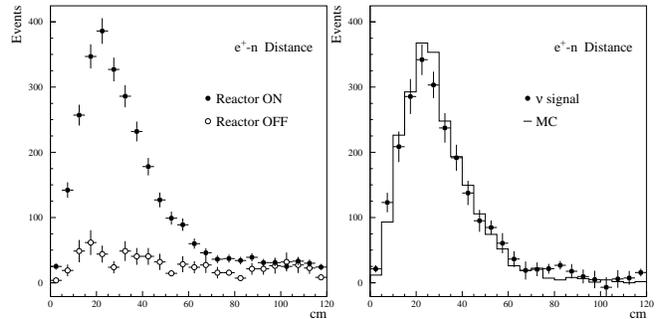

  \begin{center}
    \mbox{\hspace{-0.2cm}\includegraphics[width=0.55\linewidth]{fig.44a}
          \hspace{-0.7cm}\includegraphics[width=0.55\linewidth]{fig.44b}}
    \caption{\small Same as before, for the positron--neutron distance.}
    \label{fig:distca}
  \end{center}
\end{figure}

\begin{figure}[htb]
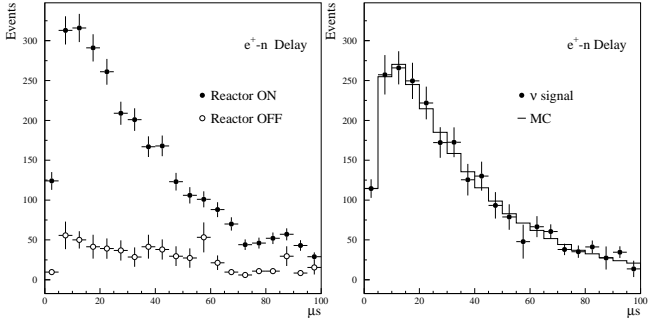

  \begin{center}
    \mbox{\hspace{-0.2cm}\includegraphics[width=0.55\linewidth]{fig.45a}
          \hspace{-0.7cm}\includegraphics[width=0.55\linewidth]{fig.45b}}
    \caption{\small Neutron delay distributions for reactor-on and reactor-off 
             periods (left) and background subtracted spectrum compared to MC
             predictions (right).}
    \label{fig:delayca}
  \end{center}
\end{figure}

Figures \ref{fig:eneuca} through \ref{fig:delayca} present the final neutrino 
candidate sample, with all the selection cuts applied. A total number of 2991 
neutrino candidates was collected, 287 of which occurred during the 
reactor-off periods. To properly compare these with expectations, the entire 
acquisition cycle was divided according to the dates 
of the threshold resetting,
when variations in the background rate were expected. 
For each resulting period,
the reactor-off background was normalized to the same 
livetime as the reactor-on spectra. 
%
%
\subsection{Measured positron spectrum}
\label{sec:posi}
\begin{figure}[htb]
  \begin{center}
    \mbox{\includegraphics[width=0.95\linewidth]{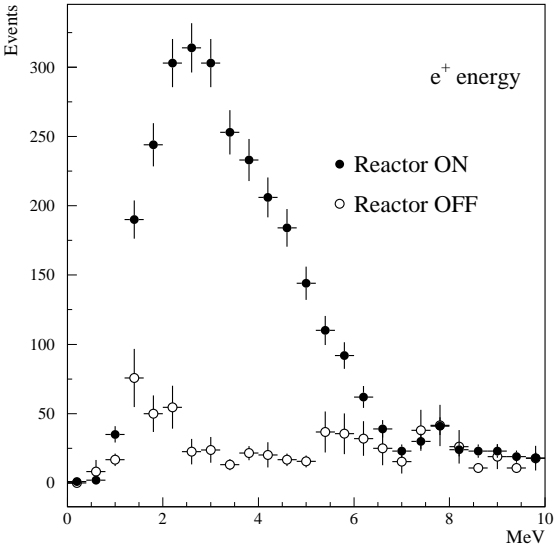}}
    \caption{\small Experimental positron spectra for reactor-on and reactor-
             periods after application of all selection criteria. The errors 
             shown are statistical.}
    \label{fig:posmea}
  \end{center}
\end{figure}

Fig.~\ref{fig:posmea} shows the complete measured spectra (reactor-on,
reactor-off), obtained by summing the spectra collected during runs relative to
different off-line periods; the resulting positron spectrum (reactor-on minus 
reactor-off) is presented in Fig.~\ref{fig:posonoff}. 
The chosen bin width ($0.4\mev$) is roughly adapted both to 
the statistics and to the 
energy resolution.
\subsection{Predicted positron spectrum}
The expected visible energy positron spectrum at the detector position, for a 
mean reactor-detector distance $L_k$, is given by:
\begin{equation}
  \begin{split}
    & S_k(E,L_k,\theta,\dmsq) = \frac{1}{4\pi L_k^2} N_p
    \int h(L,L_k) \diff L \\
    & \int \sigma(E_\Pep) S_\nu(E_\nu) 
    P(E_\nu,L,\theta,\dmsq) r(E_\Pep,E) \varepsilon(E_\Pep)
    \diff E_\Pep,
  \end{split}
  \label{posexp}
\end{equation}
where 
\begin{center}
\begin{tabular}{ll}
$E_\nu,E_\Pep$ & are related by (\ref{pos:energy}), \\
$N_p$ & is the total number of target protons \\ 
      & in the Region I scintillator, \\
$\sigma(E_\Pep)$ & is the detection cross section (\ref{inv:cross}), \\
$S_\nu(E_\nu)$ & is the $\Pagne$ spectrum, \\
$h(L,L_k)$ & is the spatial distribution function for \\ 
      & the finite core and detector sizes, \\ 
$r(E_\Pep,E)$ & is the detector response function \\
      & providing the visible $\Pep$ energy, \\
$P(E_\nu,L,\theta,\dmsq)$ & is the two-flavour survival probability, \\
$\varepsilon(E_\Pep)$ & is the combined detection efficiency.\\
\end{tabular}
\end{center}
%

\begin{figure}[htb]
  \begin{center}
    \mbox{\includegraphics[width=0.95\linewidth]{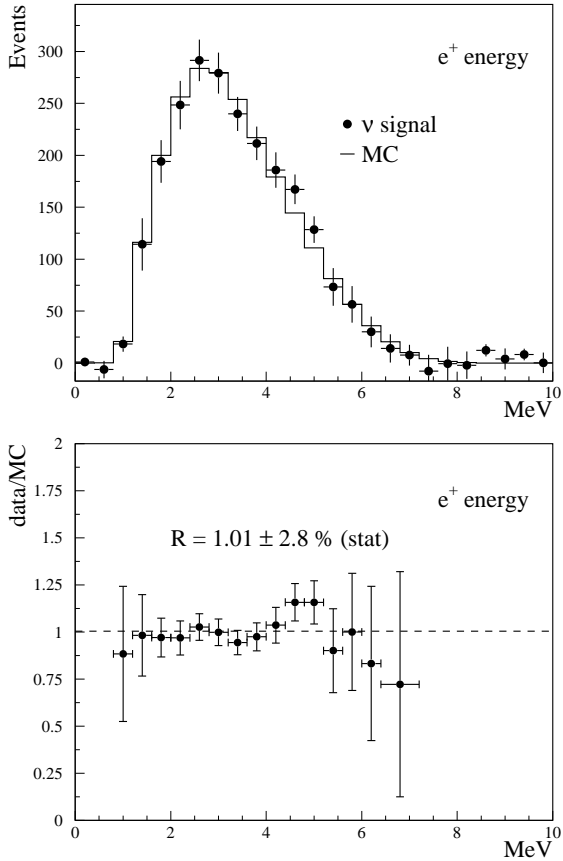}}
    \caption{\small (above) Expected positron spectrum for the case of no 
                    oscillations, superimposed on the measured positron 
                    obtained from the subtraction of reactor-on and reactor-off 
                    spectra; (below) measured over expected ratio. The errors
                    shown are statistical.}
    \label{fig:posonoff}
  \end{center}
\end{figure}

We first computed the positron spectrum in the absence of neutrino 
oscillations, by using the Monte Carlo code to simulate both reactors and
the detector. The composite antineutrino spectrum was generated for each of the 
$205$ fuel elements of each reactor core; each antineutrino was assigned a 
weight according to the prescriptions given in \S\ref{sec:reasim}. 
The interaction points 
were randomly chosen in the target; the positron and resulting annihilation 
photons were tracked in the detector, and scintillator saturation effects
were included to correctly evaluate the positron visible energy. The resulting
spectrum, summed over the two reactors, is superimposed on the one measured in 
Fig.~\ref{fig:posonoff} to emphasize the fact that the data fit with the 
no-oscillation hypothesis; the Kolmogorov-Smirnov test for compatibility of
the two distributions gives a $82\%$ probability. The measured vs. expected 
ratio, averaged over the energy spectrum (also presented in 
Fig.~\ref{fig:posonoff}) is
\begin{equation}
R = 1.01 \pm 2.8\% ({\rm stat}) \pm 2.7 \% ({\rm syst})
\end{equation}
In the next section we will discuss how this ratio constrains the oscillation
hypothesis.
\subsection{Background}
One of the main features of the experiment, the low background, 
represents a difficulty in that it is hard to measure.
Apart from 
the low statistics, another difficulty is that the background rate 
depends on the trigger conditions which, as discussed above, 
changed with time due to scintillator aging and the positron 
threshold adjustments. Separate estimates of the background are then needed for
each data taking period. Only 34 events were collected during the
1997 runs (periods 1 \& 2, until 12 January 1998, 
date of the last threshold resetting: see Tab.~\ref{tab:poilik} below) with 
reactors off, the total live time being $577$~h; this implies a background rate
of $1.41 \pm 0.24$ events per day. Most of the reactor-off statistics, amounting
to 253 events, was collected during the 1998 run when the cumulative reactor 
power was very low; the reactor-off live time
is $2737$~h. The resulting background rate is $2.22 \pm 0.14$, about twice as 
large as in the 1997 run. 

The explanation for this variation relies on
the lowering of the NSUM threshold associated with the L1lo trigger. A lower
number of hit PMTs implies a larger fiducial volume extended towards the PMT
boundary, where the event rate is dominated by the natural radioactivity.
Moreover, we saw that the reconstruction algorithm overestimates the energy of
events near the PMTs (see Fig.~\ref{fig:cfevsz}). Therefore a larger
fraction of radioactivity events is shifted into the neutrino candidate event 
window shown in Figs.~\ref{fig:esvsepon}, \ref{fig:esvsepoff}, 
\ref{fig:esvsepcut}. As a result, the accidental component of the background is
greatly enhanced. On the other hand, the correlated
background is virtually unaffected by the low trigger 
conditions, since the neutron signal is much higher than the threshold. We 
verified that the correlated background did not significantly change throughout
the experiment.

As a further cross-check of the reactor-off estimates, we measured the 
background by extrapolating to zero the candidate yield versus reactor power. 
This will be the subject of \S\ref{sec:nuyie}.
\subsection{Correlated background}
\begin{figure}[htb]
  \begin{center}
    \mbox{\includegraphics[width=0.8\linewidth]{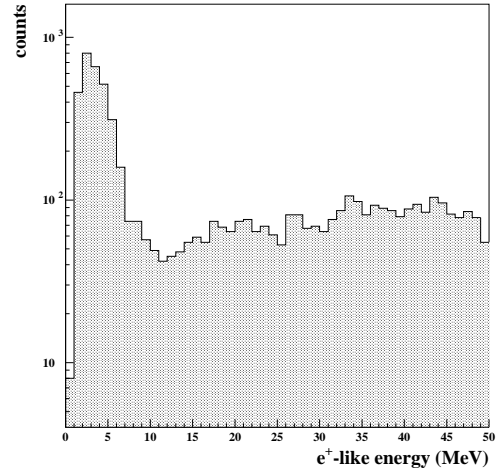}}
    \caption{\small Energy distribution of $\Pep$-like signals associated with 
      the correlated background.}
    \label{fig:corpri}
  \end{center}
\end{figure}

Fig.~\ref{fig:corpri} shows the energy distribution of $\Pep$-like signals
associated with the correlated background; these 
events are selected by applying all 
the criteria for neutrino candidate selection, except $\Pep$-like energy.
Apart from a low-energy peak (due to the neutrino events and to the 
accidental background), the spectrum has a roughly flat distribution extending 
beyond $30\mev$, with a slight increase at higher energies caused by the NNADC
saturation. A Monte Carlo 
simulation of this spectrum is very difficult, since no reliable transport code
is available for spallation neutrons in the $10\div 100\mev$ kinetic energy 
range. Moreover, the observed spectrum is affected by the scintillator 
saturation (which is relevant at low recoil proton energies). 

To verify the stability of the correlated background rate, we divided 
the complete run into three different periods (see Tab.~\ref{tab:poilik} below)
corresponding to the dates of the threshold resetting. The spectra for each 
period were fitted in the $10 <E<30\mev$ range. 
We divided them by the live times (also listed in Tab.~\ref{tab:poilik}) of the
three periods, and finally found:
\begin{equation}
  B_{corr} = 
  \begin{cases}
      (0.156 \pm 0.01)\mev^{-1}\days^{-1} & \text{for the 1st period}, \\
      (0.158 \pm 0.01)\mev^{-1}\days^{-1} & \text{for the 2nd period}, \\
      (0.151 \pm 0.01)\mev^{-1}\days^{-1} & \text{for the 3rd period},
  \end{cases}
  \label{corrback}
\end{equation}
confirming the stability of the correlated background rate throughout the
experiment.

A rough evaluation of the correlated background can be
obtained by extrapolating the rate
in the positron window ($E_{thr}< E < 8\mev$). Taking the average value from 
(\ref{corrback}) and assuming $E_{thr} = 1.5\mev$, we obtained a rate of
$(1.01 \pm 0.04(stat) \pm 0.1(syst)) {\rm events}\days^{-1}$. 
\subsection{Accidental background}
The accidental background rate was determined by separate estimates of the
singles rate for both $\Pep$-like and n-like signals. To minimize 
possible biases due to the trigger, both estimates were 
performed by looking at ``isolated old'' events, \ie  at the hits recorded
in the NNADC event buffer (storing up to 9 events) for which the previous 
and following events were more than $1 \msec$ distant (isolated) and
which occurred  at least $2\msec$ before
the L2 trigger (old). 
The event selection was operated by applying
the same cuts (energy and distance from edge) used for the candidate selection;
in the 1997 run we found $R_\Pep = (64.8 \pm 0.1)\s^{-1}$ for the $\Pep$-like 
event rate and $R_\Pn = (45 \pm 2)\hou^{-1}$ for the n-like one. The rate of 
accidental coincidences $\Pep-\Pn$ in the $2\div 100\usec$ 
time window therefore turns out to be 
$(3.4 \pm 0.15)\days^{-1}$. 
An additional reduction factor $f_d = (0.12 \pm 0.01)$
was applied to account for the selection operated by the $\Pep-\Pn$ distance 
cut. The resulting background rate is $(0.42 \pm 0.05)\days^{-1}$, which
fits with the previous determinations of the total and the correlated 
background for the 1997 run.

We checked that the measured single event rates $R_\Pep$ and $R_\Pn$ are in 
agreement with estimates based on the radioactive contaminants present in the
detector and its surroundings~\cite{Propo,SNOback}.
\subsection{Neutrino yield versus power}
\label{sec:nuyie}
The CHOOZ experiment gave us the unique opportunity of measuring the neutrino 
flux before either reactor started working and also after the reactors were 
shut down.
We were therefore able to collect enough reactor--off data to determine
the background rate 
and could further measure
the neutrino flux while the reactors were ramping up to full power. By fitting 
the slope of the measured rate versus reactor power it is possible to determine
the neutrino rate at full power, which can then be compared with the predicted 
neutrino production rate at full power. This 
procedure provides not only a very powerful tool to 
determine the neutrino deficit and test the oscillation hypothesis,
but also gives an independent estimate of the background rate to be compared 
with the values quoted above.

The fitting procedure is as follows. 
For each run $i$ the predicted number of
neutrino candidates $\overline{N}_i$ is derived from the sum of a 
signal term, linearly dependent on
reactor power $W_{1i}$ and $W_{2i}$, with the background $B$, 
assumed to be constant and independent of
power:
\begin{equation}
  \overline{N}_i = (B + W_{1i}Y_{1i} + W_{2i}Y_{2i}) \Delta t_i,
  \label{nrun}
\end{equation}
where $\Delta t_i$ is
live time of run i and $(Y_{1i},Y_{2i})$ are the positron yields
induced by each
reactor. These yields still depend on the reactor index (even in the
absence of neutrino oscillations), because of the different distances, and
on run number, as a consequence of their different fissile isotope composition. 
It is thus convenient to factor $Y_{ki}$ into a function $X_k$ (common to 
both reactors in the no-oscillation case) and distance dependent terms, as 
follows:
\begin{equation}
  Y_{ki} = (1 + \eta_{ki})\frac{L_1^2}{L_k^2} X_k,
  \label{nfact}
\end{equation}
where $k=1,2$ labels the reactors and the $\eta_{ki}$ corrections contain the
dependence of the neutrino counting rate on the fissile isotope composition of 
the reactor core. We are thus led to define a cumulative ``effective'' power 
according to the expression
\footnote{The ``effective'' power is the thermal power necessary to make the 
same number of events with a single reactor located at the reactor 1 site. It
must provide $9.55\gw$ at full operating conditions at the start of reactor 
operation.}
\begin{equation}
  W_i^\ast \equiv \sum_{k=1}^2 W_{ki} (1 + \eta_{ki})\frac{L_1^2}{L_k^2}; 
  \label{weff}
\end{equation}
Eq.(\ref{nrun}) then becomes 
\begin{equation}
  \overline{N}_i = (B + W_i^\ast X) \Delta t_i,
  \label{nrun2}
\end{equation}
$X$ being the positron yield per unit power averaged over the two reactors. 

The burn--up corrections $\eta_{ki}$ must be evaluated on a run by run basis. 
Since the GEANT routines would have been very time-consuming for such a 
task, we preferred to follow a simpler approach 
than the procedure explained at \S\ref{sec:posi}. 
10000 $\Pagne$'s per run were 
generated at each reactor according to the respective fuel composition;
the kinetic positron energy 
was obtained by the simple relation $E_\nu = T_\Pep + 1.804\mev$ (which comes 
from (\ref{pos:energy}) in the limit of infinite nucleon mass). The cross 
section
(\ref{inv:cross}) was thus multiplied by a function $\delta(E_\nu)$ to correct 
for the shift in the positron energy scale due to the finite neutron recoil 
effect~\cite{Vogel}; this correction can be parametrized as
\begin{equation}
  \delta(E_\nu) = 1 - 0.155 \times \exp \big(\frac{E_\nu/\mev  - 8}{1.4} \big) 
  \label{delrec}
\end{equation}
We then applied the detector response function (evaluated by 
Monte Carlo simulations at several positron energies ranging from $0.5\mev$ up
to $10\mev$) to the positron kinetic energy and weighted each event according 
to the positron threshold efficiency. 

We built the
likelihood function ${\cal L}$ 
for a set of n runs
as the joint Poissonian probability 
of detecting
$N_i$ neutrino candidates 
in the i-th run
when $\overline{N}_i$ are expected:
\begin{equation}
  -\ln {\cal L} = - \sum_{i=1}^{n} \ln P(N_i;\overline{N}_i) 
  \label{wlike}
\end{equation}
Searching for the maximum likelihood to determine the parameters $X$ and $B$ is
then equivalent to minimizing (\ref{wlike}). The minimization procedure is 
similar to that used in the event reconstruction (see \S\ref{sec:reco}).

Both the average positron yield $X$ and the background rate $B$ are assumed to 
be constant. This is true, by definition, for the positron yield, since the
effect of the threshold resetting on the positron efficiency is already included
in the $\eta_{ki}$ correction factors. On the other hand, we find that the 
background rate changes significantly when the trigger threshold are reset.
The complete run sample therefore needs to be divided into three periods, 
according to the dates of the threshold resetting, and the fit parameters 
for each period drawn separately. 
The results are listed in Tab.~\ref{tab:poilik}.
\begin{table}[htb]
  \caption{\small Summary of the likelihood fit parameters for the three data 
    taking periods.}
  \label{tab:poilik}
  \begin{center}
    \begin{tabular}{||l|c|c|c||}
      \hline
      \hline
      period & 1 & 2 & 3 \\
      \hline
      \hline
      starting date   & 97/4/7 & 97/7/30 & 98/1/12 \\
      \hline
      runs   & 579$\rightarrow$1074 & 1082$\rightarrow$1775 & 
               1778$\rightarrow$2567 \\
      \hline
      live time (h) & $1831.3$ & $2938.8$ & $3268.4$ \\
      \hline
      reactor-off & $38.9$ & $539.5$ & $2737.2$ \\
      time (h) & & & \\
      \hline
      $\int W \diff t$ (GWh) & $7798$ & $10636$ & $2838$\\
      \hline
      $B$ ($\days^{-1}$)& $1.25\pm 0.6$ & $1.22\pm 0.21$ &$2.2\pm 0.14$\\
      \hline
      $X$ ($\days^{-1}\gw^{-1}$)  & $2.60 \pm 0.17$ & $2.60 \pm 0.09$ &
                                          $2.51 \pm 0.17$ \\
      \hline
      $\chi^2/dof$ & $136/117$ & $135/154$ & $168/184$ \\
      \hline
      $n_\nu$ ($\days^{-1}$) & $24.8 \pm 1.6$ & $24.8 \pm 0.9$ &
                                      $24.0 \pm 1.6$ \\
      (at full power) & & &\\
      \hline
      \hline
    \end{tabular}
  \end{center}
\end{table}
%

By averaging the signal $X$ over the three periods, we obtain
\begin{equation}
  \langle X \rangle = (2.58 \pm 0.07) \,\,\,{\rm counts}\days^{-1}\gw^{-1},
  \label{xave}
\end{equation}
corresponding to $(24.7 \pm 0.7)$ daily neutrino interactions at full 
power.
%

The background rate was stable throughout the 1997 run, then 
it suddenly increased by close to a factor of 2 in coincidence with the last 
threshold resetting (see period 3 column). This can be explained in terms of a
larger accidental background rate, as a result of threshold lowering.

\begin{figure}[htb]
  \begin{center}
    \mbox{\includegraphics[bb= 23 133 565 685, 
       width=0.9\linewidth]{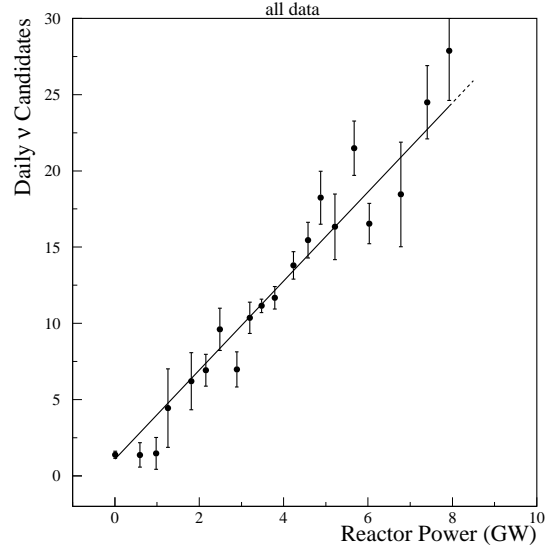}}
    \caption{\small Daily number of $\Pagne$ 
           candidates, as a function of the reactor power.}
    \label{fig:nuvspower}
  \end{center}
\end{figure}

Fig.~\ref{fig:nuvspower} shows the daily number of 
neutrino candidates as a function of reactor power 
for all the data sets together. The superimposed line corresponds
to the average fitted signal and background values.

\subsection{Neutrino yield for individual reactors}
\label{sec:nyield}
\begin{figure}[htb]
  \begin{center}
    \mbox{\includegraphics[width=0.9\linewidth]{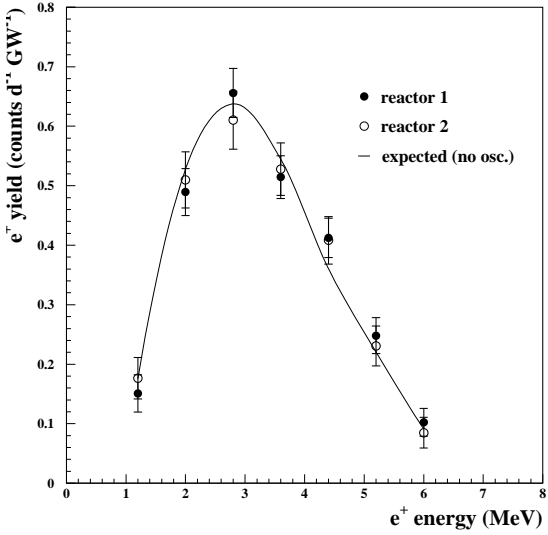}}
    \caption{\small Positron yields for the two reactors, as compared with 
             expected yield for no oscillations.}
    \label{fig:yieboth}
  \end{center}
\end{figure}

The same fitting procedure can be used to determine the 
contribution to the neutrino yield from each reactor and for each energy bin of
the positron spectra. After splitting 
the signal term into separate yields and introducing a dependence on the 
positron energy, Eq.(\ref{nrun2}) can be rewritten as
\begin{equation}
  \overline{N}_i(E_j) = (B(E_j) + W_{1i}^\ast(E_j) X_1(E_j) + 
                                  W_{2i}^\ast(E_j) X_2(E_j)) \Delta t_i
  \label{nruni}
\end{equation}
The spectrum shape is expected to vary, due to fuel aging, throughout the 
reactor cycle. Burn-up correction factors $\eta_{ki}$ then need to be 
calculated for each bin of the positron spectrum. The fitted yields, averaged
over the three periods, are listed in Tab.~\ref{tab:yieboth} and plotted in
Fig.~\ref{fig:yieboth} against the expected yield in absence of neutrino
oscillations. 
\begin{table}[htb]
  \caption{\small Experimental positron yields for both reactors ($X_1$ and 
           $X_2$)and expected spectrum ($\tilde{X}$) for no oscillation. The
           errors ($68\%$ C.L.) and the covariance matrix off-diagonal elements
           are also listed.}
  \label{tab:yieboth}
  \newcommand{\Rule}{\rule[-.7ex]{0ex}{2.9ex}}
  \begin{center}
    \begin{tabular}{||c|c|c|c|c||}
      \hline
      \hline
      \Rule
      $E_\Pep$ & $X_1\pm\sigma_1$ & $X_2\pm\sigma_2$ & $\tilde{X}$ & 
      $\sigma_{12}$ \\
      MeV& \multicolumn{3}{c}{(counts $\days^{-1}\gw^{-1}$)} \vline & \\
      \hline
      \hline
      \Rule
      $1.2$ & $0.151\pm 0.031$ & $0.176\pm 0.035$ & $0.172$ &  
              $-2.2~10^{-4}$ \\
      \Rule
      $2.0$ & $0.490\pm 0.039$ & $0.510\pm 0.047$ & $0.532$ &
              $-1.5~10^{-4}$ \\
      \Rule
      $2.8$ & $0.656\pm 0.041$ & $0.610\pm 0.049$ & $0.632$ &
              $-3.5~10^{-4}$ \\
      \Rule
      $3.6$ & $0.515\pm 0.036$ & $0.528\pm 0.044$ & $0.530$ & 
              $-3.3~10^{-4}$ \\
      \Rule
      $4.4$ & $0.412\pm 0.033$ & $0.408\pm 0.040$ & $0.379$ & 
              $-2.0~10^{-4}$ \\
      \Rule
      $5.2$ & $0.248\pm 0.030$ & $0.231\pm 0.034$ & $0.208$ & 
              $-0.7~10^{-4}$ \\
      \Rule
      $6.0$ & $0.102\pm 0.023$ & $0.085\pm 0.026$ & $0.101$ & 
              $-1.3~10^{-4}$ \\
      \hline
      \hline
    \end{tabular}
  \end{center}
\end{table}
The yield parameters $X_1,X_2$ are slightly correlated, as shown in 
Tab.~\ref{tab:yieboth}; this correlation (which does not exceed $20\%$)
is always negative, since, at given candidate and background rates, an increase
in reactor 1 yield corresponds to a decrease in reactor 2 yield (and 
{\it vice versa}). $X_1$ and $X_2$ are expected to be the same in the case of 
no-oscillation.
%
\subsection{Neutrino yield versus fuel burn-up}
As previously remarked, the contributions of the main fissile isotopes to
the thermal power change in the course of an operating period; this should 
produce a corresponding decrease in the total 
neutrino counting rate as well as 
a modification in the spectral shape. 
The magnitude of this variation amounts to
$\approx 10\%$ throughout the first cycle of the CHOOZ reactors, as we have 
already shown (see \S\ref{sec:uncert}), which by far exceeds 
the statistical and systematic
accuracy of the neutrino flux. We were then 
forced to follow the dynamics of the
fuel burn-up and to apply daily corrections to the thermal power in order to 
restore the linearity with the positron yield 
(see eqs.(\ref{nfact}),(\ref{weff})). 

\begin{figure}[htb]
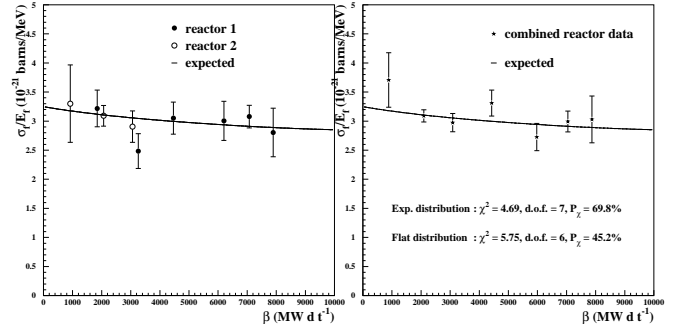

  \begin{center}
    \mbox{\hspace{-0.1cm}\includegraphics[width=0.55\linewidth]{fig.51}
          \hspace{-0.7cm}\includegraphics[width=0.55\linewidth]{fig.52}}
    \caption{\small Variation of the measured neutrino counting rate, as a 
             function of the fuel burn-up for separate (left) and combined 
             (right) reactor data and comparison with predictions. Error bars
             include only statistical uncertainties.}
    \label{fig:yieburn}
  \end{center}
\end{figure}

We set out to establish whether the neutrino counting rate varies with the
reactor burn-up according to predictions, with the purpose of improving the 
reliability and internal consistency of our results. The number $N_\nu$ of 
neutrino events, recorded in a time interval $\Delta t$, by a detector at a 
distance $L$ from the core of a reactor working at a power $W$ can be derived 
from Eq.(\ref{rate:nu}); by inverting we find
\begin{equation}
  \frac{\sigma_f}{E_f} = 4 \pi \frac{L^2}{W \Delta t}
                         \frac{N_\nu}{N_p \varepsilon}
  \label{intrate}
\end{equation}
$\sigma_f$ and $E_f$ being respectively the reaction cross section
(\ref{inv:cross}) and the average energy absorbed in the core in a single 
fission. The ratio $\sigma_f/E_f$, which contains the dependence on the fuel 
composition, can be evaluated for either reactor by considering the runs where 
reactors were alternatively {\it on}; runs are selected 
if the thermal energy is 
$\int W \diff t > 850 \gw \days$ (which is the energy released in one day by
a reactor at $20\%$ of full power). The number of neutrino events is obtained 
by subtracting the number of background events (as determined in the previous 
Section) from the number of candidate events collected in a run. The resulting 
$\sigma_f/E_f$ values are finally grouped in $1000 \mw\days$ burn-up intervals 
and plotted vs. reactor burn-up in Fig.~\ref{fig:yieburn}(left).

A more significant test is performed by using 
combined reactor information. Each
run is assigned an average burn-up defined as
\begin{equation}
  \overline{\beta} \equiv \frac{\beta_1 W_1/L_1^2 + \beta_2 W_2/L_2^2}{W_1/L_1^2
                    + W_2/L_2^2};
  \label{avebur}
\end{equation}
where $\beta_1$ and $\beta_2$ are the burn--up values for the two
reactors.

Similarly, $\sigma_f/E_f$ is obtained by
\begin{equation}
  \frac{\sigma_f}{E_f} = 4 \pi \big(\frac{L_1^2}{W_1 \Delta t} +
                   \frac{L_2^2}{W_2 \Delta t} \big)
                         \frac{N_\nu}{N_p \varepsilon}
  \label{averate}
\end{equation}
The resulting values are plotted in Fig.~\ref{fig:yieburn}(right)
and compared with expectations; the compatibility
is excellent, since $\chi^2 = 4.69$ with $7$
d.o.f., corresponding to $69.8\%$ $\chi^2$-probability. The data compatibility
with a flat distribution is also good ($\chi^2 = 5.75$ with $6$
d.o.f., $P_{\chi^2} = 45.2\%$), but lower than in the previous comparison.
\subsection{Neutrino direction}
The use of reaction (\ref{invbet}) to detect low energy antineutrinos in large
volume scintillator detectors is an ideal tool for measuring the antineutrino 
energy spectrum. Moreover, it provides a good determination of the antineutrino
incoming direction~\cite{venice,Beacom}. This determination
is based on the neutron boost in the forward direction, as a result of the 
kinematics of the reaction; the neutron then retains a memory of the
source direction, which survives even after collisions with the protons in the
moderating medium.

\begin{figure}[htb]
  \begin{center}
    \mbox{\includegraphics[width=0.7\linewidth]{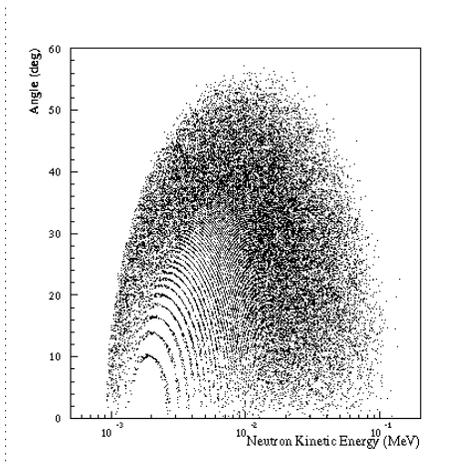}}
    \caption{\small Neutron emission angle (with respect to the incident 
      $\Pagne$ direction) vs. its kinetic energy; the discrete structure of 
      lower-left part of the picture is an effect of the logarithmic scale for 
      the abscissa combined with the $\Pagne$ energy binning.} 
    \label{fig:neuang}
  \end{center}
\end{figure}

In CHOOZ we exploited this neutron recoil technique to locate the reactor 
direction\cite{nudir}, with the twofold objective of 
testing our event reconstruction method
and tuning our Monte Carlo simulations of the slowing down and capture of 
neutrons in the scintillator. We then studied a possible extension of this 
technique to much larger scintillation detectors, such as Kamland, and their 
capabilities in locating astrophysical neutrino sources, such as Supernov\ae. 
%
The neutron emission angle with respect to the incident 
neutrino direction is limited to values 
below $\sim 55^\circ$; this results from
Fig.~\ref{fig:neuang}, where the neutron angle $\theta_\Pn$ is plotted vs. the
neutron kinetic energy $T_\Pn$ (which extends up to $\sim 100\kev$). Moreover,
the neutron moderation maintains some memory of the initial neutron 
direction~\cite{Amaldi}: in each elastic scattering the 
average cosine of the outgoing neutron is $\langle \cos\theta_\Pn \rangle= 2/3 
A$, $A$ being the mass number of the scattering nucleus. The direction is
thus best preserved by collisions on protons, which is also the most effective
target nucleus at energies below $1\mev$. The neutron mean free path 
rapidly reduces during moderation, since the scattering cross section rapidly 
increases at increasingly lower neutron energies; so the bulk of the neutron
displacement is due to the first two or three collisions. The isotropic 
diffusion of thermalized neutrons does not affect the average neutron 
displacement along the neutrino direction. 

The average neutron displacement in CHOOZ has been calculated 
to be $1.7\cm$. Since the
experimental position resolution is $\sigma_x\approx 19\cm$ for the neutron and
the collected neutrino statistic is $\approx 2500$, the precision of the method
is $\approx 0.4\cm$; so neutron displacement can be observed at 
$\sim 4\sigma$ level. The average direction (in spherical coordinates) of the 
two reactors in the CHOOZ detector frame was measured by standard surveying 
techniques to be $\phi = (-50.3 \pm 0.5)^\circ$ and 
$\theta = (91.5\pm 0.5)^\circ$.

The neutrino direction resulting from the data and
the associated uncertainty, is presented in Tab. \ref{tab:avedir}; 
the Monte Carlo
predictions, for a sample with the same statistics, is also listed for 
comparison. The measured direction has a $16\%$ probability of being compatible
with the expected value, while the probability of a fluctuation of an 
isotropic distribution is negligible.
\begin{table}[htb]
    \caption{\small Measurement of neutrino direction: data and Monte Carlo.
	The angle $\delta$ is the $1\sigma-$uncertainty on $\Pagne$ direction.}
    \label{tab:avedir}
  \begin{center}
    \begin{tabular}{|c|c|c|c|}
      \hline
       & $\phi$ & $\theta$ & $\delta$ \\
      \hline
       Data & $-70^\circ$ & $103^\circ$ & $18^\circ$ \\ 
       MC   & $-56^\circ$ & $100^\circ$ & $19^\circ$ \\ 
      \hline
    \end{tabular}
  \end{center}
\end{table}
The average neutron displacement was found to be $(1.9\pm 0.4)\cm$, in 
agreement with what expected.
\section{Neutrino oscillation tests}
Three different approaches were adopted to analyse the results of the 
experiment. The first one (``analysis A'' in what follows) uses the predicted 
positron spectrum in addition to the measured spectra 
for each reactor. This approach uses the absolute normalization 
and is therefore sensitive to the 
mixing angle even for large $\dmsq$
values, where the oscillation structure could no longer be resolved in the 
energy spectrum and the oscillation 
limits exclusively depend on the uncertainty
in the absolute normalization.

The second approach (``analysis B'') uniquely relies on
a comparison of measurements taken at different reactor core--detector 
distances. We are led to limits on the oscillation parameters which 
are practically independent of the uncertainties in the reactor antineutrino 
flux and spectrum; other major sources of systematic uncertainties, such as 
detection efficiencies and the reaction cross section, also cancel out. The 
result of this analysis can thus be regarded as free from all these systematic 
uncertainties.

The third approach (``analysis C'') is somewhat intermediate between the first
two analyses. It uses the shape of the predicted positron spectrum, while
leaving the absolute normalization free. The only contribution to the systematic
shape uncertainty comes from the precision of the neutrino spectrum extraction 
method~\cite{Schreck1}.
\subsection{Global test (Analysis A)}
In a simple two-neutrino oscillation model, the expected positron yield 
for the k-th reactor and the j-th energy spectrum bin, can be 
parametrized as follows:
\begin{multline}
  \overline{X}(E_j,L_k,\theta,\dmsq) = \tilde{X}(E_j) 
  \overline{P}(E_j,L_k,\theta,\dmsq), \\
  (j=1,\ldots,7 \quad k=1,2)
  \label{xosc}
\end{multline}
where $\tilde{X}(E_j)$ is the distance-independent positron yield in the absence
of neutrino oscillations defined in the previous section, $L_k$ is the 
reactor-detector distance and the last factor represents the survival 
probability averaged over the energy bin and the finite detector and reactor 
core sizes. The procedure to compute such a probability at varying oscillation
parameters is similar to the method
used to calculate the burn-up corrections to the positron yields (see 
\S\ref{sec:nuyie}); the positron 
spectrum is obtained by Eq.(\ref{posexp}) after
adding the detector response function and the size function; the
same procedure was applied to obtain the spectrum for no oscillations; the 
probability $\overline{P}(E_j,L_k,\theta,\dmsq)$ results then from the 
ratio of the j-th bin contents of the two spectra. 

In order to test the compatibility of a certain oscillation hypothesis $(\theta,
\dmsq)$ with the measurements, we built a $\chi^2$ statistic 
containing the experimental yields for the 7 energy bins
at the two positions $L_k$  
are listed in Tab.~\ref{tab:yieboth}. We grouped 
these values into a 14-element array $X$ arranged as follows:
\begin{equation}
  \vec{X} = (X_1(E_1),\ldots,X_1(E_7),X_2(E_1),\ldots,X_2(E_7)),
  \label{xarray}
\end{equation}
and similarly for the associated variances. These components are not 
independent, as already noted in \S\ref{sec:nyield}
By combining the statistical
variances with the systematic uncertainties related to the neutrino spectrum,
the $14\times 14$ covariance matrix can be written in a compact form as follows:
\begin{multline}
  V_{ij} = \delta_{i,j}(\sigma_i^2 + \tilde{\sigma}_i^2) + 
           (\delta_{i,j-7} + \delta_{i,j+7})\sigma^{(i)}_{12}\\
              (i,j=1,\ldots,14),
  \label{cormat}
\end{multline}
where $\sigma_i$ are the statistical errors associated with the yield array
(\ref{xarray}), $\tilde{\sigma}_i$ are the systematic uncertainties and 
$\sigma^{(i)}_{12}$ are the covariance of reactor 1 and 2 yield contributions
to the i-th energy bin (see Tab.~\ref{tab:yieboth}). These systematic errors,
including the statistical error on the measured $\beta$-spectra measured at 
ILL~\cite{Schreck1} as well as the bin-to-bin systematic error inherent in
the conversion procedure, range from $1.4\%$ at $2\mev$ (positron energy) to 
$7.3\%$ at $6\mev$ and are assumed to be uncorrelated\footnote{
The extraction of the neutrino spectra from the $\beta$ measurement at ILL 
should introduce a slight correlation of the bin-to-bin error systematic error.
Nevertheless, the overall uncertainties on the positron yields are dominated by
statistical errors, so that neglecting the off-diagonal systematic error
matrix does not affect the oscillation test significantly. Also previous
reactor experiments, even with much lower statistical errors, did not take this
correlation into account.}.

\begin{figure}[htb]
  \begin{center}
    \mbox{\includegraphics[width=0.9\linewidth]{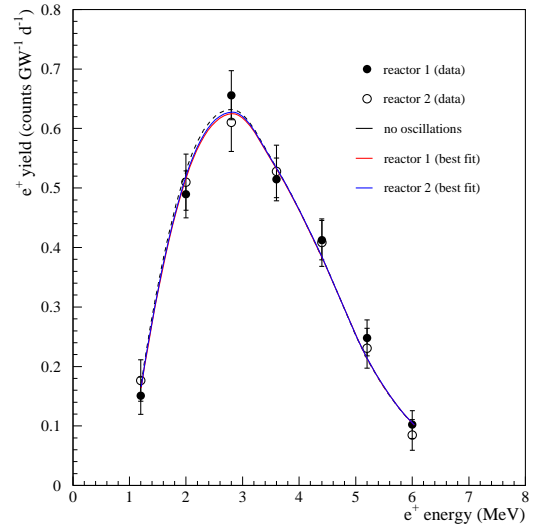}}
    \caption{\small Positron yields for reactor 1 and 2; the solid curves
             represent the predicted positron yields corresponding to the
             best-fit parameters, the dashed one to the predicted yield for the
             case of no oscillations.}
    \label{fig:yiemin}
  \end{center}
\end{figure}

We still have to take into account the systematic error related to the
absolute normalization; combining all the contributions listed in 
Tab.~\ref{tab:sysnorm}, we obtain an overall normalization uncertainty of
$\sigma_\alpha = 2.7\%$. We may define the following $\chi^2$ 
function
\begin{multline}
  \chi^2 \bigl(\theta,\dmsq,\alpha,g \bigr) = \\
   \sum_{i=1}^{14} \sum_{j=1}^{14}
  \Bigl( X_i - \alpha \overline{X} \bigl( g E_i,L_i,\theta,\dmsq \bigr)
  \Bigr) \\ V_{ij}^{-1}
  \Bigl( X_j - \alpha \overline{X} \bigl( g E_j,L_j,\theta,\dmsq \bigr)
  \Bigr) + \\
  \left( \frac{\alpha-1}{\sigma_\alpha} \right)^2 + \left( \frac{g-1}{\sigma_g}
  \right)^2,
  \label{chiA}
\end{multline}
where $\alpha$ is the absolute normalization constant, $g$ is the 
energy-scale calibration factor, $L_{i,j}= L_1 \text{ for } i,j\leq 7$ and 
$L_{i,j}= L_2 \text{ for } i,j> 7$. The 
uncertainty on $g$ is $\sigma_g=1.1\%$, resulting from the
accuracy on the energy scale calibration ($16\kev$ at the $2.11\mev$ visible 
energy line associated with the n-capture on Hydrogen) and the $0.8\%$ drift in 
the Gd-capture line, as measured throughout the acquisition period with 
high-energy spallation neutrons (see Fig.~\ref{fig:enestab}). 
\begin{table}[htb]
  \caption{\small Contributions to the overall systematic uncertainty on the 
    absolute normalization factor.}
  \label{tab:sysnorm}
  \begin{center}
    \begin{tabular}{lc}
      \hline
      parameter & relative error $(\%)$ \\
      \hline
      reaction cross section & $1.9\%$ \\
      number of protons & $0.8\%$ \\
      detection efficiency & $1.5\%$ \\
      reactor power & $0.7\%$ \\
      energy released per fission & $0.6\%$ \\
      \hline
      combined & $2.7\%$ \\
      \hline
    \end{tabular}
  \end{center}
\end{table}
The $\chi^2$ in (\ref{chiA}) thus contains $14$ 
experimental errors with 2 additional parameters, yielding altogether $16$
variances. The $\chi^2$ value for a certain parameter set 
$(\theta,\dmsq)$ is determined by minimizing (\ref{chiA}) with respect to 
the gain factor $g$ and to the normalization $\alpha$; the minimization then 
leads to $12$ degrees of freedom. The minimum value $\chi^2_{min} = 5.0$, 
corresponding to a $\chi^2$ probability $P_{\chi} = 96\%$, is 
found for the parameters $\sin^2(2\theta) = 0.23$, $\dmsq=8.1\cdot 10^{-4}
\ev^2$, $\alpha = 1.012$, $g=1.006$; the resulting positron yields are shown 
by solid lines in Fig.~\ref{fig:yiemin} and superimposed on the data.
Also the no-oscillation hypothesis, with
$\chi^2(0,0)=5.5$, $\alpha=1.008$ and $g=1.011$, is found to perfectly fit  
with the data ($P_\chi = 93\%$).

To test a particular oscillation hypothesis ($\theta$, $\dmsq$) against the
parameters of the best fit and to determine the $90\%$ confidence belt, we 
adopted the Feldman \& Cousins prescription~\cite{Feldman}. 
The ``ordering'' principle is 
based on the logarithm of the ratio of the likelihood functions for the two 
cases:
\begin{equation}
  \lambda(\theta,\dmsq) = \chi^2(\theta,\dmsq) - \chi^2_{min}
  \label{loglikrat}
\end{equation}
where the mimimum $\chi^2$ value must be searched for within the physical 
domain ($0<\sin^2(2\theta)<1$, $\dmsq>0$).
Smaller $\lambda$ values imply better compatibility of the hypothesis with the 
data. The $\lambda$ distribution for the given parameter set was evaluated by 
performing a Monte Carlo simulation of a large number (5000) of experimental 
positron spectra whose values are scattered around the predicted positron yields
$\overline{X}(E_i,L_i,\theta,\dmsq)$ with Gaussian-assumed variances 
$\sigma_i$ and correlation coefficients given by (\ref{cormat}). For each set
we extracted the quantity $\lambda_c(\theta,\dmsq)$ such that $90\%$ of the
simulated experiments have $\lambda<\lambda_c$. The $90\%$ confidence domain 
then includes all points in the $(\sin^2(2\theta),\dmsq)$ plane such that
\begin{equation}
  \lambda_{exp}(\theta,\dmsq) < \lambda_c(\theta,\dmsq),
  \label{clord}
\end{equation}
where $\lambda_{exp}$ is evaluated for the experimental data for each point in 
the physical domain. 

The acceptance domain at the $90\%$~C.L. (solid line) and $95\%$~C.L. are shown
in Fig.~\ref{fig:exclplotC}; all 
the parameters lying to the right of the curves are 
excluded by CHOOZ with the indicated confidence level, while the parameter
regions on the left are still compatible with our data. The region allowed by 
Kamiokande for the $\Pgngm\rightarrow \Pgne$ oscillations is also shown for 
comparison; this hypothesis, a possible explanation for the $\Pgngm$ deficit in
the atmospheric neutrino flux, is excluded. The $\dmsq$ limit at full
mixing is $7\cdot 10^{-4} \ev^2$; the limit for the mixing angle in the 
asymptotic range of large mass differences is $\sin^2(2\theta) = 0.10$.
%
%
\subsection{Two-distance test (analysis B)}
\begin{figure}[htb]
  \begin{center}
    \mbox{\includegraphics[width=0.9\linewidth]{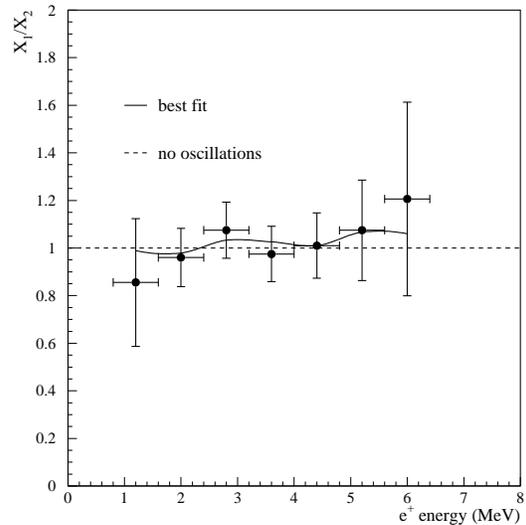}}
    \caption{\small Measured ratio of experimental positron yield, compared with
     the predicted ratio in the best oscillation hypothesis (solid line) and 
      in the case of no oscillations (dashed line).}
    \label{fig:ratmin}
  \end{center}
\end{figure}

%
The predicted ratio of the two--reactor positron yields equals the ratio of
the corresponding survival probabilities. At full mixing ($\sin^2(2\theta) = 
1$) and at low mass differences ($\dmsq \approx 10^{-3}\ev^2$), this ratio
can be approximated by
\begin{multline}
  \overline{R} \approx \big[ 1 - \big( \frac{1.27 \dmsq L_1}{E_\nu} \big)^2
  \big] \big[ 1 + \big(\frac{1.27 \dmsq L_2}{E_\nu} \big)^2 \big]\\
  \approx
  1 - 2 \big(\frac{1.27 \dmsq}{E_\nu}\big)^2 L \delta L,
  \label{ratprob}
\end{multline}
where $L$ is the average reactor-detector distance and $\delta L$ is the 
difference of the two distances. Therefore an experiment 
which measures this ratio with
an uncertainty $\sigma$ is sensitive to 
oscillations down to mass-difference values as low as
\begin{equation}
  \dmsq \approx \frac{E_\nu}{1.27} \sqrt{\frac{k\sigma}{2 L \delta L}},
  \label{dmlow}
\end{equation}
$k$ being the number of standard deviations corresponding to the chosen 
confidence level.
This value can be compared
to the sensitivity limit $\dmsq_0$ inherent in analysis A by 
noting that
\begin{equation}
  \dmsq \approx \sqrt{\frac{L}{2\delta L}}\dmsq_0\approx 2\dmsq_0
  \approx 1.5 \cdot 10^{-3} \ev^2
  \label{dm0}
\end{equation}
Although twice as large, this limit is lower than the lowest $\delta
m^2$ value allowed by Kamiokande (see Fig.~\ref{fig:exclplotC}). 
%
The ratio $R(E_i)\equiv X_1(E_i)/X_2(E_i)$ of the measured positron yields must 
be compared with the expected values; since the expected 
yields are the same for 
both reactors in the case of no-oscillations, 
the expected ratio for the i-th energy bin reduces to the 
ratio of the average survical probability in that bin. We can then 
build the following $\chi^2$ statistic:
\begin{equation}
  \chi^2 = \sum_{i=1}^7 \left( \frac{R(E_i)-\overline{R}(E_i,\theta,\dmsq)}{
    \delta R(E_i)} \right)^2
  \label{chiB1}
\end{equation}
where $\delta R(E_i)$ is the statistical uncertainty on the measured ratio. The
minimum $\chi^2$ value ($\chi^2_{min} = 0.78$ over 5 d.o.f.) occurs at 
$\sin^2(2\theta)=1$ and $\dmsq=0.6\ev^2$; the compatibility 
of the no-oscillation 
hypothesis is still excellent (see Fig.~\ref{fig:ratmin}), as 
$\chi^2(0,0) = 1.29$.

We adopted the same procedure described 
in the previous section to determine the
confidence domain in the $(\sin^2(2\theta),\dmsq)$ 
plane and for each point in
this plane we simulated the results of 5000 
experiments. 
If the 
positron yields of both reactors are Gaussian distributed around the predicted 
values $\overline{X}_{1,2}$, and if $\overline{X}_2/\sigma_2$ is sufficiently 
large so that the $X_2$ generated have only positive values, then the variable
\begin{equation}
  Z = \displayfrac{\frac{\overline{X}_1}{X_1}- \frac{\overline{X}_2}{X_2}}{
      \sqrt{\frac{\sigma_1^2}{X_1^2}+\frac{\sigma_2^2}{X_2^2} -
      2\frac{\sigma_{12}}{X_1 X_2}}} = 
      \frac{\overline{X}_1 - \overline{X}_2 R}{\sqrt{\sigma_1^2 +
          \sigma_2^2 R^2 -2\sigma_{12} R}}
  \label{zr}
\end{equation}
is normally distributed with zero mean and unit variance~\cite{Eadie}. 
We could then
use a normal random generator to extract $Z$ and invert (\ref{zr}) to determine
the ratio $R(E_i)$.

The exclusion plot obtained at $90\%$~C.L. is also shown in 
Fig.~\ref{fig:exclplotC}. Although less powerful 
than the previous analysis, the region excluded by this oscillation test almost
completely covers the one allowed by Kamiokande.
\subsection{Shape test (Analysis C)}
This test is similar to analysis A, the only difference being related
to the hypothesis on the absolute normalization. In analysis A we fixed the
integral counting rate to be distributed around the predicted value ($\alpha = 
1$), with $\sigma_\alpha = 2.7\%$ systematic uncertainty; in the shape test,
on the contrary, we gave up any constraint on the 
normalization parameter (which is equivalent to 
having $\sigma_\alpha= \infty$).
The $\chi^2$ statistic for this test is the same expression as (\ref{chiA})
without the term depending on the normalization, so we can write
\begin{multline}
  \chi^2(\theta,\dmsq,\alpha,g) = \\ 
  \sum_{i=1}^{14} \sum_{j=1}^{14}
  \Bigl(X_i - \alpha \overline{X}\bigl(g E_i,L_i,\theta,\dmsq\bigr)\Bigr) 
  \\ V_{ij}^{-1} 
  \Bigl(X_j - \alpha \overline{X}\bigl(g E_j,L_j,\theta,\dmsq\bigr)\Bigr) 
  + \\ \left( \frac{g-1}{\sigma_g} \right)^2,
  \label{chiC}
\end{multline}

\begin{figure}[htb]
  \begin{center}
    \mbox{\includegraphics[width=\linewidth]{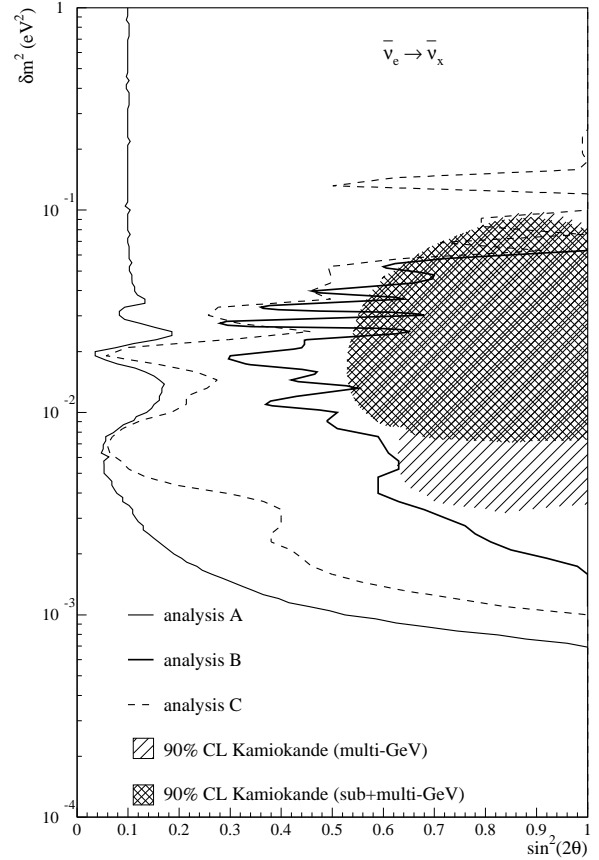}}
    \caption{\small Exclusion plot contours at $90\%$~C.L. obtained by the 
three analyses presented. Analysis A uses both shape and normalization of the
background-subtracted positron spectrum; analysis B uses the baseline 
difference between the two reactors; analysis C uses the spectrum shape. The
Kamiokande $\Pgngm \leftrightarrow \Pgne$ allowed region to atmospheric 
neutrino oscillations is also shown for comparison.}
    \label{fig:exclplotC}
  \end{center}
\end{figure}

This $\chi^2$ has a minimum value $\chi^2_{min}=2.64$ (over 11 degrees of 
freedom) at $\sin^2(2\theta) = 0.23$, $\dmsq = 2.4\cdot 10^{-2} \ev^2$ and
$g = 1.008$; the null hypothesis gives instead $\chi^2(0,0) = 5.5$ with
$g=1.006$. The exclusion plot, obtained according to the Feldman--Cousins
prescriptions, is shown in Fig.~\ref{fig:exclplotC} 
and compared to the results 
of the other tests. As in the case of the two-distance test 
it is not sensitive to oscillations
at large squared mass-difference values ($\dmsq 
\gsim 0.1 \ev^2$), where the oscillation length
\begin{equation}
  L_{osc}(\m) = \frac{2.48 E_\nu(\mev)}{\dmsq(\ev^2)} \,\lsim\, 100\m
\end{equation}
becomes much lower than the average reactor--detector distance. The 
$\sin^2(2\theta)$ limit at the maximum oscillation probability is similar to 
that obtained with analysis A.
\subsection{Implications of the CHOOZ results}
The importance of the CHOOZ results on neutrino oscillations has been pointed 
out by many 
authors~\cite{Sbilenky,Fogli,Perkins,Barbieri,Gonzales,Petcov,Smirnov}. In a 
three-flavour neutrino mixing frame, the see-saw mechanism indicates a mass 
hierarchy ($m_1 \ll m_2 \ll m_3$) from which
\begin{equation}
\delta m_{12}^2 \ll \delta m_{23}^2.
\label{dege}
\end{equation}
\begin{figure*}[htb]
  \begin{center}
    \mbox{\includegraphics[bb=70 140 510 760,width=0.7\linewidth]{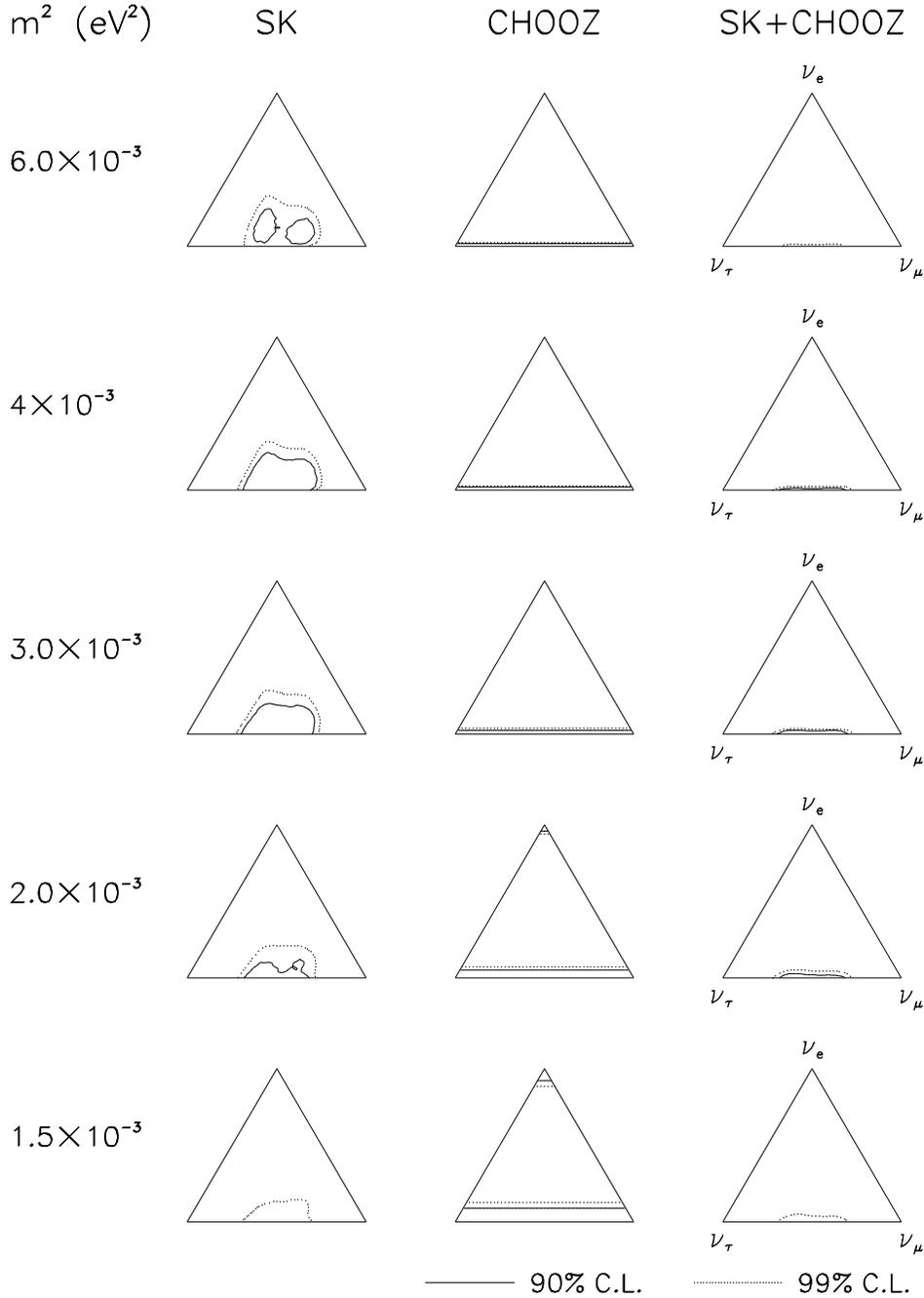}}
    \caption{\small Results of a three flavour mixing analysis of separate and
      combined Super-Kamiokande and CHOOZ data, for five representative values
      of $m^2 \equiv \delta {m^2}_{23}$. The analysis concerns the 79.5 
      kTy data sample for Super-Kamiokande and the first CHOOZ result.}
    \label{fig:fogli}
  \end{center}
\end{figure*}
The only available possibile explanation for solar and atmospheric
neutrino anomalies through neutrino oscillations is that $\delta m_{12}$ is
relevant for the transitions of solar neutrinos and $\delta m_{13}$ is the one 
probed by atmospheric and L-B accelerator neutrino experiments. By 
approximation (\ref{dege}) it is possible to show that the CP-violating phase 
in the $3\times 3$ mixing matrix does not give rise to observable effects, and 
that the mixing angle $\theta_{12}$ (associated with the lower mass states) can
be rotated away in the atmospheric neutrino analysis. Therefore, condition 
(\ref{dege}) implies that the transition probability in atmospheric and L-B
experiments depends only on the largest mass squared difference $\delta m_{23}$
and the elements $U_{\alpha 3}$ connecting flavour neutrinos with $\nu_3$; it
assumes then the simple form 
\begin{equation}
  P_{\nu_\alpha \rightarrow \nu_\beta} = 
  \begin{cases}
    1 - 4 U_{\alpha 3}^2 (1-U_{\alpha 3}^2)
    \sin^2\left(\displayfrac{1.27 \dmsq(\ev^2) L(\m)}{
      E(\mev)}\right) \\ \text{if $\alpha = \beta$}, \\
    4 U_{\alpha 3}^2 U_{\beta 3}^2
    \sin^2\left(\displayfrac{1.27 \delta m^{2}(\ev^2) L(\m)}{
      E(\mev)}\right) \\ \text{if $\alpha \neq \beta$}.
  \end{cases}
  \label{probatm}
\end{equation}
In particular, the survival probability reduces to the usual two-flavour 
formula (\ref{prob:trans}) with $\sin^2 2\theta = 4 U_{\alpha 3}^2 
(1-U_{\alpha 3}^2)$. Therefore, information on the parameter $U_{e3}$ can be
obtained from the CHOOZ exclusion plot. The upper limit for $\sin^2 2\theta$ 
implied by the exclusion plot in Fig.~\ref{fig:exclplotC} is $\sin^2 2\theta$
$\lsim 0.1$ for $\delta m_{13}$ $\gsim 2\cdot 10^{-3}$; it follows that
\begin{equation}
  U_{e3}^2 < 0.03 \qquad {\text or} \qquad U_{e3}^2>0.97
\label{ue3}
\end{equation}
Large values of values of $U_{e3}$, those allowed by the second inequality in
Eq.(\ref{ue3}), are excluded by the solar neutrino data~\cite{Sbilenky};
the solar $\Pgne$ survival probability would be larger than $0.95$, which is
incompatible with the deficit observed in all solar neutrino experiments. The 
CHOOZ result then constrains the mixing of electron neutrinos with all other 
flavours, in the atmospheric and L-B range, to small values.

It has been remarked~\cite{Fogli} that even the most
recent Super-Kamiokande data alone do not completely 
exclude sizable $\Pgne$ mixing. This can be seen in Fig.~\ref{fig:fogli}, where
the confidence regions for separate and combined Super--Ka\-mio\-kan\-de 
and CHOOZ 
data are shown for different values of $m^2 \equiv \delta {m^2}_{23}$ (the 
mass difference relevant for atmospheric neutrinos). Each set of mixing 
parameters $(U_{e3},U_{\mu 3}, U_{\tau 3})$ 
is associated with a point embedded 
in a triangle graph (equilateral due to the unitarity condition), whose corners
represent the flavour eigenstates; by convention $\Pgne$ is assigned the upper
corner. The distance from the three sides are equal
to $U_{e3}^2,\,U_{\mu 3}^2,\,U_{\tau 3}^2$; in a two-flavour scheme 
($U_{\alpha 3} = 0$), the point is bound on the side connecting the two mixed
flavour states (and the mean point of that side is associated with the maximum
mixing hypothesis).

The CHOOZ result excludes large horizontal stripes in the triangle plot, 
according to Eq.(\ref{probatm}); the stripe becomes 
increasingly narrower at lower 
$\delta {m^2}_{23}$ values, as a consequence of the reduced sensitivity to 
$\sin^2 2\theta$. The Super-Kamiokande allowed region lies on the triangle base
and protrudes towards the centre, 
which implies a non-negligible $\Pgngm \rightarrow \Pgne$ oscillation 
probablility. Yet in the combined analysis graph the allowed region is 
significantly flattened on the base, thus indicating a dominance of the 
$\Pgngm \rightarrow \Pgngt$ maximum mixing hypothesis.

\begin{figure}[htb]
  \begin{center}
    \mbox{\includegraphics[bb=188 331 414 606,clip='true',
     width=\linewidth]{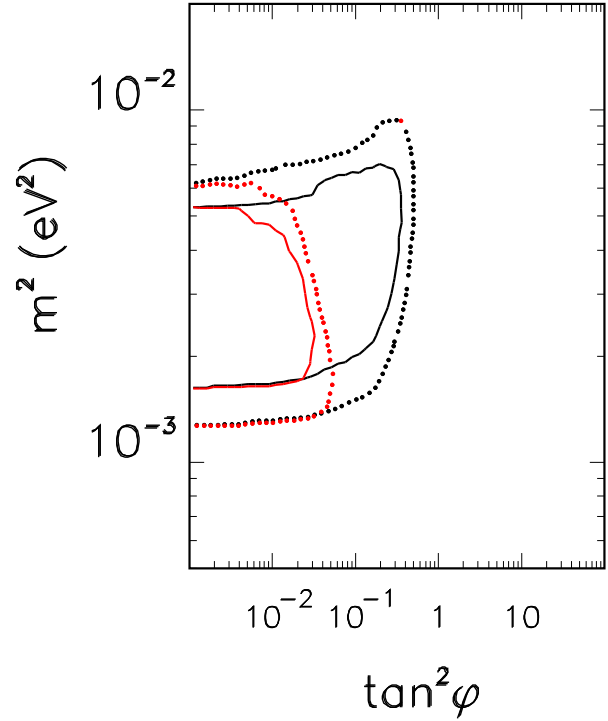}}
    \caption{\small 90\% (solid line) and 99\% C.L. (dotted line) limits in the
          $m^2$ ($\equiv \delta {m^2}_{23}$), $tan^{2}\phi$
          ($\phi \equiv \theta_{13}$) parameter space obtained with
          the use of Super--Kamiokande alone (less restrictive limits) 
         and CHOOZ +Super--Kamiokande.}
    \label{fig:fogli2}
  \end{center}
\end{figure}

The impact of CHOOZ in constraining the $\Pgne$ mixing is more evident
in the bilogarthmic plot of Fig.~\ref{fig:fogli2} where the allowed 
regions (90 and 99 \% C.L.) in the  
$m^2$ ($\equiv \delta {m^2}_{23}$), $tan^{2}\phi$
($\phi \equiv \theta_{13}$) parameter space are shown for the Super--Kamiokande
data only (less restrictive limits) and for the combined 
Super--Kamiokande and CHOOZ data. The improvement obtained by using
the CHOOZ data is of about one order of magnitude.
\def\Pagnx{\relax\ifmmode{\overline{\nu}_{\rm x}}%
            \else$\overline{\nu}_{\rm x}$\fi}%
\def\ssq{\relax\ifmmode{\sin^2(2  \theta)}%
            \else$\sin^2(2\theta)$\fi}%
\def\dm{\relax\ifmmode{\Delta     m^2}%
            \else$\Delta m^2$\fi}%
\renewcommand{\vec}[1]{\boldsymbol{#1}}
\newcommand{\eg}{{\sl e.g.}\ }
\section{A novel method for the derivation of confidence regions}
The derivation of confidence regions for parameters, 
in neutrino oscillation 
experiments, has been the focus of attention and debate in recent years.
There has been a general consensus in applying the Feldman and Cousins method
(FC), which, apart from its intrinsic merit, allows a simpler 
comparison between experiments.
Discussions about the statistical analysis are far from 
conclusive. We therefore tested an interesting new 
scheme which may represent a further step towards an objection-free procedure.
It tends to produce confidence regions which are 
somewhat larger in size and, similarly, its 
application would more or less  affect the results of all experiments. 
The new prescription~\cite{Punzi} in confidence interval estimate
is based on the concept of {\sl strong Confidence Level} ({\sl sCL}), to make 
an inference 
on neutrino oscillation parameters. This new method is purely frequentist,
just like the FC ``unified approach'', yet it seems to be free from any 
pathologies inherent in every frequentist procedure proposed to date. Moreover,
the definition of strong limits can be generalized in a natural way to
include exact frequentist treatment of systematics~\cite{SCL}. 
Previous limits (Fig.~\ref{fig:exclplotC}) were computed by replacing the 
likelihood function with the one obtained by maximizing with respect to the
systematic parameters (``profile'' likelihood). This is an 
approximation~\cite{Kendall} often used to overcome computational difficulties.
It has been shown~\cite{SCL} that this approximation no longer  holds 
in cases where the systematic error is of the same 
order of magnitude of the statistical one. 
\subsection{Application of the method}
In this method we need to construct 
the probability density function ({\sl pdf}) of
our experiment in terms of the data (energy spectra) and the parameters 
(oscillation + normalization). For practical reasons, we decided
to condense all the information (shape + normalization)
contained in a spectrum $\vec{X}$ into the three parameters
($\hat{\ssq}$, $\hat{\dmsq})$, $\hat{\alpha}$) for which the likelihood is maximum;
any parameter set uniquely identifies a particular energy spectrum.
The set of maximum likelihood estimators (MLE) is our observable.

In order to obtain strong confidence limits,
we computed the probability density function 
numerically. The domain of oscillation parameters  ($0  < \ssq  < 1$,
$10^{-4}$ $< \dmsq  <$ $1 \ev^2$) was sampled by a $100 \times$  $100$ grid
(with a constant binning in $\log \dmsq$);  the range  $0.95$ $ < \alpha <$
$1.05$ was divided into 10 cells%
\footnote{We verified that  values of $\alpha$ outside the  considered
range give no further  contribution to the  projection of the band  on
$(\ssq,\dmsq)$-space.}. For each set of
($\ssq$, $\dmsq$, $\alpha$) we generated $5\cdot 10^4$ spectra scattered around
$\alpha \vec{\overline{X}}(\ssq,\dmsq)$ with uncertainties given by the
covariance matrix. MLE for each experiment are searched for in the same
space spanned by the parameters but with a coarser grid ($20\times 20
\times 10$).
\begin{figure}[htb]
  \includegraphics[width=1.\columnwidth]{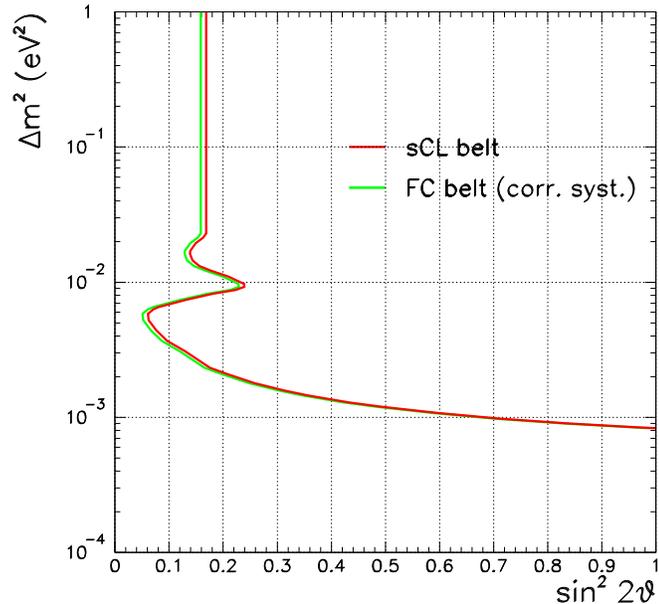}
  \caption{Exclusion plot at $90$\% sCL for the oscillation parameters
    based on the differential energy spectrum; the FC contour,
    obtained with ``correct systematics'' treatment, is also shown.}
  \label{fig:mlx}
\end{figure}

The results of our computation are shown in Fig.~\ref{fig:mlx}. The
confidence bounds obtained are significantly higher than those obtained
by the procedure explained in the previous section. In fact, oscillations
$\Pagne \rightarrow \Pagnx$ are excluded for  $\dmsq \geq 8\cdot 10^{-4}
\ev^2$ at maximum mixing and $\ssq \geq 0.17$ at large $\dmsq${} values.
It should be noted that the limits quoted are only slightly looser than 
those obtained by using the FC prescription with the correct 
inclusion of systematics, as shown in Fig.~\ref{fig:mlx}.
\section{Conclusions}
The CHOOZ experiment stopped taking data 
in July 1998, about 5 years after the submission of the proposal for approval.
With more than 1-year data taking, the statistical error ($2.8\%$) on the 
neutrino flux matched the goal ($3\%$) of the proposal.
Accurate estimatea of the detection efficiencies 
as well as precise measurements of the detector parameters also allowed us to 
keep the systematic uncertainty ($2.7\%$) below expectations ($3.2\%$).

We found (at $90\%$ confidence level) no evidence for neutrino oscillations 
in the $\Pagne$ disappearance mode, for the parameter region given 
by approximately  $\dmsq > 7 \cdot 10^{-4}\ev^2$ for maximum mixing, 
and $\sin^2 2\theta = 0.10$ for large $\dmsq$. Less sensitive results, 
based only on the comparison of the positron spectra from the two
different-distance nuclear reactors (and therefore independent of the absolute
normalization of the $\Pagne$ flux, the number of protons and the detector 
efficiencies) were also presented. 

Our result does not allow the atmospheric neu\-tri\-no a\-no\-ma\-ly to be 
explained in terms of $\Pgngm \rightarrow \Pgne$ oscillations, 
thus leaving, in a 
three-flavour mixing scheme, the $\Pgngm \rightarrow \Pgngt$ possibility. 

Many cross-checks were performed on the data to 
test the internal consistence 
and improve the reliability of our results. As a by-product, we have
shown that the use of reaction (\ref{invbet}) allowed us to 
locate the $\Pagne$ source within a cone of half-aperture $\simeq 18^\circ$ at
$68\%$ confidence level. 
\begin{acknowledgement}
Construction of the laboratory was funded by \'Electricit\'e de France
(E.D.F.). Other work was supported in part by IN2P3--CNRS (France), INFN
(Italy), the United States Department of Energy, and by RFBR (Russia).
We are very grateful to the Conseil G\'en\'eral des Ardennes for 
providing us with the facilities for the experiment. At
various stages during the organization and management of the experiment, we
were assisted by the efficient staff at SENA (Soci\'et\'e
Electronucl\'eaire des Ardennes) and by the E.D.F. CHOOZ B nuclear plant.
Special thanks to the technical staff of our laboratories for their
excellent work in designing and building the detector.

We would like to thank Prof.~Erno Pretsch and 
his group at ETH Zurich, for some 
precise measurements of the target scintillator Hydrogen content. 
\end{acknowledgement}

\end{document}